\documentclass[]{aa}
\makeatletter
\renewcommand*\aa@pageof{, page \thepage{} of \pageref*{LastPage}}
\makeatother

\newcommand\swi{{\it Swift}}
\newcommand\xrt{{\it Swift}--XRT}

\newcommand\cha{\textit{Chandra}}
\newcommand\xmm{{XMM-{\it Newton}}}

\newcommand\lat{{\it Fermi}-LAT}

\usepackage{natbib}
\usepackage{float}
\usepackage{color}
\usepackage{xcolor}
\usepackage{graphicx}
\usepackage{textcomp,gensymb}
\usepackage{array,amsmath}
\usepackage{mathtools}
\usepackage{multirow,makecell}
\usepackage{enumitem}
\usepackage{longtable}
\usepackage{hyperref}
\usepackage{ltablex}
\usepackage{ulem}

\usepackage[switch,modulo]{lineno}
\keepXColumns

\usepackage{placeins}
\usepackage{pdflscape}
\usepackage{geometry}
\usepackage{booktabs}
\definecolor{amethyst}{rgb}{0.6, 0.4, 0.8}

\begin{document} 

    \title{Constraining the synchrotron peak and estimating the VHE brightness of a sample of extreme high synchrotron peak blazars}

    \author{Federica Sibani\inst{1}, Stefano~Marchesi\inst{1,2,3}, Ettore~Bronzini\inst{1,2}, Marco~Ajello\inst{3},
    Michele~Doro\inst{4}, 
    Lea~Marcotulli\inst{5,3},
    Elisa~Prandini\inst{4},
    Cristian~Vignali\inst{1,2}
    }
    \institute{Dipartimento di Fisica e Astronomia (DIFA) Augusto Righi, Università di Bologna, via Gobetti 93/2, I-40129 Bologna, Italy
    \and INAF-Osservatorio di Astrofisica e Scienza dello Spazio (OAS), via Gobetti 93/3, I-40129 Bologna, Italy
   \and Department of Physics and Astronomy, Clemson University, Kinard Lab of Physics, Clemson, SC 29634, USA
    \and INFN Sezione di Padova and Università degli Studi di Padova, Via Marzolo 8, I-35131 Padova, Italy
    \and Deutsches Elektronen-Synchrotron DESY, Platanenallee 6, 15738 Zeuthen, Germany
               }

\titlerunning{Constraining the synchrotron peak and estimating the VHE brightness of a sample of EHSP blazars}
\authorrunning{Sibani et al.}

\abstract
{We present the results of a multi-wavelength study of a population of X-ray bright ($\rm log(F_{0.2-12 \ keV})>-12.5$), non-$\gamma$-ray detected high and extreme high synchrotron peak (HSP, EHSP; $\rm log(\nu_{\rm peak,\ Hz})>16$) BL Lacs to $i$) put stronger constraints on the synchrotron peak location and shape and $ii$) model their expected behaviour in the very high-energy band. First, we performed an X-ray spectral analysis, using \xmm, \cha, \xrt, and eROSITA data, and fitting the spectra using both a power law and a log parabola model. Out of 78 sources in the initial sample, 17 were best described by a log parabola model, a result that supports a scenario where the synchrotron peak falls in the X-ray band. Among these 17 sources, we further selected the 10 objects dominated by the jet emission, with no significant contamination of the host galaxy. We performed a $\gamma$-ray analysis of \lat\ data for these objects, obtaining upper limits providing information on their flux in the 100 MeV - 300 GeV energy range. We then modelled the broadband SED of these objects with JetSeT using two models: one assuming a log parabola for the electron distribution and the other one with a broken power law electron distribution, using parameters consistent with those describing the emission of the prototypical EHSP 1ES 0229+200. We found the models to be generally consistent with the available multi-wavelength detections and upper limits. Furthermore, they confirmed that a subsample of sources could display relevant emission in the TeV energy range, even potentially reaching the threshold for detectability by the Cherenkov Telescope Array Observatory.
}


\maketitle

\section{Introduction}\label{sec:intro}
Blazars are accreting supermassive black holes (SMBHs), or active galactic nuclei (AGN), whose relativistic jets are pointed in the direction of the observer. This causes a significant enhancement of the source luminosity due to the relativistic speed of the particles causing the emission, an effect known as ``Doppler boosting''. Such objects are usually classified on the basis of their optical features in Flat Spectrum Radio Quasars (FSRQ) and BL Lacs: while the former tend to display strong, broad emission lines ($\rm >5 \ \mathring{A}$), the latter
show at most weak lines and in many cases are actually completely featureless. More in general, the spectral energy distribution (SED) of such objects is characterised by two clear bumps  \citep[e.g.,][]{abdo10}. The one at lower frequencies, known as synchrotron bump, is usually attributed to synchrotron emission caused by the relativistic electrons in the jets. The one at higher frequencies, the so-called inverse Compton (IC) bump, is instead sometimes assumed to be produced by the interaction and subsequent up-scattering in frequency of the synchrotron--produced photons with the same relativistic electrons \citep[in the so--called synchrotron self-Compton scenario; e.g.,][]{kirk98}.
The frequency of these two peaks can vary significantly depending on several characteristics, most prominently the luminosity of the blazars and their optical class \citep[e.g.,][]{padovani95,ghisellini98,giommi99,prandini22}. For example, the synchrotron peak can be found, especially in FSRQs, at frequencies as low as $\rm \nu_{synch}$ = 10$^{12}$ Hz, which is in the far-infrared band \citep[e.g.,][]{chen09}, and, in BL Lacs, as high as $\rm \nu_{synch}$ = 10$^{18}$ Hz, which corresponds roughly to 4.1 keV and belongs to the X-ray band \citep[which is, conventionally, between 0.1 and several hundred keV; see, e.g.,][]{chang19}. 

In these latter extreme objects, namely High Synchrotron peaked BL Lacs \citep[HSP, $\rm \nu_{synch} \ge 10^{15}$ Hz; e.g.,][]{massaro11b,arsioli15,chang17,chang19} or Extreme High Synchrotron peaked BL Lacs (EHSP, $\rm \nu_{synch} \ge 10^{17}$ Hz), the observational evidence supports a scenario where the X-ray emission is correlated with the very high-energy (VHE; $>$ 100 \ GeV) emission. 
A study by \citealt[][]{marchesi25} of all the known 77 VHE blazars reported in the TevCAT\footnote{The TeVCAT is a regularly updated catalogue of sources detected in the VHE band, at energies above 50\,GeV. \url{https://tevcat.org/}} \citep{wakely18} quantified this relation, showing that a direct correlation exists between X-ray and TeV fluxes in the EHSP population, and such correlation is broadly consistent with a 1:1 trend, although with a fairly large dispersion.
As a consequence of this correlation, previous works focused on using blazars X-ray emission to select promising VHE emitters. This is often done by combining the X-ray information with information from at least another band (such as the infrared band – see, e.g., \citealt{massaro13,giommi24, Metzger25}, or the radio band – see, e.g., \citealt{bonnoli15}). So far, however, the overwhelming majority of the sources analysed in these works were known $\gamma$-ray emitters, and in particular were sources detected by the \textit{Fermi} Large Area Telescope (LAT). Indeed, predictions for detections of extragalactic sources in VHE surveys fairly often use as a starting point extrapolations based on \lat\ luminosity functions in the MeV to GeV band. While this is certainly a tested and reliable method to estimate VHE fluxes, it could nonetheless miss a population of sources undetected in the $\gamma$-ray band and still visible at larger energies \citep[e.g.,][]{costamante20}.

A different approach to the study of the VHE emission of currently $\gamma$-ray-undetected blazars could therefore exploit the aforementioned correlation to their X-ray emission in combination with a broadband modelling of their entire SED. Multi-wavelength modelling is an already well-known tool for the study of blazars and AGNs jets, and multiple softwares have been designed to do so \citep[e.g.,][]{tramacere20,klinger24,stathopoulos24}. By reproducing the broadband spectra and temporal behaviour of these sources in a way that can be compared with observational data, it is possible to test the validity of models relying on different acceleration processes, and constrain key physical parameters of the emitting regions  \citep{bottcher07,celotti08,ghisellini09}. An additional strength of multi-wavelength modelling lies in its predictive power: once a source model is established, the flux expected in regions of the electromagnetic spectrum not yet observationally sampled can be compared with telescopes sensitivity curves. This allows us to evaluate whether sources are within reach of current facilities, or represent promising targets for upcoming observatories.

Such a process is particularly relevant in the current context of VHE astrophysics, since we are now entering into a new era for Cherenkov telescopes, thanks to the upcoming Cherenkov Telescope Array Observatory (CTAO; \citealt{CTAscience18,hofmann23}). Using a series of arrays of telescopes located in two sites, one in the Northern, the other in the Southern hemisphere, CTAO will provide full-sky coverage, improving sensitivity by an order of magnitude over current Cherenkov instruments in the 20 GeV to 300 TeV energy range. CTAO will be instrumental in answering a variety of questions that are still open in blazar studies, such as the exact mechanism responsible for jet particle acceleration, the connection between jets and accretion onto the SMBH, and the jets composition, among others. In particular, even though VHE emission from blazars is expected at energies around 100 GeV, surprisingly, some of these objects have recently already been observed even up to 10 TeV \citep{biteau20}. This has led to the distinction between extreme-synchrotron blazars and extreme-TeV blazars, which are not always coincident, and could be originated from different processes. It is therefore important to focus on VHE-emitting blazars, looking for new potential candidates that could help to better characterise the properties of such a population and investigate its emission mechanisms.

Within this framework, we therefore present a multi-wavelength study of a sample of X-ray bright, non $\gamma$-ray detected HSP/EHSP blazars \citep[based on the sample by][]{marchesi25}, to model their broadband SED and estimate their emission in the VHE band.
The work is organised as follows: in Section~\ref{sec:sample} we present the sample used in this work. In Section~\ref{sec:xray} we present 
 the  X-ray catalogues from which the data were collected and we discuss in details the steps of the X-ray analysis that led to the selection of the sources expected to have relevant emission in the VHE band. In Section~\ref{sec:Gammaray} we describe how we used \lat\ data to obtain upper limits to constrain the emission in the MeV to GeV band for the selected sources. Then, in Section~\ref{sec:SEDmod} we present the modelling of the selected sources and discuss the expected emission in the VHE band and CTAO detectability. Finally, we summarise the main results of this work and discuss possible future developments in Section~\ref{sec:conclusions}. 
Throughout the rest of the work, we assume a flat $\Lambda$CDM cosmology with H$_0$=69.6\,km\,s$^{-1}$\,Mpc$^{-1}$, $\Omega_m$=0.29 and $\Omega_\Lambda$=0.71 \citep{bennett14}. Errors are quoted at the 90\,\% confidence level, unless otherwise stated.

\section{Sample selection}\label{sec:sample}

All the sources analysed in this work are initially selected from a subsample of the 5th ROMABZCAT -- Roma--BZCAT Multi--Frequency Catalogue of Blazars \citep[hereafter 5BZCAT,][]{massaro15}.
Specifically, we worked with the 1007 sources without a detection in the \lat\ 14-year Source Catalogue \citep[hereafter 4FGL-DR4,][]{abdollahi22, ballet23} but with at least one counterpart in either one of the main X-ray catalogues currently available: the \xmm\ Survey Science Catalogue \citep{webb20}, the second release of the \cha\ Source Catalogue \citep{evans24}, the second \swi\ X-ray Point Source catalogue of detections by \xrt\ \citep{evansP20_swi}, or in the recently released eROSITA-DE Data Release 1 catalogue \citep{merloni24}. A broader description of these sources and their multi-wavelength properties is presented in \citealt{marchesi25}.

\begin{figure} 
 \centering 
 \includegraphics[width=0.49\textwidth,clip]{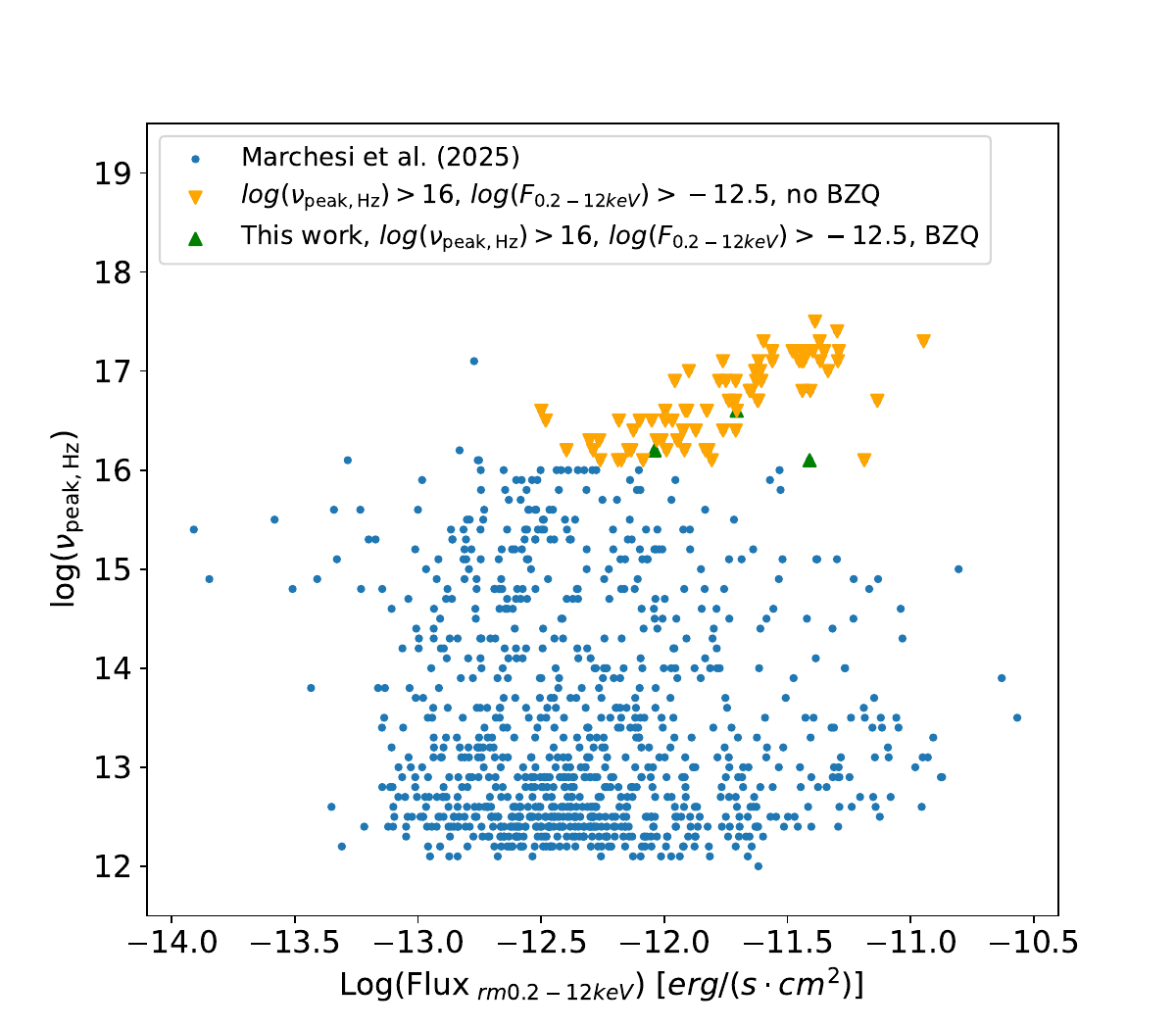} 
\caption{\normalsize 
Synchrotron peak frequency as a function of 0.2-12 keV flux for the 1007 blazars with X-ray counterpart and without a counterpart in the 4FGL catalogue (\citet{marchesi25}). The subsample of sources analysed in this work is plotted in orange; the three FSRQs excluded from the analysis are instead plotted in green. Finally, the rest of the \citet{marchesi25} sample is shown in blue.
}\label{fig:sample}
\end{figure}

Then, to focus on a subsample of objects that have the highest chance of detection in the TeV range, we select sources with a high synchrotron peak frequency ($\rm log(\nu_{\rm peak,Hz})>16$), and a high flux in the X-ray, set as $\rm log(F_{0.2-12 \ keV})>-12.5$. Since the correlation between the X-ray and the VHE flux has been established mainly for BL Lacs, we also excluded the FSRQs from our sample. Additionally, we excluded the source 5BZUJ1459-1018, which was reported in \citet{green17} as displaying an X-ray emission dominated by cluster emission, but was not removed from \cite{marchesi25} due to a change in nomenclature. In this way, we obtained a sample of 78 sources that meet all the aforementioned requirements. The targets with a reported redshift are in the range 0.09$< z <$ 0.68 (median redshift: 0.28). 2 sources had no redshift reported. In Figure \ref{fig:sample}, we report the location of the targets with respect to the overall parent sample in the synchrotron peak versus  0.2-12 keV flux parameter space.

\section{X-ray data collection and analysis}\label{sec:xray}

The first step of this work is to perform X-ray analysis on the sources in our sample to select blazars with likely appreciable TeV emission. 

\subsection{X-ray Data collection}
We retrieved the calibrated spectra and response files from publicly accessible catalogues for the X-ray analysis of each of the 78 targets. As a rule, when a source had been observed by more than one telescope among \xmm, \cha, and \swi-XRT, the analysis was performed for all observations. Instead, we used eROSITA data for sources which were only observed by this telescope, or if the count statistic of the spectrum derived using one of the other facilities led to a spectral fit with inadequate quality (i.e., with fewer than 25 degrees of freedom). When available, we also inspected the sources light curves to check for variability. We describe in detail the handling of sources with indications of strong variability in the coming sections for every telescope specifically.

\subsubsection{\xmm\ and the 4XMM-DR14 catalogue}\label{sec:XMM_an}
The \xmm\ telescope was launched on December 10, 1999, and has been operational since February 2000 \citep{jansen01}. \xmm\ combines an excellent effective area in the 0.2–12 keV band, a large ($\sim30$' diameter) field of view, and a good angular resolution ($\sim5$'' on-axis point-spread function, PSF).
For this work, we used the most recent release of the \xmm\ Survey Science Catalogue, 4XMM-DR14, which was generated from 13864 \xmm\ EPIC observations, covering an energy interval from 0.2 keV to 12 keV and made between the $\rm 3^{rd}$ of February 2000 and the $\rm 31^{st}$ of December 2023.

The datasets for the six sources in our sample that have a counterpart in the  4XMM-DR14 catalogue were retrieved directly from the online catalogue\footnote{\url{http://xmm-catalog.irap.omp.eu/}}. The majority of the sources were observed only once. The source 5BZGJ1510\allowbreak+3335 had 3 observations, so each dataset was fitted separately, getting consistent results between different observations, as well as with the ones obtained from two different \cha\ datasets. The source 5BZBJ0333$-$3619 was observed 12 times but, since a good fit was obtained already from a fit to \xrt\ data, we analysed only the observation with the longest exposure among the 12 \xmm\ ones.

\subsubsection{\cha\ and the CSC 2.1 catalogue}\label{sec:cha_an}
The \cha\ telescope was launched on July 23, 1999 \citep{weisskopf02}. \cha\ covers the 0.3–7 keV energy range and is the only X-ray instrument with a subarcsecond on-axis angular resolution (PSF$\sim$0.5'') and a field of view of $\sim30$' in diameter. 

We retrieved the data products from the most recent release of the \cha\ Source Catalog, CSC version 2.1, which was released on April $\rm2^{nd}$ 2024, with a further minor update (v2.1.1 in October 2024). It includes measured properties for 407,806 unique point-like and extended X-ray sources in a total sky coverage of $ \sim 730 \ \text{deg}^2 $. For the 9 sources of our sample in the catalogue,  we retrieved data through the application \texttt{cscview}. All the sources were only observed once, besides the blazar 5BZGJ1510+3335. This target had been observed 6 times: once in 2001, once in 2007 and multiple times in November 2010, so we analysed the data products from the only analysis performed in 2001 and 2007, and for the longest observation performed in 2010. As reported in Table~\ref{tab:result_xanalysis}, the best-fit properties and the flux of this source changed between the observations, thus suggesting that some blazar variability took place over the nine-year timespan dividing observations.

\subsubsection{\xrt\ and the 2SXPS catalogue}\label{sec:swift_an}
The Neil Gehrels \swi\ Observatory was launched on November 20, 2004, and its three different telescopes observe the sky in the optical and ultraviolet (Ultraviolet/Optical Telescope, or UVOT), in the hard X-rays (14–195 keV; Burst Alert Telescope, or BAT), and in the soft X-rays (0.2–10 keV; X-ray Telescope, or XRT) \citep{Gehrels04}. The XRT has an on-axis angular resolution of $\sim$18'' and a 23.6 $\times$ 23.6 arcmin$^2$ field of view.
For this work, we used the \swi\ X-ray Point Source (2SXPS) catalogue of detections, obtained with the telescope used in Photon Counting (PC) mode in the 0.3-10 keV energy range. This catalogue detected a total of 206,335 unique sources over an area of $ \sim 3790 \ \text{deg}^2 $ in the period from the $\rm 1^{st}$ of January 2005 to the $\rm 1^{st}$ of August 2018.

The data for the 58 sources in our sample with a counterpart in the 2SXPS  were retrieved through the web tool\footnote{\url{https://www.swift.ac.uk/user\_objects/}} available on the telescope website, which produces X-ray images, spectra and light curves using the software HEASOFT \citep[v6.32;][]{Evans07,Evans09}. The majority of the sources were observed more than once, so we always visually inspected their light curve to decide how to handle the data from different observations. When the different observations did not display significant flux variability, we treated all observations as a single dataset. When, instead, the source displayed significant flux variability, observations were split in multiple datasets. 

Exceptions to this approach are sources 5BZBJ0851\allowbreak+0549 and 5BZBJ1410+0515, since observations were too short and the web tool could not provide a light curve, so the data was treated as a single dataset, despite the lack of information concerning flux variability. Therefore, the results for these sources should be treated with additional caution. Finally, for 5 sources (5BZBJ0951+0102, 5BZGJ1157+2822, 5BZBJ1228$-$0221, 5BZGJ1324+5739, 5BZBJ1554+2011), the dataset quality was too low for the fitting, so they were discarded. Some of them are nevertheless present in the list of analysed sources, because they were also observed by other telescopes.

\subsubsection{eROSITA and the eROSITA-DE Data Release 1 catalogue}\label{sec:eROS_an}
eROSITA \citep[extended ROentgen Survey with an Imaging Telescope Array;][]{predehl21} is an X-ray telescope consisting of seven identical and co-aligned X-ray mirrors, each equipped with a charge-coupled device (CCD) camera in its focus. It is mounted on the Spektrum Roentgen Gamma (SRG) orbital observatory \citep{sunyaev21}, launched on July 13, 2019, and covers the 0.2–8 keV energy range, albeit with a significant decline in effective area at energies of $>$2.3 keV. eROSITA is a whole-sky survey instrument: however, in the present work, we make use of the eRASS1 catalogue of the Western hemisphere (which is 359.9442$>$l$>$179.9442 in Galactic coordinates, covering $ \sim 20626.5 \ \text{deg}^2 $), released on the $\rm 31^{st}$ of January 2024 and based on data taken in the first six months of eROSITA observations, completed in June 2020.

We used eROSITA data for 19 sources (11 observed by eROSITA only and 8 with low data quality from other observations) out of the 27 objects in our sample with a counterpart in the eRASS1 catalogue, as explained in Section \ref{sec:sample}. We downloaded the spectra from the eROSITA-DE website\footnote{\url{https://erosita.mpe.mpg.de/dr1/erodat/catalogue/search/}}. For each source, the calibrated products were downloadable in a single dataset, combining the data from all cameras of the telescope. However, for the three sources 5BZBJ0920+3910, 5BZBJ0930+3933, 5BZUJ0933+0003 the data available was not sufficient to properly fit the spectra, so we did not include them in our analysis.

\subsection{X-ray data analysis}
As mentioned in Section~\ref{sec:sample},we selected our HSPs and EHSPs based on their synchrotron peak frequency as computed by \citet{marchesi25} using BlaST \citep[Blazar Synchrotron Tool;][]{glauch22}. BlaST is a deep neural network-based tool developed from the work of \citet{lakshminarayanan16}, and is used to obtain an estimate of blazars synchrotron peak frequency from a broadband SED analysis. While the synchrotron peak values from BlaST are generally reliable, the neural network estimates imply fairly large uncertainties (of the order of $\sim$0.5 dex), which can significantly affect any prediction one can make on the position, and intensity, of the VHE peak. For this reason, we performed a detailed X-ray spectral analysis of our sources, with the goal of determining how many of them are best-fitted with a model that diverges from a simple power law one, by showing relevant spectral curvature. In these objects, such a  phenomenological result would indirectly support the hypothesis that the synchrotron peak is indeed in the X-ray band, thus strengthening the claim on their HSP and EHSP nature.  

\subsubsection{Spectral analysis}
We perform our X-ray spectral fit using  \texttt{XSPEC} \citep[v.12.15.0;][]{Arnaud96}. Specifically, we fit each source in our sample with both a power law (\texttt{zpowerlaw} in XSPEC) and a curved log-parabolic model (\texttt{logpar} in XSPEC). Both the models are multiplied by the Galactic absorption, computed for each source using the \texttt{nh} tool \citep{kalberla05} in HEASOFT.
We work under the assumption that sources with a change in slope clearly present in the data would favour the fit with the log parabola model with respect to that with a power law model. We use C-statistic \citep{cash79,kaastra17} to quantify this hypothesis: having defined the parameter $\Delta$Cstat as Cstat$_{\rm pl}-$ Cstat$_{\rm logpar}$, we consider the log parabola fit as better representing the data when favoured at the 90\% confidence level, namely when $\Delta$Cstat $>2.7$. Additionally, for sources favouring the log parabola fit, we also compute the confidence intervals for the parameters $\alpha$ and $\beta$: for consistency with previous works, in the following sections we consider true log parabola best-fits only those with \(-1 <\beta < 1.5 \) \citep{middei22}.

\subsubsection{Results}

\begin{figure*} 

\begin{minipage}{0.33\textwidth} 
 \centering 
 \includegraphics[angle=0,  width=0.9\textwidth, height=0.2\textheight]{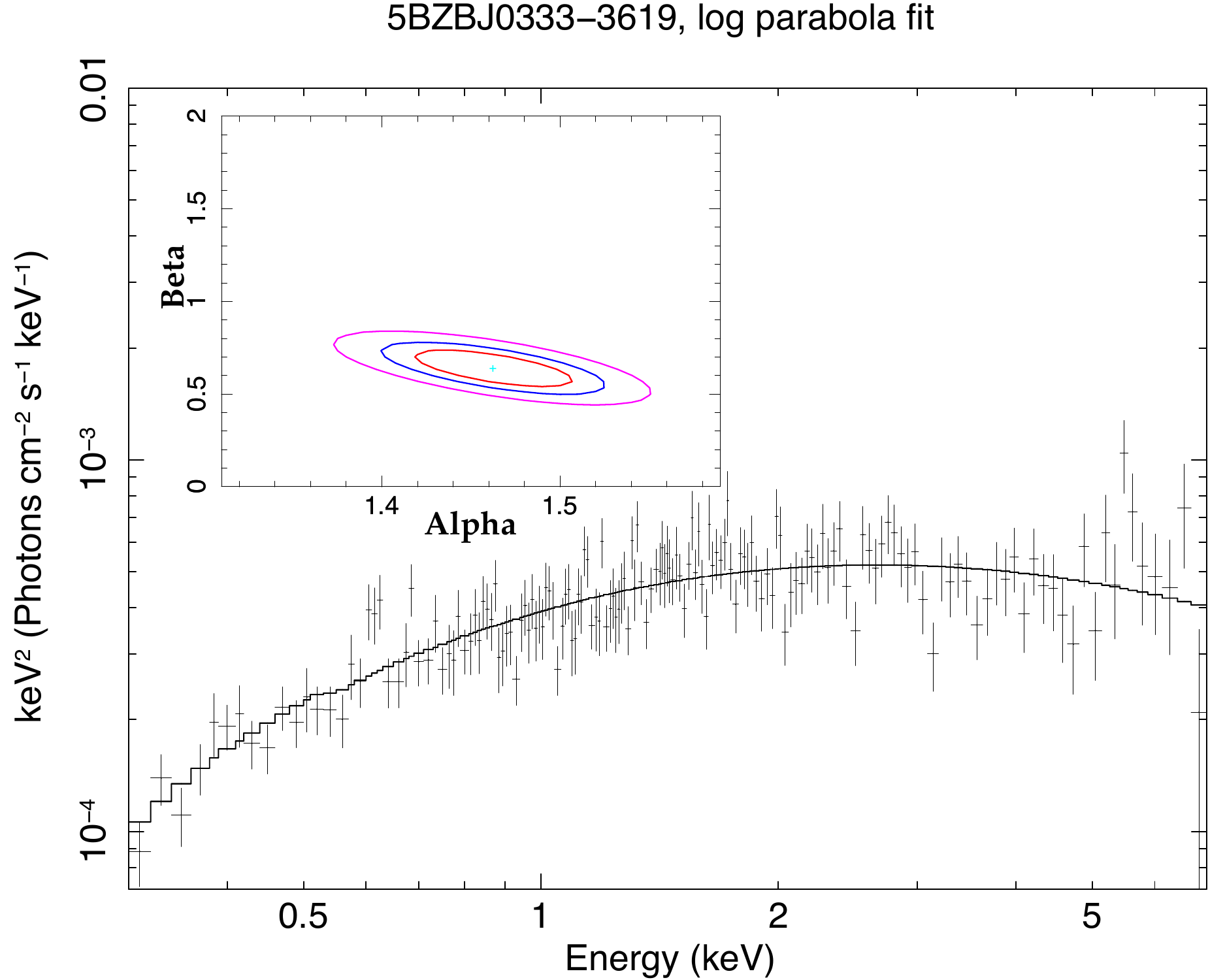} 
\end{minipage}
\begin{minipage}{0.33\textwidth} 
 \centering 
 \includegraphics[angle=0, width=0.9\textwidth, height=0.2\textheight]{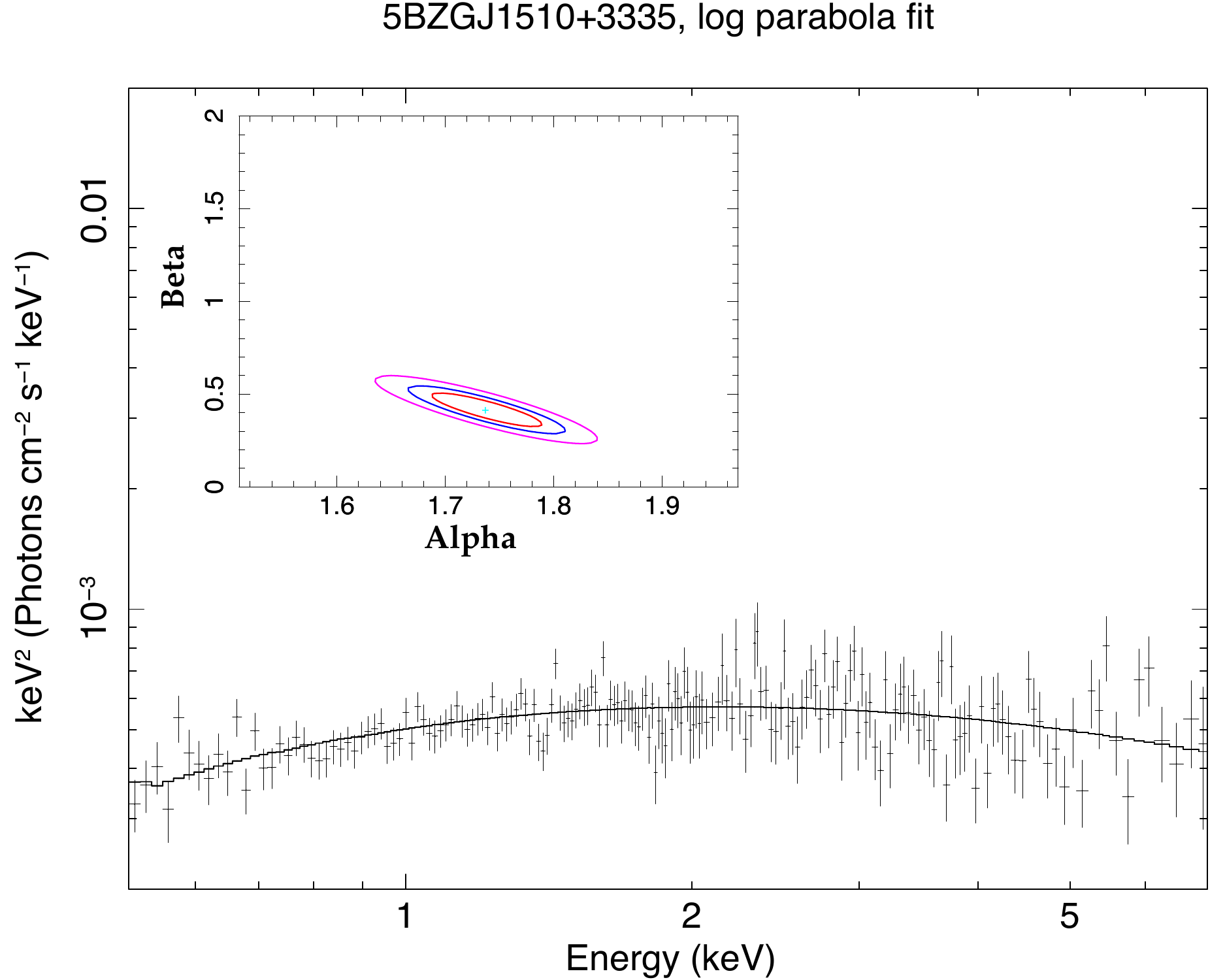} 
\end{minipage}
\begin{minipage}{0.33\textwidth} 
 \centering 
 \includegraphics[angle=0, width=0.9\textwidth, height=0.2\textheight]{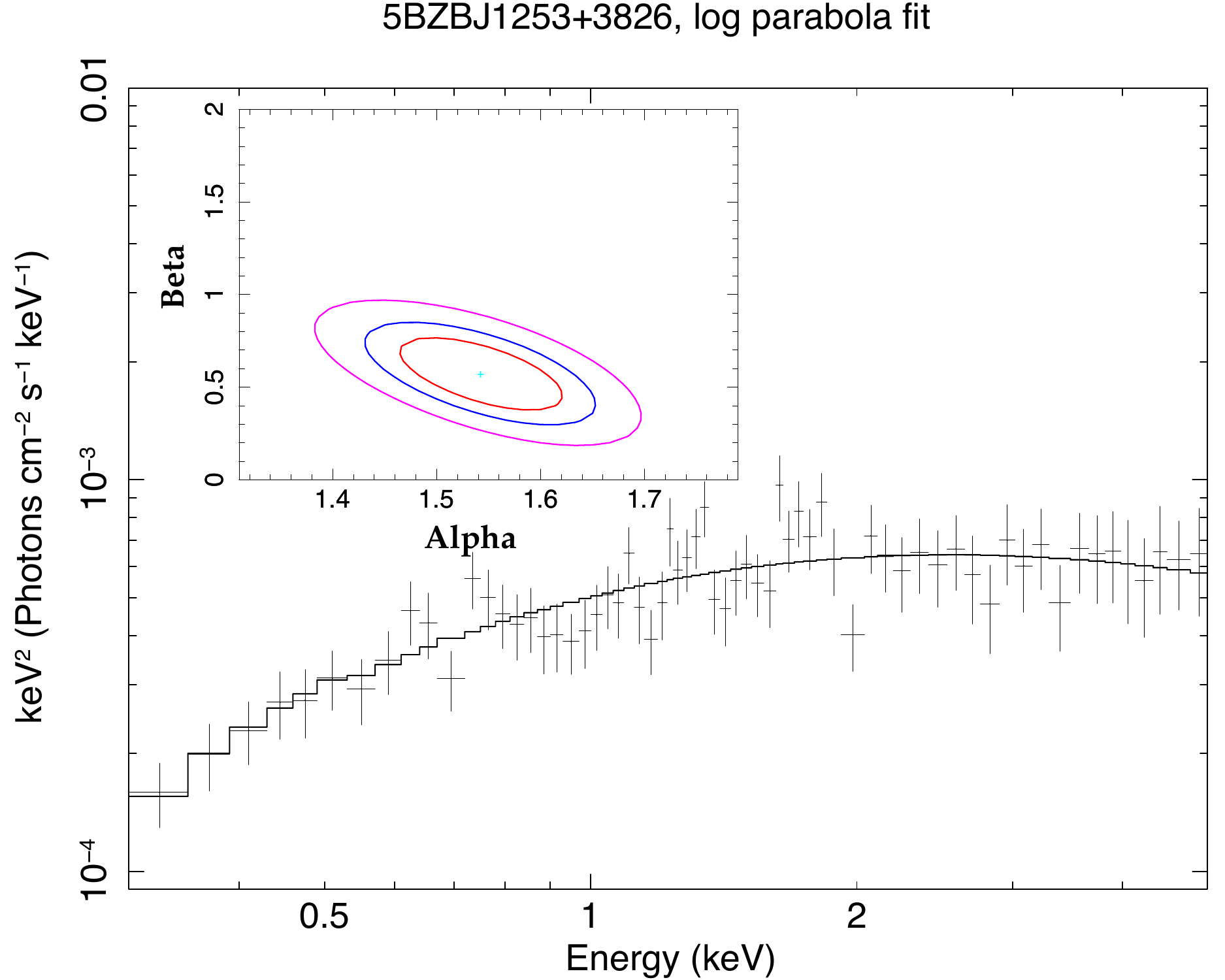} 
\end{minipage}
\begin{minipage}{0.33\textwidth} 
 \centering 
 \includegraphics[angle=0, width=0.9\textwidth, height=0.2\textheight]{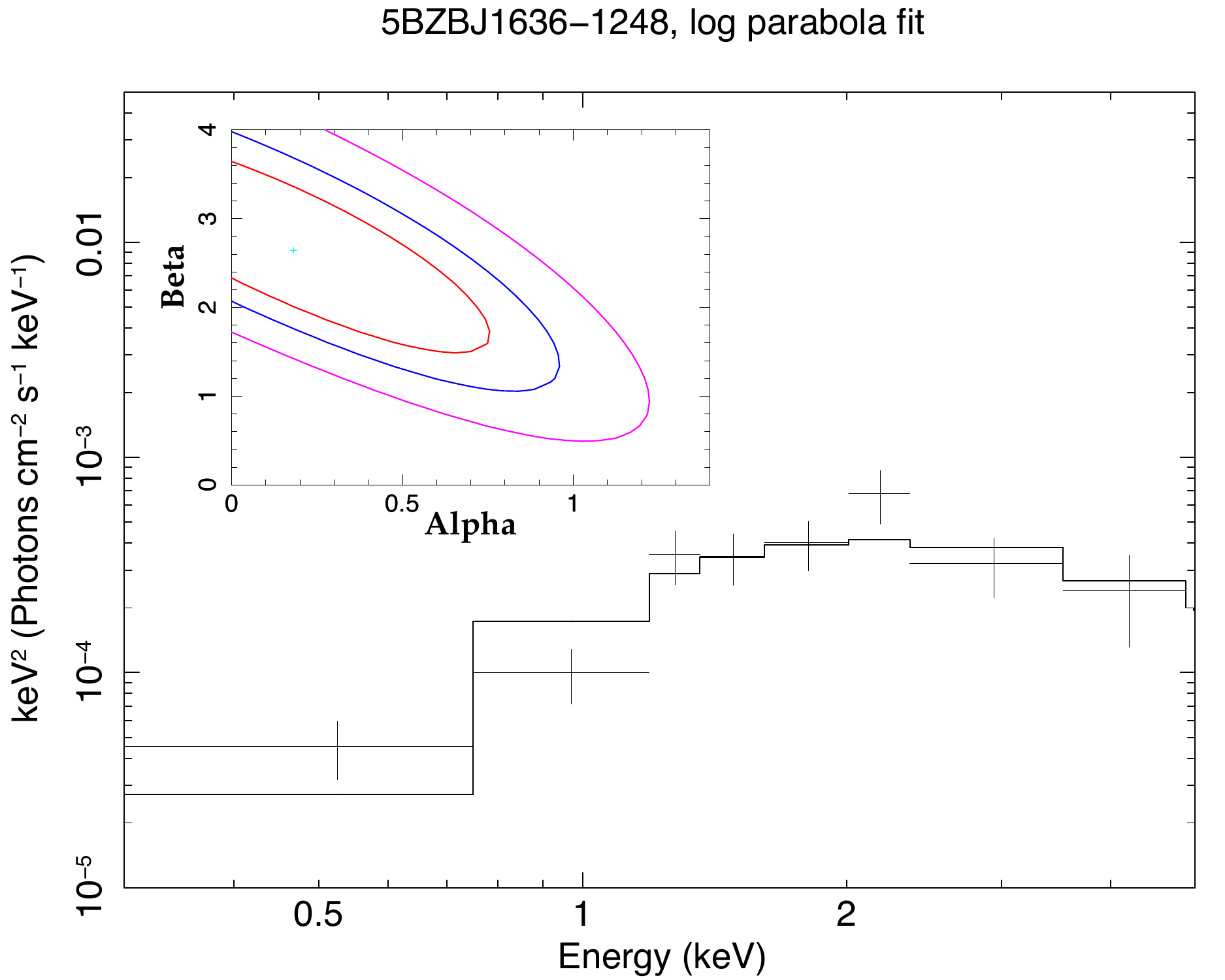} 
\end{minipage} 
\begin{minipage}{0.33\textwidth} 
 \centering 
 \includegraphics[angle=0, width=0.9\textwidth, height=0.2\textheight]{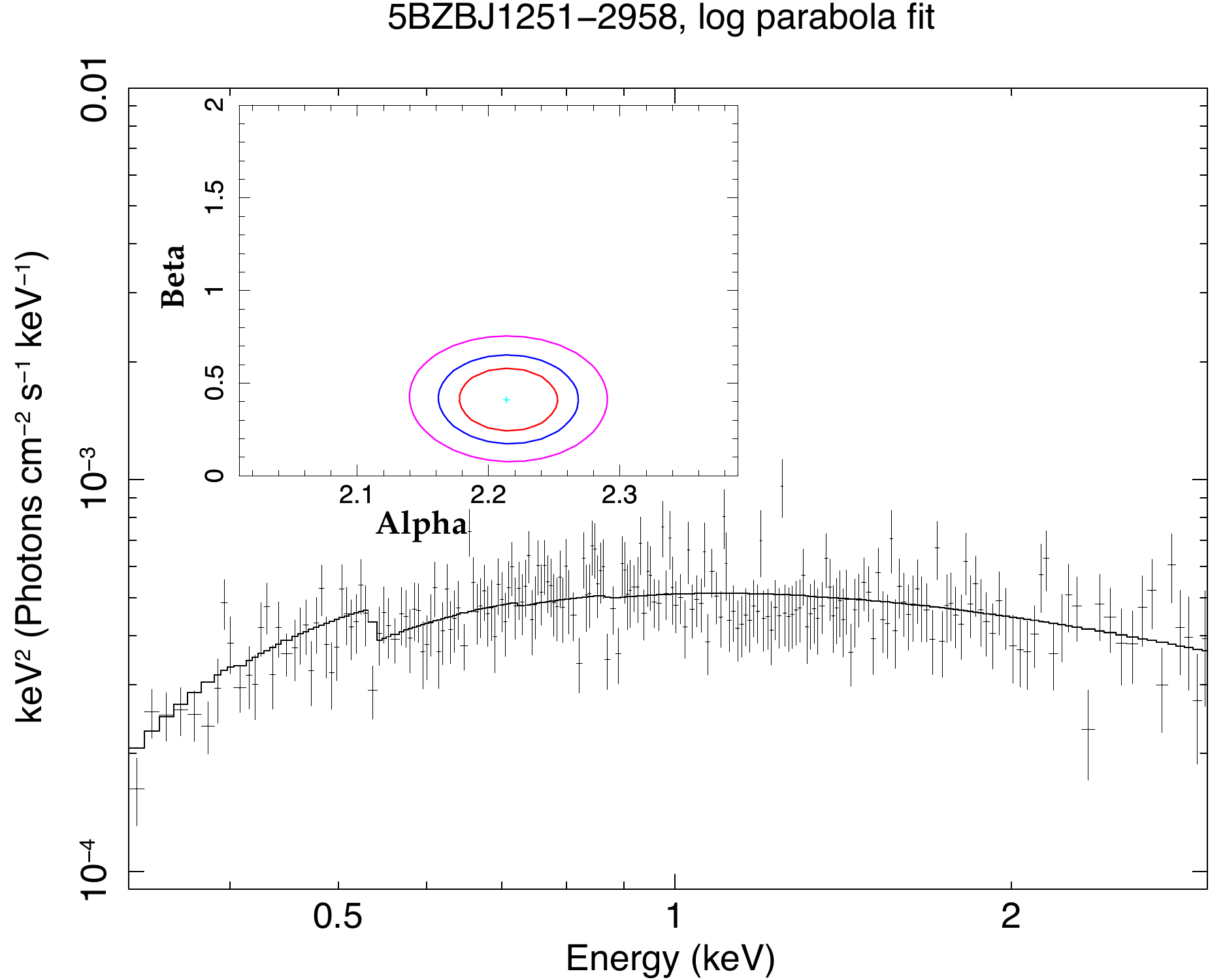} 
\end{minipage}
\begin{minipage}{0.33\textwidth} 
 \centering 
 \includegraphics[angle=0, width=0.9\textwidth, height=0.2\textheight]{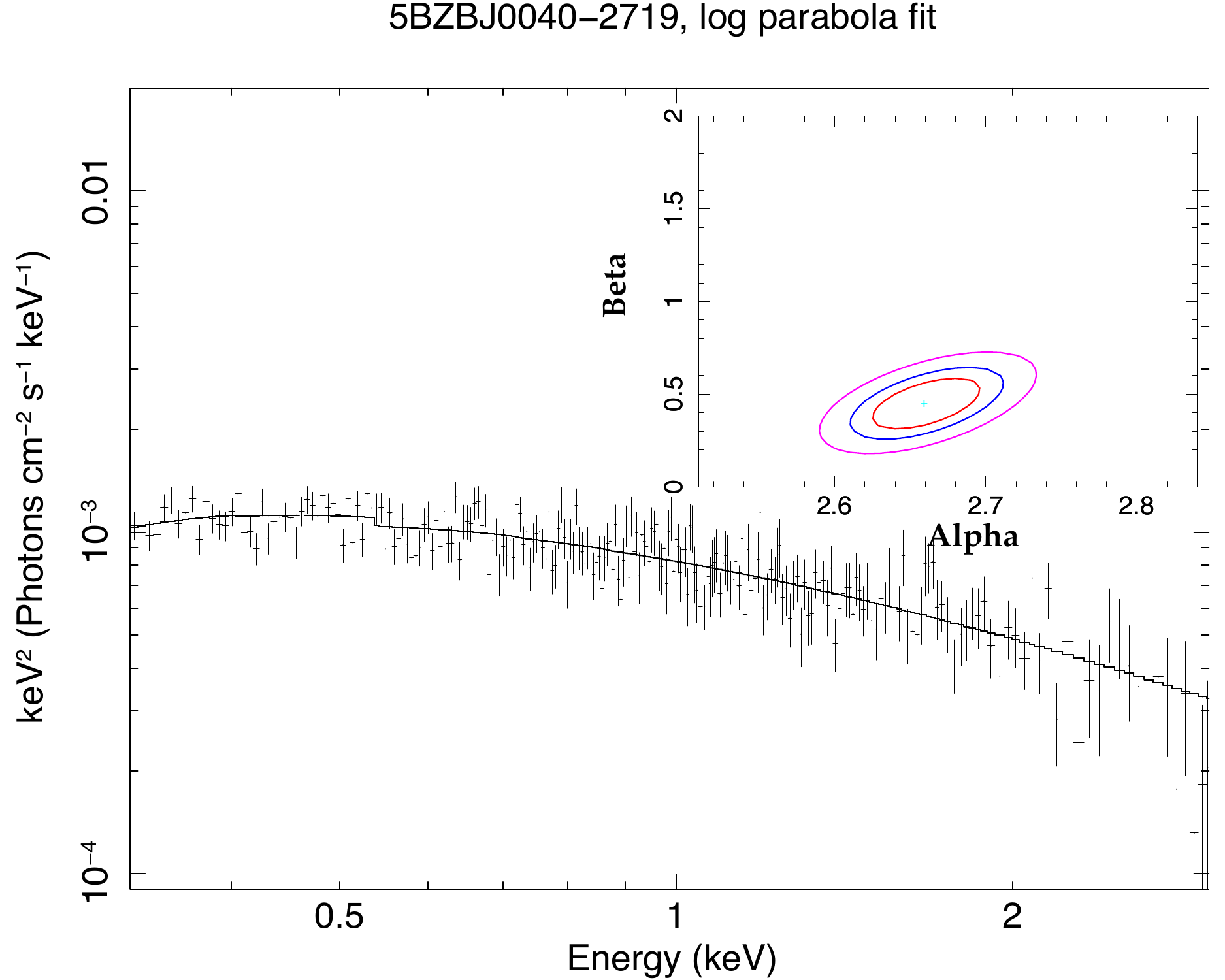} 
\end{minipage}
\begin{minipage}{0.33\textwidth} 
 \centering 
 \includegraphics[angle=0, width=0.9\textwidth, height=0.2\textheight]{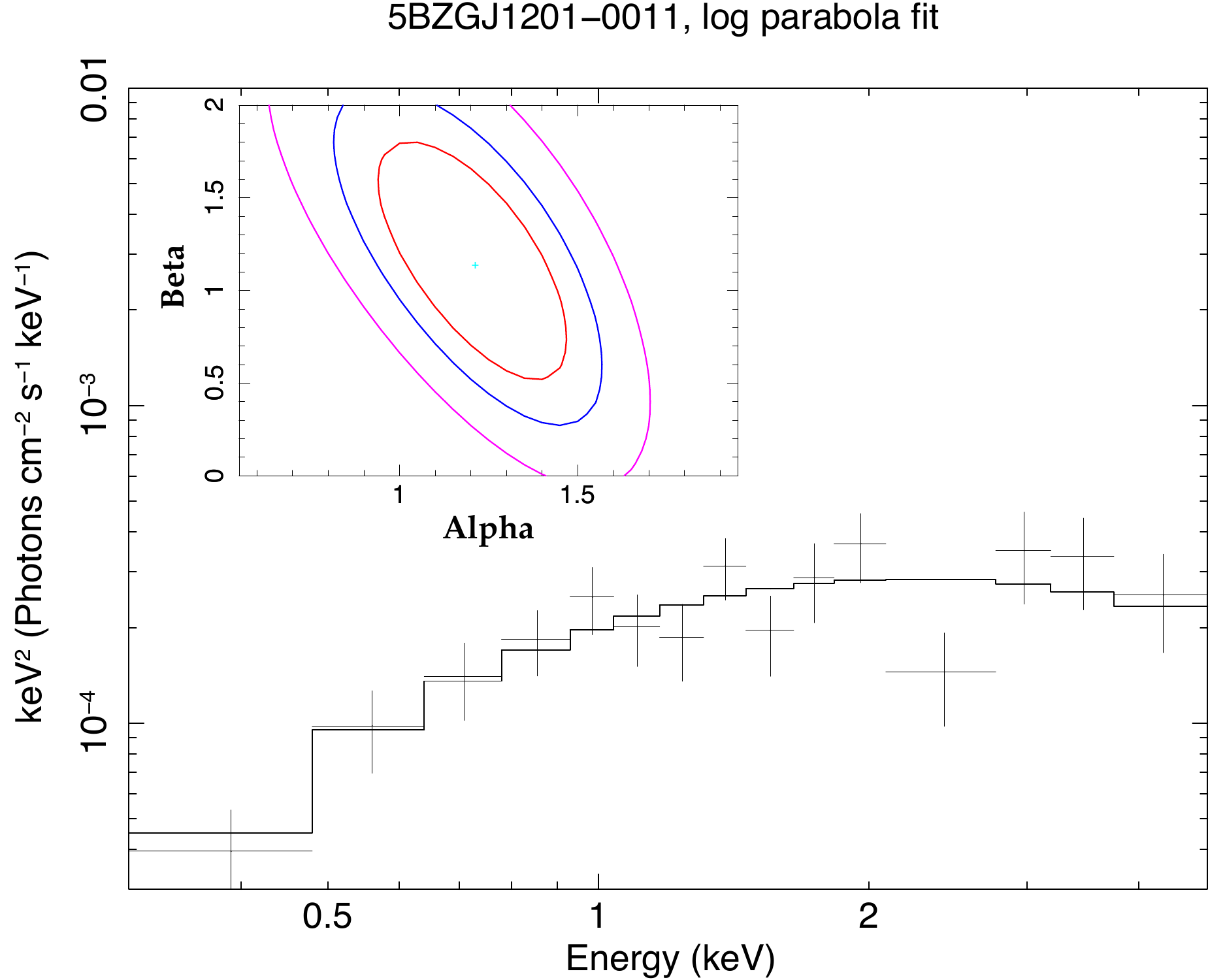} 
\end{minipage}
\begin{minipage}{0.33\textwidth} 
 \centering 
 \includegraphics[angle=0, width=0.9\textwidth, height=0.2\textheight]{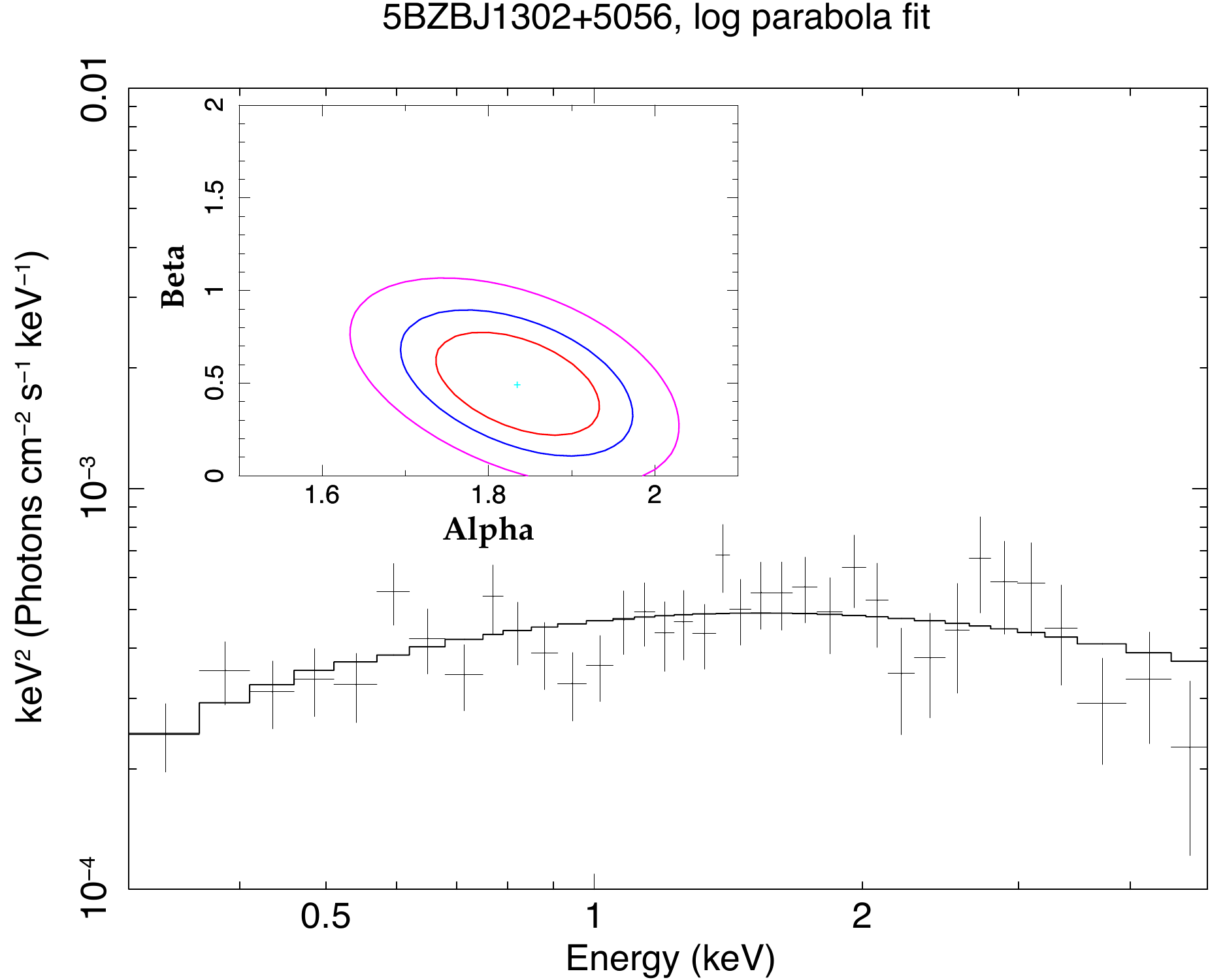} 
\end{minipage}
\begin{minipage}{0.33\textwidth} 
 \centering 
 \includegraphics[angle=0, width=0.9\textwidth, height=0.2\textheight]{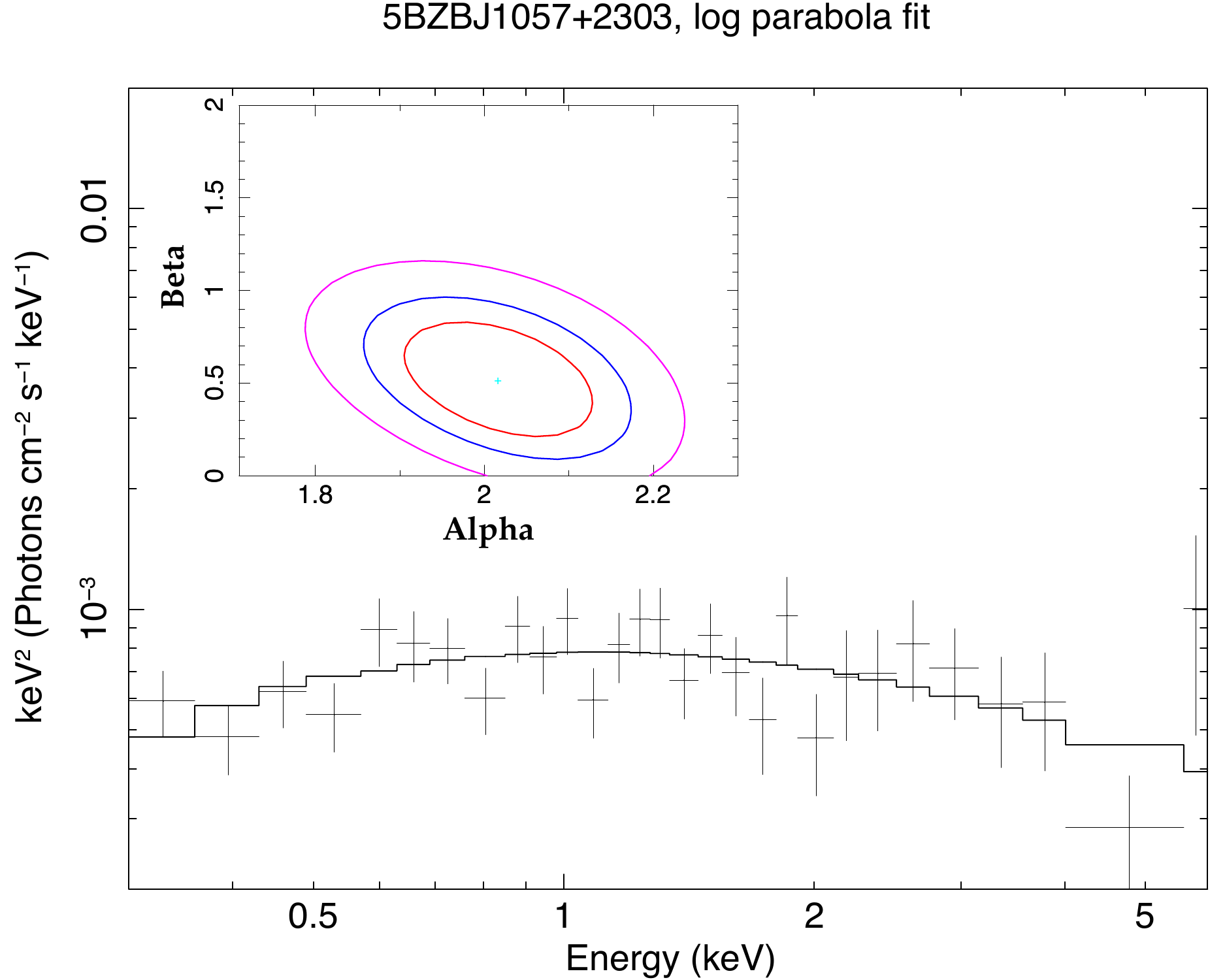} 
\end{minipage}
\caption{ \label{fig:xray1}  From left to right, top to bottom: Results of the log parabola fitting of the blazars 5BZBJ0333-3619 (\swi-XRT data), 5BZGJ1510+3335 (\cha\ data), 5BZBJ1253+3826 (\swi-XRT data), 5BZBJ1636--1248 (\swi-XRT data), 5BZBJ1251-2958 (\xmm\ data), 5BZBJ0040-2719 (\xmm\ data), 5BZGJ1201-0011 (\swi-XRT data), 5BZBJ1302+5056 (\swi-XRT data) and  5BZBJ1057+2303 (\swi-XRT data). In the inset, we report the confidence regions at 1$\sigma$ (red), 2$\sigma$ (blue) and 3$\sigma$ (pink) for the two parameters of the fit, $\alpha$ and $\beta$.}
\end{figure*}

\begin{figure*} 

\begin{minipage}{0.33\textwidth} 
 \centering 
 \includegraphics[angle=0, width=0.9\textwidth, height=0.2\textheight]{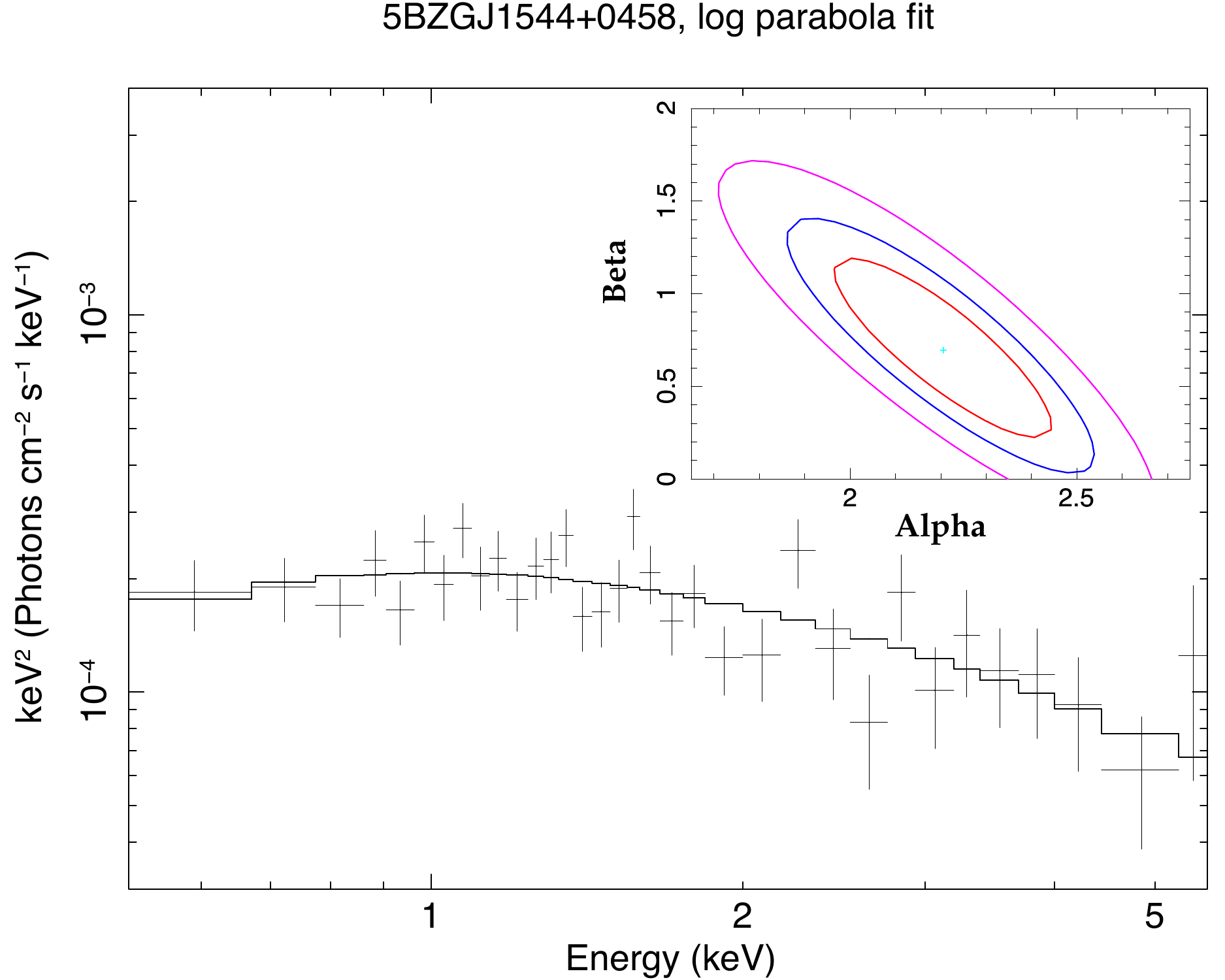} 
\end{minipage}
\begin{minipage}{0.33\textwidth} 
 \centering 
 \includegraphics[angle=0, width=0.9\textwidth, height=0.2\textheight]{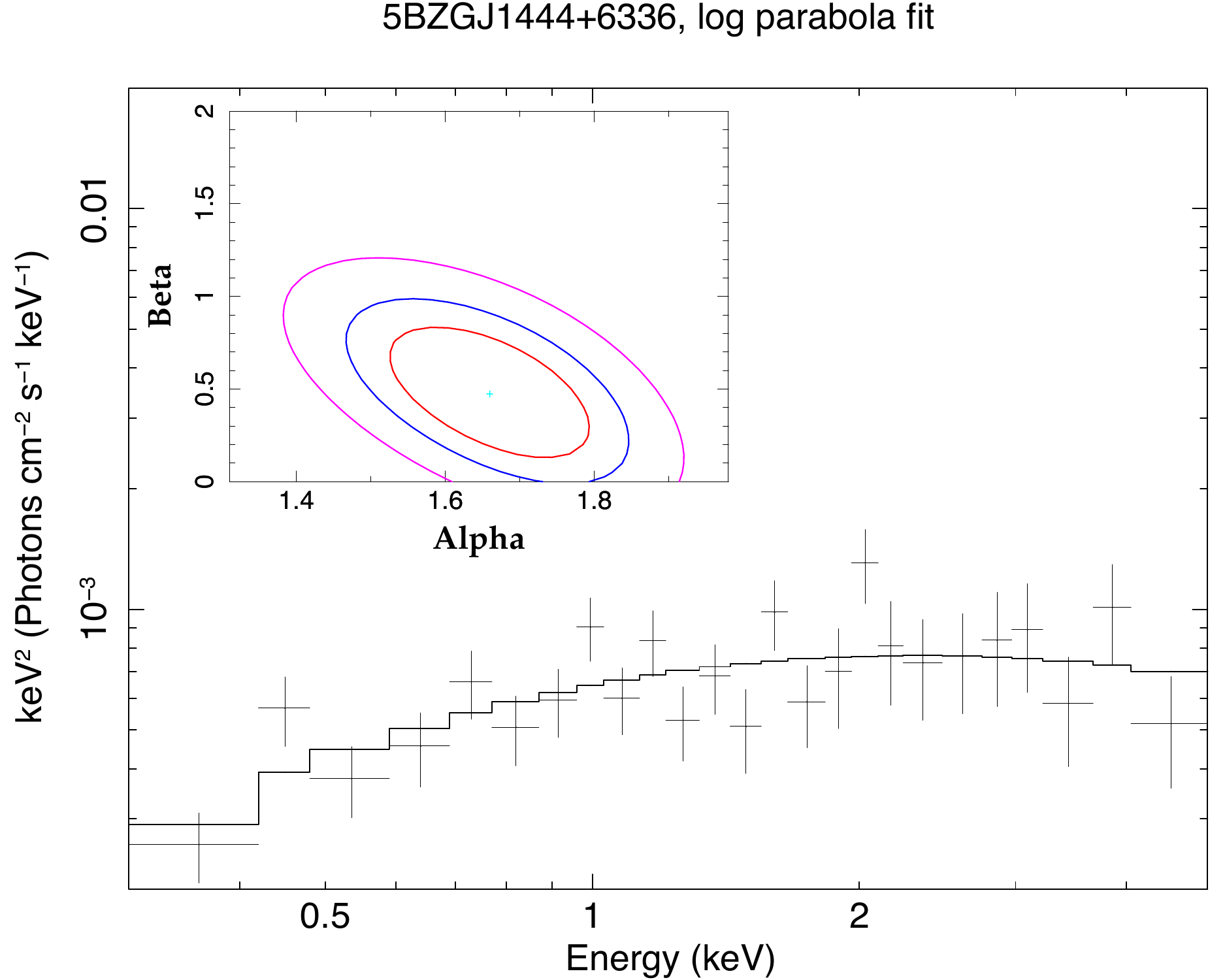} 
\end{minipage}
\begin{minipage}{0.33\textwidth} 
 \centering 
 \includegraphics[angle=0, width=0.9\textwidth, height=0.2\textheight]{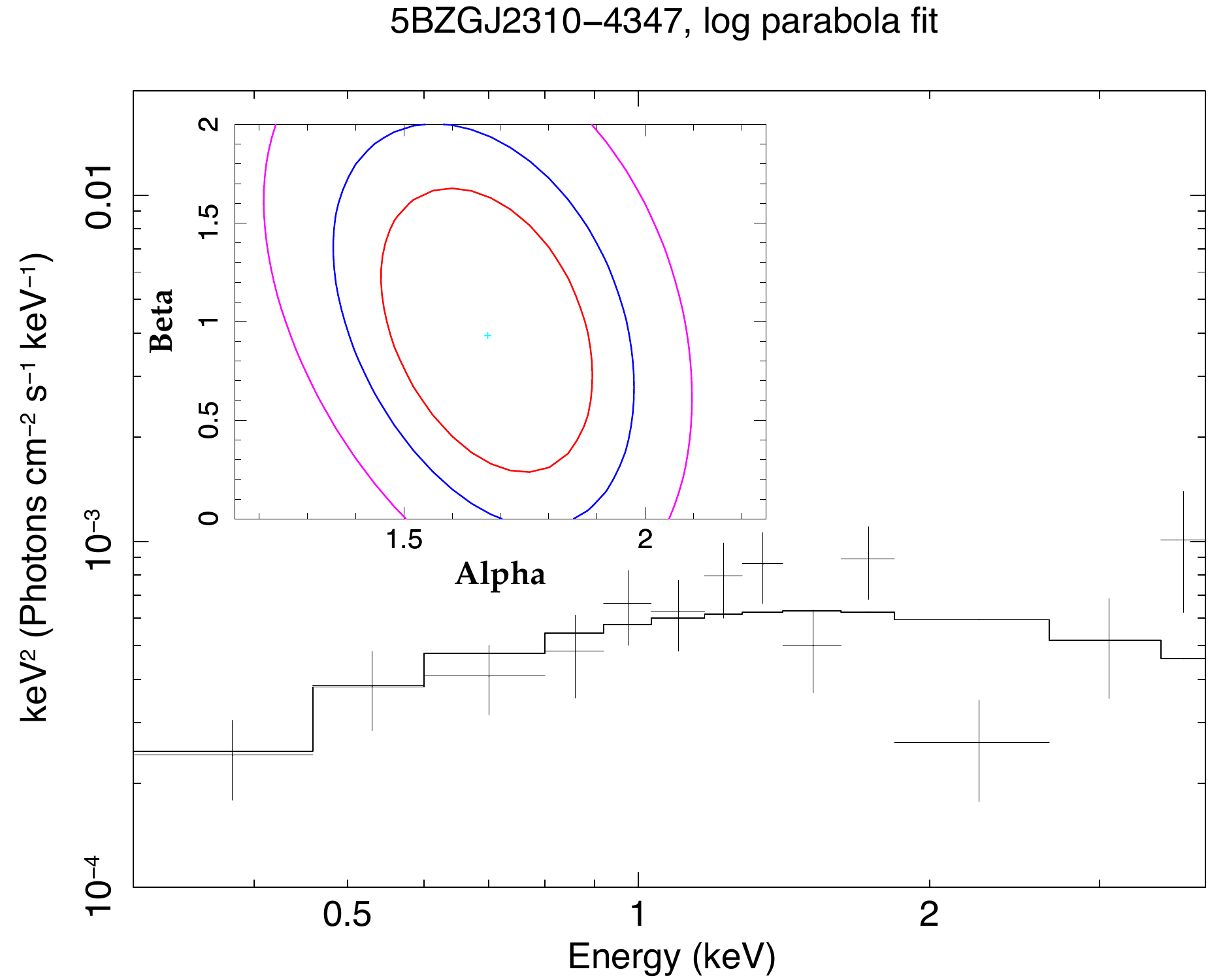} 
\end{minipage}
\begin{minipage}{0.33\textwidth} 
 \centering 
 \includegraphics[angle=0, width=0.9\textwidth, height=0.2\textheight]{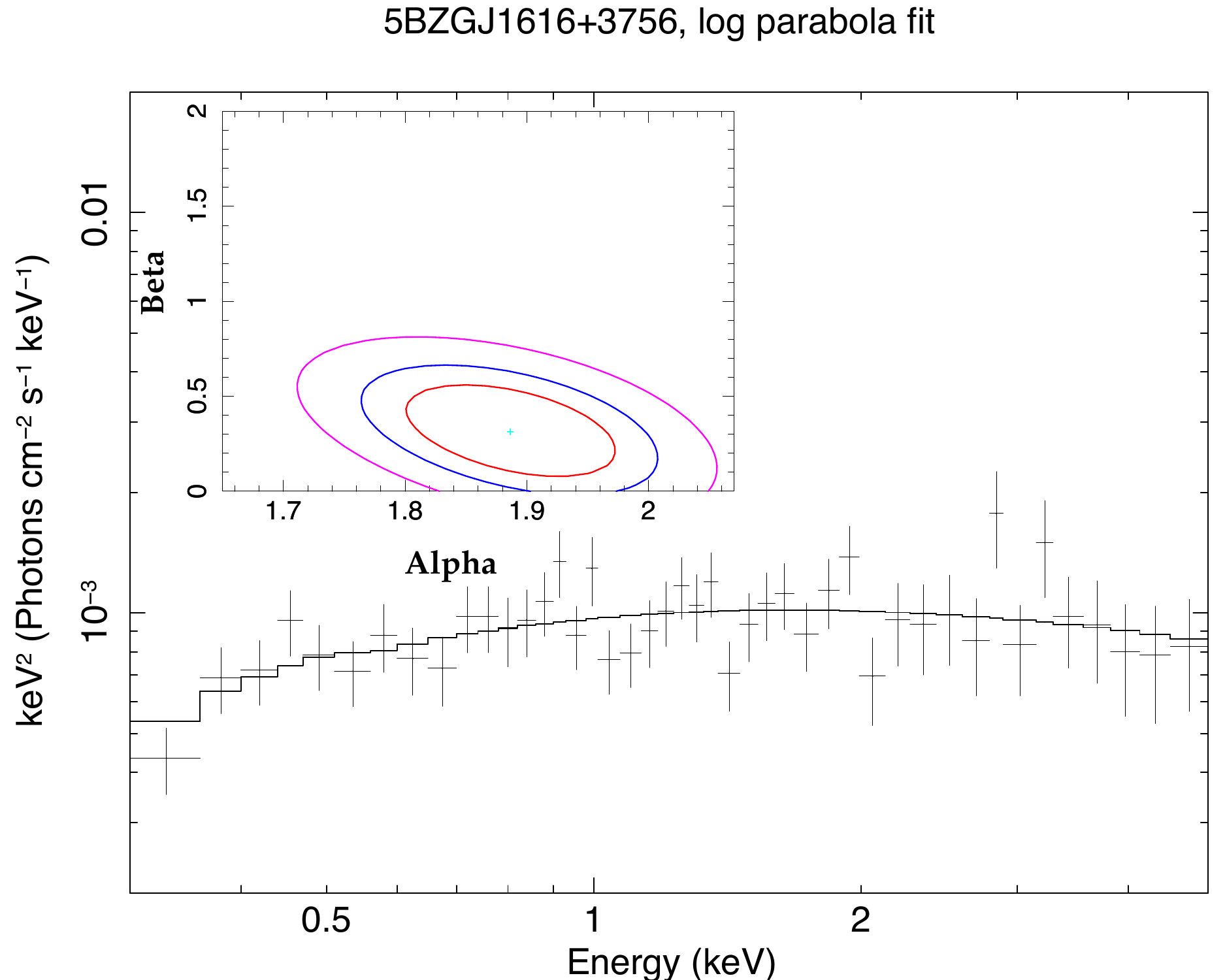} 
\end{minipage}
\begin{minipage}{0.33\textwidth} 
 \centering 
 \includegraphics[angle=0, width=0.9\textwidth, height=0.2\textheight]{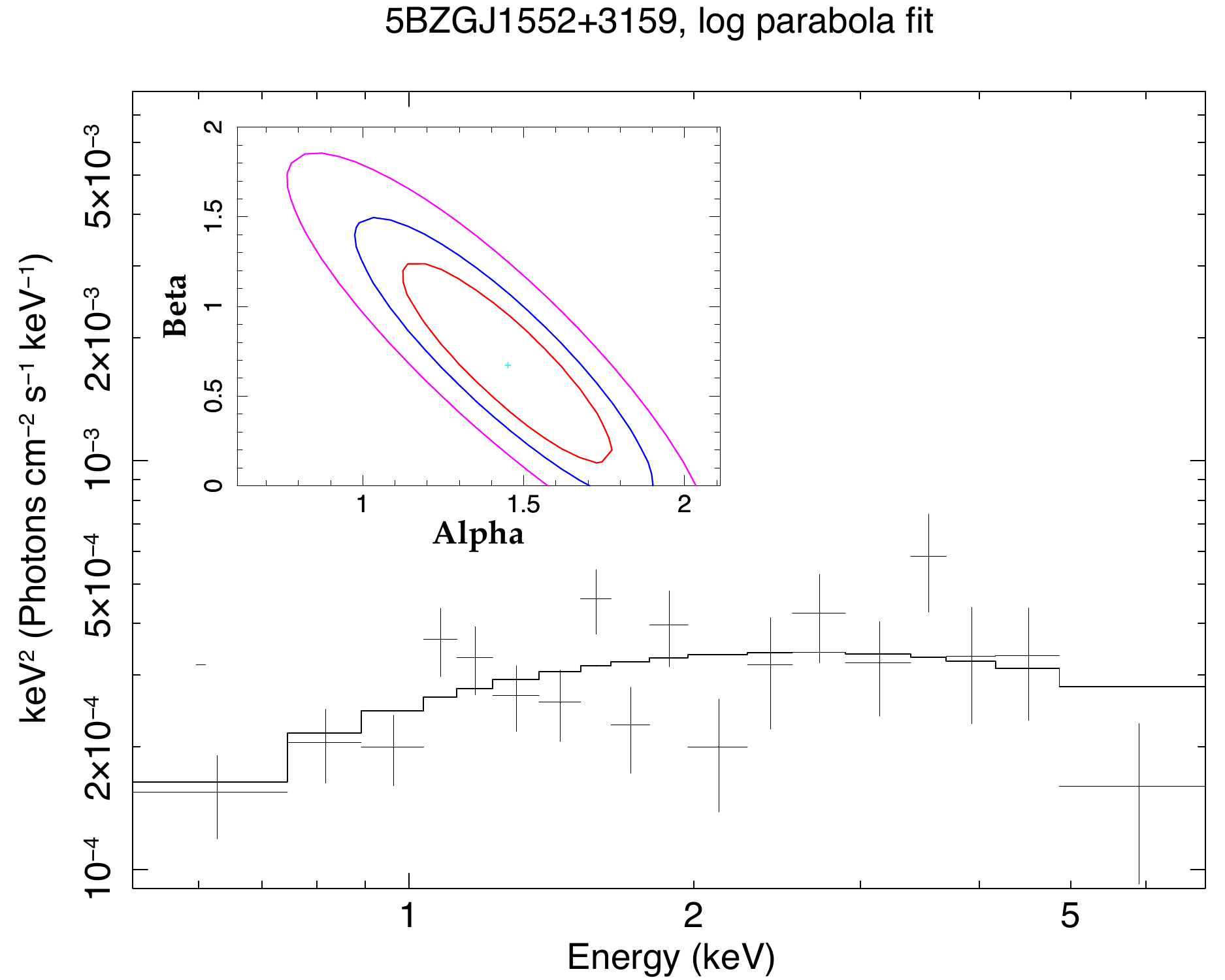} 
\end{minipage}
\begin{minipage}{0.33\textwidth} 
 \centering 
 \includegraphics[angle=0, width=0.9\textwidth, height=0.2\textheight]{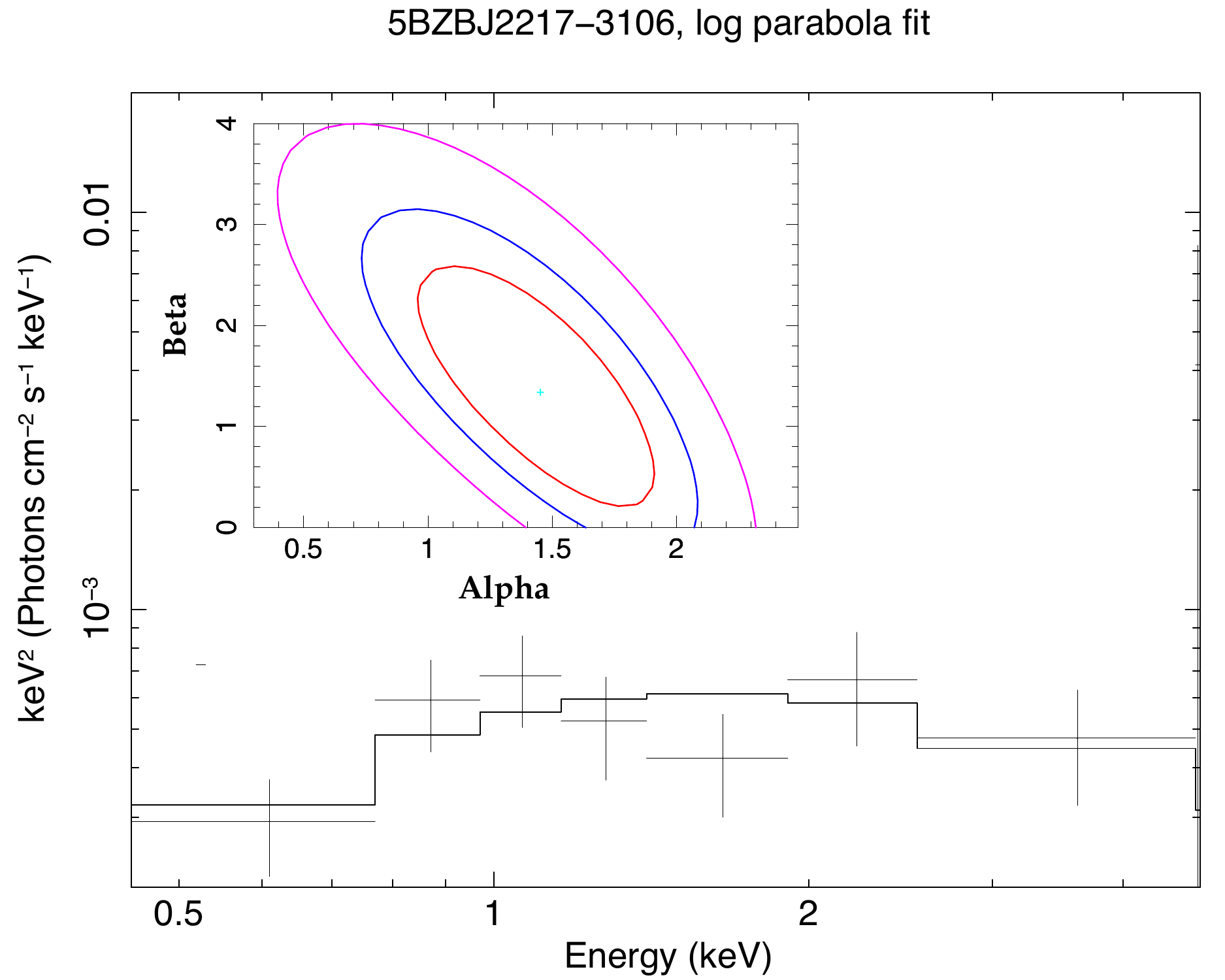} 
\end{minipage}
\begin{minipage}{0.33\textwidth} 
 \centering 
 \includegraphics[angle=0, width=0.9\textwidth, height=0.2\textheight]{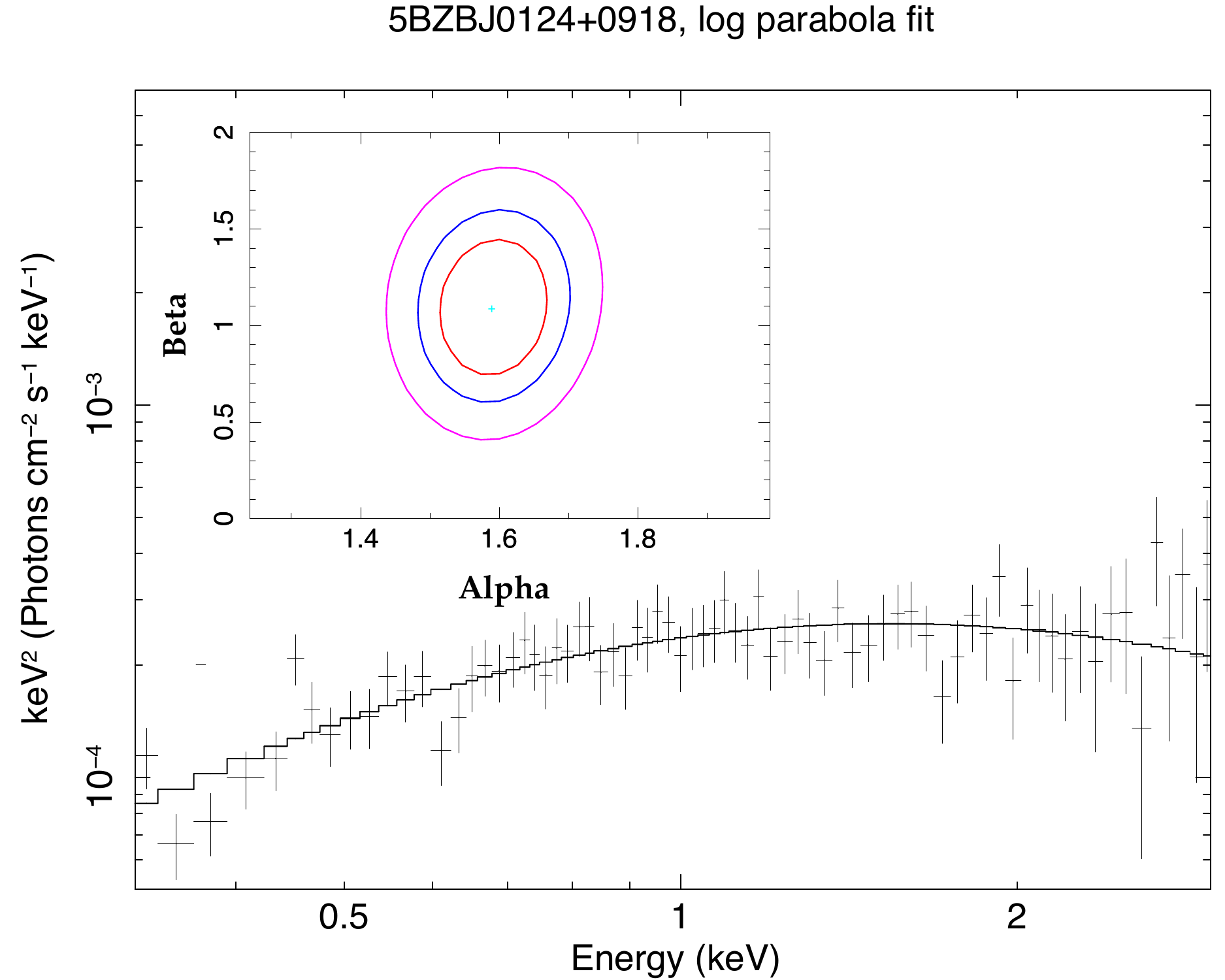} 
\end{minipage}
\begin{minipage}{0.33\textwidth} 
 \centering 
 \includegraphics[angle=0, width=0.9\textwidth, height=0.2\textheight]{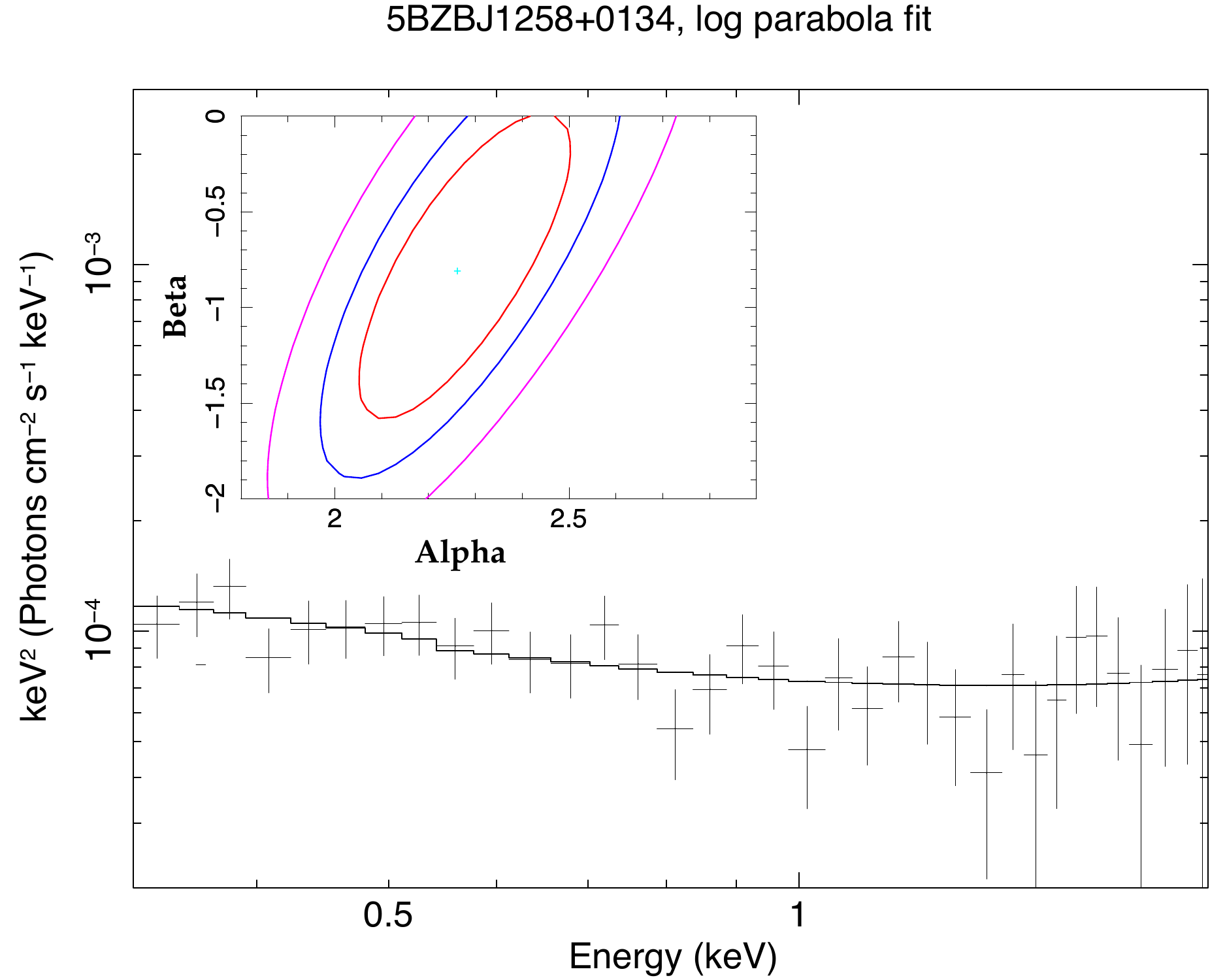} 
\end{minipage}
    \caption{ \label{fig:xray3} From left to right, top to bottom: Results of the log parabola fitting of the blazars 5BZGJ1544+0458 (\cha\ data), 5BZGJ1444+6336 (\swi-XRT data), and 5BZGJ2310--4347 (\swi-XRT data), 5BZGJ1616+3756 (\swi-XRT data), 5BZGJ1552+3159 (\cha\ data), 5BZBJ2217--3106 (\swi-XRT data), 5BZBJ0124+0918 (\xmm\ data), and 5BZBJ1258+0134 (\xmm\ data). In the inset, we report the confidence regions at 1$\sigma$ (red), 2$\sigma$ (blue) and 3$\sigma$ (pink) for the two parameters of the fit, $\alpha$ and $\beta$.}
\end{figure*}

Overall, we analysed 73 out of 78 total sources in the initial sample, since for the other five (5BZBJ0920+31910, 5BZGJ1324+5739, 5BZBJ1554+2011, 5BZBJ0930\allowbreak+3933, 5BZUJ0933+0003) the data quality was too low to allow fitting.
In Appendix~\ref{app:xresults} we report the results of our analysis, and in particular we report the best-fit parameters for each source in Table~\ref{tab:result_xanalysis}. Some of the sources present more than one entry because they were observed by more than one telescope, or were observed in multiple epochs, which we have fit separately.

In the table, the 17 sources we select for additional analysis are reported at the top, before the splitting line. 16 out of 17 objects are  best fitted with log parabola model and meet the requirements mentioned above (i.e., $\Delta$Cstat $> 2.7$ and \(-1 <\beta < 1.5 \)). 
One object, 5BZBJ1636--1248, instead, had $\beta$=2.64$^{+0.95}_{-1.25}$, outside the range reported in \citet{middei22}: this source, however, has a significantly hard photon index ($\Gamma\sim$1.2) when fitted with a simple power law, a result that supports a scenario where the synchrotron peak is in the X-ray band, at energies larger than the one sampled by our X-ray spectrum. For this reason, we choose to keep this object in our sample for further analysis.
Finally, we note that 5 sources would meet the requirement on $\Delta$Cstat, but we exclude them from further analysis due to the value of the $\beta$ parameter.

We report in Figures \ref{fig:xray1} and \ref{fig:xray3} the log parabola best fit models for the 17 sources best fitted with the log parabola model: for those sources with more than one observation meeting the aforementioned criteria, we show only the spectrum with the largest $\Delta$Cstat.  
In the insets of the figures, we also display the confidence regions for the parameters $\alpha$ and $\beta$.

\begin{figure} 
 \centering 
 \includegraphics[width=0.49\textwidth,clip]{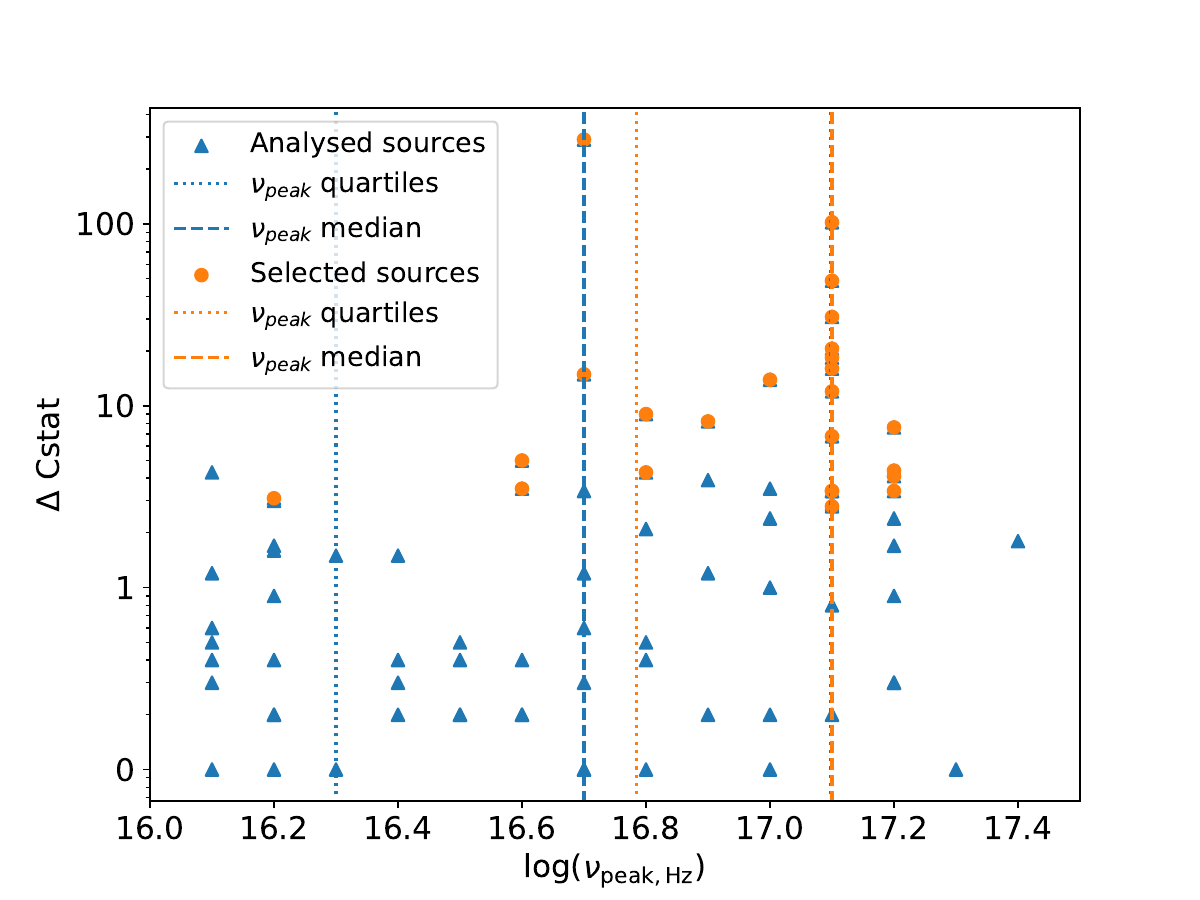} 
\caption{\normalsize 
Distribution of the $\Delta$Cstat of the analysed sources plotted against their synchrotron peak. The blue triangles represent the results for all the analysed sources, whereas the orange circles represent the sources favouring a log parabola fit, having $\Delta$Cstat $>$ 2.7 and -1 $< \beta <$ 1.5. The dotted lines represent the quartiles and median distribution for the two populations. As it can be seen, the sources best fitted with a log parabola model tend to have slightly larger synchrotron peak frequency.
}\label{fig:DeltaC-nupeak}
\end{figure}

We also test the hypothesis that sources best fitted by a log parabola tend to have their synchrotron peak at higher frequencies, and are therefore more effectively sampled by the X-ray spectra analysed here. We find a tentative evidence for such a scenario, since the sources best fitted with a log parabola consistently show higher values of $\nu_{peak}$ across lower quartiles as reported in Figure~\ref{fig:DeltaC-nupeak}. If we consider all the 78 sources we analysed, the median value for log($\nu_{\rm peak, \ Hz}$), as reported in the 5BZCAT catalogue, is 16.7, with first quartile Q1(25\%)=16.3, and third quartile Q3(75\%)=17.1. When instead we look specifically at the statistics of the 17 sources best fitted with a log parabola model the median value is shifted to 17.1 (Q1(25\%)=16.8, Q3(75\%)=17.1).

\section{\texorpdfstring{$\gamma$-}{Gamma-}ray analysis}\label{sec:Gammaray}
Since all the sources in the sample are (by original selection in the parent sample) lacking a \lat\ 4FGL-DR4 counterpart, no $\gamma$-ray spectral points are available. However, for the purposes of a broadband SED modelling, even upper limits are extremely valuable, since they allow us to put constraints on the shape of the SEDs of our targets where we expect to see the rising of the second peak. Therefore, we performed a $\gamma$-ray analysis on \lat\ data in the MeV to GeV regime to have sufficient data to constrain the broadband SED modelling of the selected sources. In particular, we did this for the 10 out of 17 sources with $\Delta$Cstat $>2.7$ and either \(-1 <\beta < 1.5 \), or $\Gamma<$1.4, which were reported to not be contaminated by host galaxy emission: in 5BZCAT terminology, we exclude from our analysis all sources flagged as BZG, and focused on the BZB subsample. BZBs (BL Lac sources where the broadband emission is dominated by the jet) are, in fact, better suited to have their non-thermal emission properly constrained from jet modelling. In future works, we will extend the analysis to the BZG subsample as well.

For each source, we use the software \texttt{easyfermi} \citep{deMenezes22} to retrieve the photon, spacecraft and diffuse files (the standard galactic interstellar emission model \texttt{gll\_iem\_v07.fits} and isotropic spectral template \texttt{ iso\_P8R3\_SOURCE\_V3\_v1.txt}) for data collected between 2008-08-04 15:43:36 UTC, and 2025-01-01 00:00:01 UTC at energies in the 100 MeV -- 300 GeV range at the radio coordinates of the sources under study. We also use \texttt{easyfermi} to perform the standard likelihood analysis on the data through \texttt{fermipy} \citep{wood17} functions, since it can be used to study even non-catalogued sources: such targets are in fact added as point-sources with a power-law spectrum ($\Gamma = 2$). The software is therefore used to first optimise the model, using default settings, to add to it absent sources with significance $>5\sigma$, and to fit the data to a power law model using the optimizer \texttt{NewMinuit}. During the fit, the radius within which the parameters of all sources are free to vary (normalisation and spectral shape) is set as half the width of the region of
interest (ROI). When the fit does not converge, \texttt{easyfermi} automatically reruns it after deleting sources with lower significance than the target source. Finally, we use the software to obtain the high-energy SED of each source in our sample, considering for each source 10 energy bins. The width of the ROI and the zenith angle ($z_{max}$) cut are set by the standard mode of the software and depended on the minimum energy set up, the ROI decreasing progressively from 17$^{\circ}$ to 10$^{\circ}$ around the source location and $z_{max}$ growing from 80$^{\circ}$ to 105$^{\circ}$ for increasing energies. We use test statistic (TS) to determine sources significance and whether the SED points (obtained at the 95\% confidence level) were to be considered as actual data points or upper limits. In fact, since the TS is defined as TS$\rm = -2 (log\mathcal{L}_{null}-log\mathcal{L}_{src})$, where $\mathcal{L}_{null}$ is the likelihood under the background-only hypothesis and $\mathcal{L}_{src}$ is the maximum likelihood of the data given a model including a source at the position of interest \citep{mattox96}, we set the relevant threshold for SED data points at TS $=25$ ($\approx5\sigma$ significance). All sources SEDs ended up being made of upper limits only. 

\section{Broadband Spectral Energy Distribution modelling}\label{sec:SEDmod}

For the broadband modelling of selected blazars ($\Delta$Cstat $>$ 2.7 and $-1$ $< \beta <$ 1.5), we generate multi-wavelength SEDs using the software JetSeT\footnote{\url{https://jetset.readthedocs.io/en/1.3.0/}}
\citep[v1.3.0,][]{tramacere09,tramacere11,tramacere20}, and compare them with the multi-wavelength data collected for each source through the software VOU-Blazar \citep{chang20}, accessible from the web tool Firmamento\footnote{\url{https://firmamento.nyuad.nyu.edu/}}~\citep{giommi25}. Firmamento is an integrated platform designed to access, explore, and characterise released astronomical data for multi-frequency sources. It brings together information from a wide range of catalogues and observatories, enabling the construction of multi-wavelength spectral energy distributions, the search for redshift information, and the identification of potential counterparts through the Error Region Counterpart Identifier (ERCI) tool; in this work, the latter component has not been used since the coordinates of the sources under investigation were already available from \citealt{marchesi25} catalogue. By combining simultaneous and non-simultaneous observations from different instruments, Firmamento also allows users to visually investigate long-term variability across the electromagnetic spectrum.

The non-detection of selected blazars in the GeV domain prevented us from performing an actual fit of the modelled SEDs. Therefore, with the goal of obtaining a more realistic range of possible characterisations of the observed SEDs, we produce multiple modellings, testing different combinations of parameters, and visually compared them with the SED data points. To generate our SEDs, we use as a reference the recent paper by \citet{hota24}, who performed one of the most detailed modelling of the SED of an EHSP blazar: 1ES 0229+200. We therefore work under the assumption that the SED of this EHSP, despite its extreme nature, can be treated as a good first approximation of the SED of other such sources.

In our analysis, the broadband emission is assumed to have jet origin and is modelled with a one-zone synchrotron self-Compton (SSC) leptonic model: such an approach has been used multiple times to explain VHE emission from blazars in general and EHSPs in particular \citep[see, e.g.,][]{maraschi92,costamante18,foffano19}. In the model, the emission is assumed to arise from a jet region of spherical shape and comoving radius R, filled by a relativistic plasma and a tangled, homogeneous magnetic field, B, which stores a magnetic field energy density of $ \rm \mathcal{U}_B = B^2 / 8 \pi$.  The region also contains a population of relativistic leptons (mainly electrons), N($\gamma$), which are the other main energetic component of the jet. The region is moving along the jet with a bulk Lorentz factor $\Gamma$, and $\theta$ is the angle between the jet axis and the observer line of sight. From these parameters we can derive the Doppler factor $\delta=[\Gamma(1-\beta cos \theta)]^{-1}$, which describes the boosting of the emission.
Due to their interactions with the jet magnetic field, the electrons in the blob radiate via synchrotron emission. The synchrotron radiation then provides the seed photons for the IC mechanism, thus justifying the SSC name. For each source, we test two empirical particle distributions for the lepton population, as in \citet{hota24}:

\begin{enumerate}
    \item A log parabola model, defined as:
    \begin{equation}
    \label{eq:logparEDISTR}
        \rm N(\gamma) = N_0 \bigg(\frac{\gamma}{\gamma_0}\bigg)^{-s -r \ log(\gamma/\gamma_0)}   \ \ \ for \ \gamma_{min} \le \gamma \le \gamma_{max}
    \end{equation}
    where $\rm N_0$ is a normalisation factor, s is the spectral slope, and r is the spectral curvature. $\rm \gamma_{min}$ and $\gamma_{max}$ are respectively the low-energy and high-energy cut-off for the leptons.
    \item A broken power law model, defined as:
    \begin{equation}
    \label{eq:bknEDISTR}
        \rm N(\gamma)= 
        \begin{cases}
            \rm N_0 \gamma^{-p_1} \ \ \ \ \ \ \ \ \ \ \ \ for \ \gamma_{min} \le \gamma \le  \gamma_{break} \\
            \rm N_0 \gamma_{break}^{p_2-p_1}\gamma^{-p_2} \ \ \ for \ \gamma_{break}<\gamma \le \gamma_{max} 
        \end{cases}
    \end{equation}
    where $\rm p_1$ and $\rm p_2$ are the low energy and high energy spectral slopes and $\gamma_{break}$ is the turn-over energy.
\end{enumerate}

Finally, in the high-energy end of the SED, the modelling accounts for $\gamma$-ray attenuation due to the interaction with the extragalactic background light (EBL), which was implemented through JetSeT following the model by \citet{saldana21} and \citet{dominguez24}. The EBL is the integrated intensity of radiation emitted by stars in the UV, optical and IR  over cosmic history, which gets redshifted and accumulated in intergalactic space and can therefore interact with photons emitted by different sources. In particular, EBL photons interacting with very energetic $\gamma$-ray photons can annihilate, leading to a significant decrease in the observed flux of sources emitting at energies above hundreds of GeV \citep[see][for further details]{cooray16,cao23}.

We also compare the final, EBL--corrected models, and in particular their predicted VHE ($>100$ GeV) flux, with the expected sensitivity of the CTAO arrays in their Alpha Configuration,  to assess the future detectability of the blazars under investigation. We determine the best array for each source (i.e., North or South) based on their coordinates: specifically, we pick the array that leads to the highest altitude above the horizon throughout the year. 
Then, we compute the array sensitivity by downloading the instrument response functions made available by the CTAO collaboration \citep[version prod5 v0.1;][]{cherenkov2021}\footnote{\url{https://www.ctao-observatory.org/science/cta-performance/ }} for a deep, 50 hours observation at a 20$^\circ$ zenith by both arrays, and then analysing them with \texttt{gammapy}\footnote{\url{https://docs.gammapy.org/2.0/tutorials/analysis-1d/cta_sensitivity.html}} \citep[v1.2;][]{gammapy:2023,gammapy:zenodo-1.2}.

\subsection{Parameters selection}\label{sec:SEDparams}

\begin{table}
\caption{\label{tab:logpar_par} Log Parabola Model Parameters}
\centering
\scalebox{0.8}{
\renewcommand*{\arraystretch}{1.2}
\begin{tabular}{|l|c|c|c|c|c|}
\hline
 & \textcolor[HTML]{1f77b4}{Epoch 1} & \textcolor[HTML]{ff7f0e}{Epoch 2} & \textcolor[HTML]{2ca02c}{Epoch 3} & \textcolor[HTML]{d62728}{Epoch 4} & \textcolor[HTML]{9467bd}{Epoch 5} \\
 &  \makecell{\scriptsize{2017} \\ \scriptsize{Sep 21–23}}  &  \makecell{\scriptsize{2017} \\ \scriptsize{Dec 9–10}}  &  \makecell{\scriptsize{2017} \\ \scriptsize{Dec 21–22}}  &  \makecell{\scriptsize{2018} \\ \scriptsize{Jan 8–9}}  &  \makecell{\scriptsize{2021} \\ \scriptsize{Aug 8–12}}  \\
\hline
\texttt{R} [cm]& \multicolumn{5}{c|}{$10^{17}$} \\
\hline
\texttt{$\delta$} & \multicolumn{5}{c|}{40} \\
\hline
\texttt{B} [mG]& 1.02 & 2.93 & 2.80 & 3.04 & 1.38 \\
\hline
\texttt{r} & 2.40 & 2.65 & 2.59 & 2.67 & 2.44\\
\hline
\texttt{s} & 0.34 & 0.27 & 0.29 & 0.26 & 0.32\\
\hline
\end{tabular}}
\tablefoot{ Parameters set for the modelling of the blazar SEDs assuming an electron distribution following the log parabola model: these are assumed to be equal for each source we analysed. The different epochs are those of \citet{hota24}, as discussed in the text.}
\end{table}

\begin{figure}[htbp]
        \centering
        \includegraphics[width=0.45\textwidth,clip]{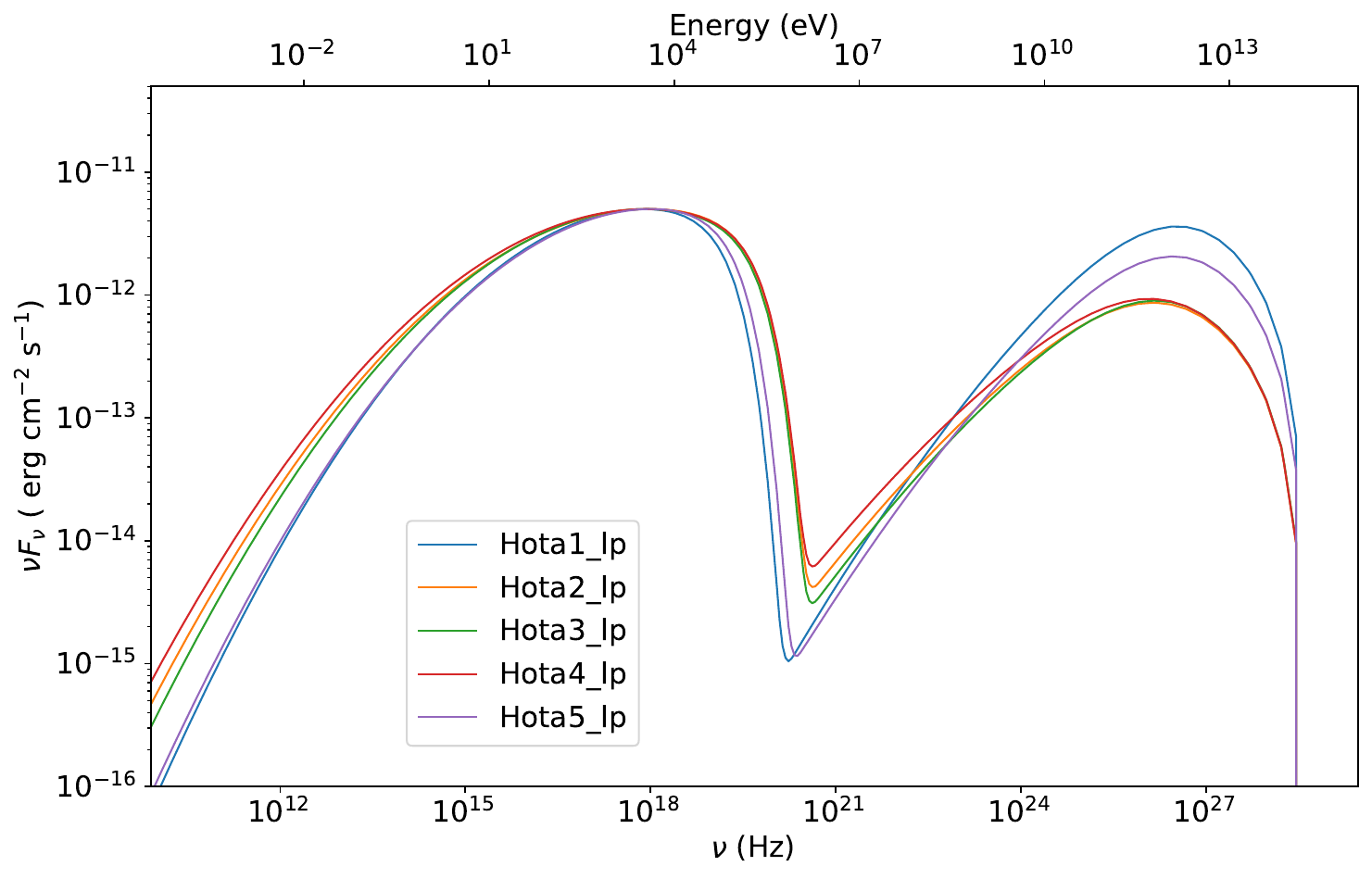}
        \caption{\label{fig:models_lp} Models for different broadband SEDs of the blazar 1ES0229+200, assuming a log parabola distribution for electrons, $z$=0.139,  and $\gamma_0$ of the order of $10^5$, varying in the range 2.8$\times 10^5$-3.9$\times10^5$, due to the different values of B used in each model, as reported in Table~\ref{tab:logpar_par}.}
        
    \end{figure}

\begin{table}
\caption{\label{tab:powlaw_par} Broken Power Law Model Parameters}
\centering
\scalebox{0.8}{
\renewcommand*{\arraystretch}{1.2}
\begin{tabular}{|l|c|c|c|c|c|}

\hline
 & \textcolor[HTML]{1f77b4}{Epoch 1} & \textcolor[HTML]{ff7f0e}{Epoch 2} & \textcolor[HTML]{2ca02c}{Epoch 3} & \textcolor[HTML]{d62728}{Epoch 4} & \textcolor[HTML]{9467bd}{Epoch 5} \\
 &  \makecell{\scriptsize{2017} \\ \scriptsize{Sep 21–23}}  &  \makecell{\scriptsize{2017} \\ \scriptsize{Dec 9–10}}  &  \makecell{\scriptsize{2017} \\ \scriptsize{Dec 21–22}}  &  \makecell{\scriptsize{2018} \\ \scriptsize{Jan 8–9}}  &  \makecell{\scriptsize{2021} \\ \scriptsize{Aug 8–12}}  \\
\hline
\texttt{R} [cm]& \multicolumn{5}{c|}{$10^{17}$} \\
\hline
\texttt{$\delta$} & \multicolumn{5}{c|}{40} \\
\hline
\texttt{$\gamma_{min}$} & \multicolumn{5}{c|}{10} \\
\hline
\texttt{$\gamma_{max}$} & \multicolumn{5}{c|}{$10^8$} \\
\hline
\texttt{B} [mG]& 1.67 & 3.12 & 2.58 & 3.40 & 1.97 \\
\hline
\texttt{$\rm p_1$} & 2.25 & 2.46 & 2.40 & 2.55 & 2.32\\
\hline
\texttt{$\rm p_2$} & \multicolumn{5}{c|}{4}\\
\hline
\end{tabular}}
\tablefoot{ Parameters set for the modelling of the blazar SEDs assuming an electron distribution following the broken power law model: these are assumed to be equal for each source we analysed. The different epochs are those of \citet{hota24}, as discussed in the text.}
\end{table}

\begin{figure}[htbp]
        \centering
        \includegraphics[width=0.45\textwidth,clip]{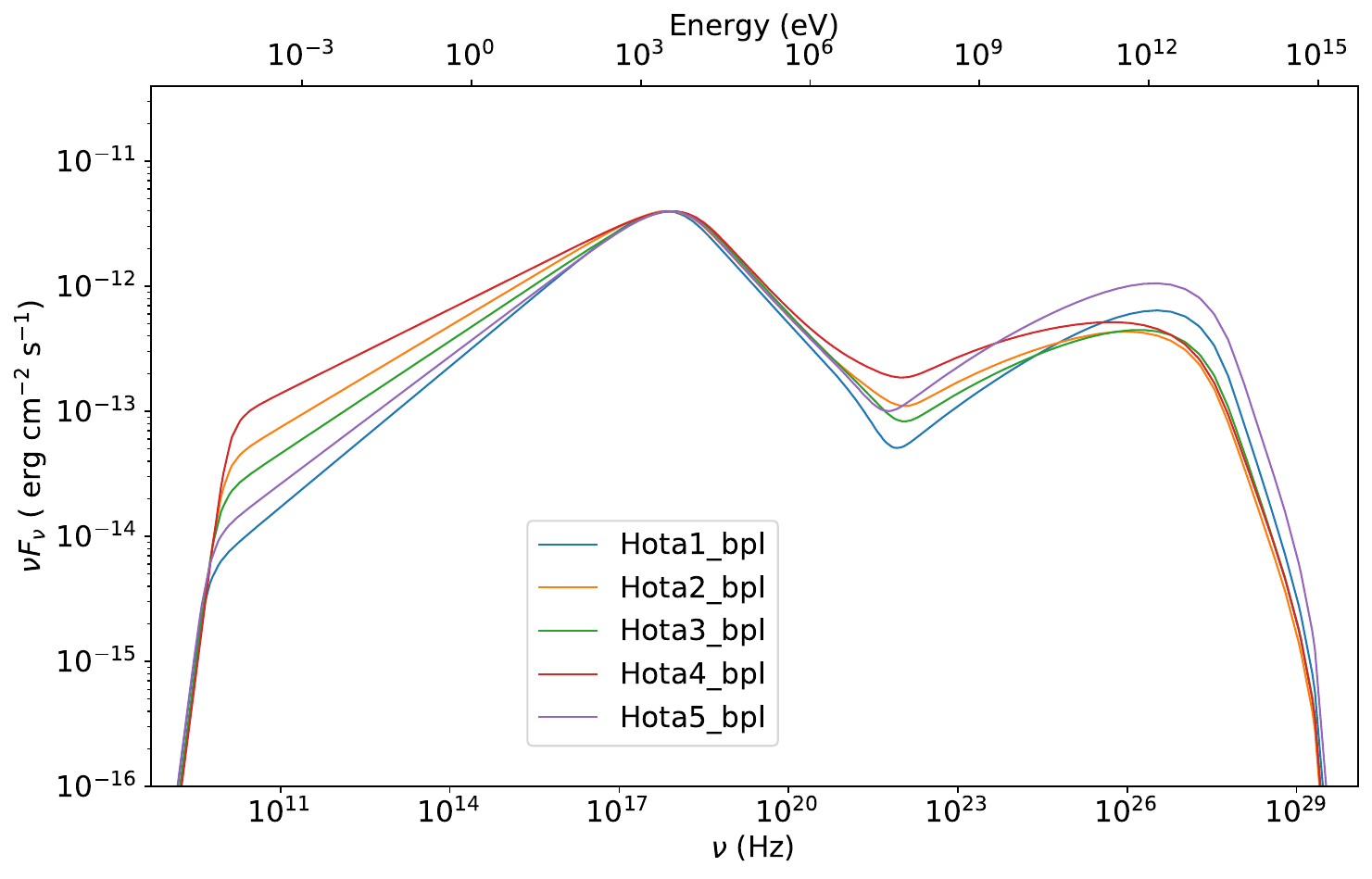}
        \caption{\label{fig:models_pl} Models for different broadband SEDs of the blazar 1ES0229+200, assuming a broken power law distribution for electrons, $z$=0.139,  and $\gamma_{break}$ of the order of $10^6$, varying in the range 1.5$\times 10^6$-2.3$\times10^6$, due to the different values of B used in each model, as reported in Table~\ref{tab:powlaw_par}.  }
        
    \end{figure}

In Tables~\ref{tab:logpar_par} and ~\ref{tab:powlaw_par}, we report the parameters, and their corresponding values, derived by \citet{hota24}: we assumed these to be equal for all the sources in our analysis, and we used them to set the environmental conditions of our blazars. To properly rescale the \citet{hota24} SED to those of our sources we input the correct redshift for each source and we use different JetSeT routines deriving for each source the expected particles emitters density, as well as the maximum $\gamma_{max}$(\texttt{gmax} in JetSeT terminology) and the break energies $\gamma_0$ and $\gamma_{break}$ (\texttt{gamma0\_log\_parab} and \texttt{gamma\_break} for the log parabola and the broken power law models in JetSeT, respectively), 
so that the synchrotron peak frequency and flux of the models match the data, as updated with the new results by the machine learning algorithm BlaST. Specifically, for 6 out of 9 targets, we use the redshift reported in the 5BZCAT. For 2 of the remaining 3 sources with no $z$ information in the 5BZCAT, we instead use the values reported in the 3HSP catalogue \citep{chang19}. Finally, no redshift is available in either of the catalogues for 5BZBJ1258+0134. Following a conservative approach, we therefore assume that it has the same redshift as the farthest source in our subsample of 10 modelled sources (5BZBJ1302+5056, $z$=0.688).
Then, we set the particle emitters density with the JetSeT routine \texttt{set\_N\_from\_nuFnu()}. Afterwards, we compute the break energy required to match the parameters by BlaST by first obtaining the expected Lorentz factor for the peak emission with  \texttt{find\_gamma\_Sync()}, and then feeding it into \texttt{find\_turn\_over()} to find the corresponding energy. This value automatically updates the jet parameters $\gamma_0$ and $\gamma_{break}$, and allows us to compute realistic values for $\gamma_{max}$.

We note that, by manually setting the luminosity of the synchrotron peak, the main driver for the expected flux of the SSC peak becomes the magnetic field, as shown in \citealt{tavecchio98,ghisellini13}: 
 \begin{equation}
        \frac{L_{SSC}}{L_{synch}}= \frac{U_{synch}}{U_B}
    \end{equation}

where $U_{synch}$ and $U_B$  are respectively the synchrotron radiation energy density and the magnetic field energy density, in the comoving frame. This effect can also be clearly distinguished in Figure~\ref{fig:models_lp}, and Figure~\ref{fig:models_pl}: in both the broken power law and log parabola distribution, Epochs 1 and 5, which have the lowest magnetic field values, have IC peaks at the highest VHE fluxes.

\subsection{Source modelling, and the use of 1ES 0229+200 as a prototypical EHBL }\label{sec:ModDescr}

We visually establish the goodness of the models by overlapping the data retrieved through Firmamento and \texttt{easyFermi} with the modelled regions (modelling-by-eye). Then, we assess the detectability by CTAO by comparing the expected VHE flux, as inferred from the models, with the sensitivity of the CTAO array best suited for the observation. 

To validate this qualitative approach, we first performed a sanity check on the EHBL 1ES 0229+200, to verify if JetSeT models could replicate the results obtained by \citet{hota24}: we thus compared the multi-wavelength data with the two models, assuming respectively a log parabolic and a broken power law electron distribution, as shown in Figure~\ref{fig:1ESmod}. 
In both models, at lower frequencies, from the radio to the optical band, the data displays an excess with respect to the model, while at higher frequencies (from the UV band), we observe a good agreement between data and models. 

The radio excess is commonly observed when modelling the SED of blazars, since radio emission in jets tends to be more extended and originated further away from the region of origin of the jet itself, and it cannot consequently be explained with the one-zone model used in this study. It has also been suggested \citep[see, e.g.,][]{massaro04} that, although produced by the same electron population, the optical emission can include a contribution that might not be originated by the same component responsible for the X-ray emission. For these reasons, we do not take into account the SED behaviour in the radio and optical band when comparing data and models in our subsample of sources.

Moving to the high and VHE band, the X- and $\gamma$-ray data of 1ES 0229+200 is nicely reproduced by the JetSeT models, within the observational uncertainties. This confirms that the parametrisation adopted from \citet{hota24} can be reliably implemented in JetSeT and used with confidence for the analysis of the target sample.

\subsection{Results}

We report in Appendix~\ref{app:modelling} the individual analysis of each of the 10 sources in our sample: here we summarise the main results we infer from the sample as a whole. 
In general, we observe that the synchrotron peaks are usually well constrained in frequency, with the data following slightly better the expected behaviour from a log parabola electron distribution model, as visible for example in the modelling of the sources 5BZBJ0040-2719 (Figure~\ref{fig:0040mod}) and 5BZBJ1258+0134 (Figure~\ref{fig:1258mod}). Only for source 5BZBJ1302+5056 (Figure~\ref{fig:1302mod}) the model seems inconsistent with the data, likely because the synchrotron peak flux derived by BLaST is underestimated with respect to the data. At higher energies, in the MeV to GeV regime, the models are generally consistent with the derived upper limits, and only some singular upper limits remain below the expected flux. However, being them isolated cases, they could reasonably be attributed to effects of Poissonian noise.

Overall, for almost all sources (the only exception being 5BZBJ0124+0918, Figure~\ref{fig:0124mod}) the VHE intrinsic emission is well above the sensitivity threshold of CTAO. However, accounting for EBL causes strong attenuation, significantly affecting our chances of detection.  More specifically, assuming the model with log parabolic electron distribution, the sources 5BZBJ1636$-$1248 (Figure~\ref{fig:1636mod}), 5BZBJ1251$-$2958 (Figure~\ref{fig:1251mod}), 5BZBJ0040$-$2719 (Figure~\ref{fig:0040mod}) and 5BZBJ1057+2303 (Figure~\ref{fig:1057mod}) are those with the highest flux value in correspondence of the VHE peak, so that it slightly overcomes the sensitivity threshold. Additionally, for 5BZBJ1253+3826 (Figure~\ref{fig:1253mod}) and 5BZBJ1258+0134 (Figure~\ref{fig:1258mod}) the modelled flux remains just below the sensitivity threshold so they might actually be observable as well in cases of enhanced emission or with longer-lasted observations. However, we caution against the validity of the SED of 5BZBJ1258+0134, since no information concerning the redshift of this source was available, and therefore it was simply assumed to have the same redshift as the farthest object in our sample (5BZBJ1302+5056, $z= 0.688$). This, combined with the fact that this source is the one with the lowest log($\rm \nu_{peak}$)=16.2 among those analysed in this work, makes the estimate of the IC peak less reliable than the ones we derived for the other sources. 

Overall, log parabola models tend to be more diversified and account for a larger range of possible fluxes, resulting in \lat\ upper limits often being more consistent with these models than with the broken power law ones. Indeed, if one assumes the broken power law electron distribution models, all sources display lower ranges of values for the IC peak flux, and tend to have lower absolute fluxes at the peak with respect to the log parabola models. Consequently, log parabola models produce higher chances of observation of the target sources by CTAO. In fact, with the broken power law model no source is expected to have fluxes significantly above CTAO sensitivities, with only 5BZBJ1636$-$1248 (Figure~\ref{fig:1636mod}), 5BZBJ1251$-$2958 (Figure~\ref{fig:1251mod}) and 5BZBJ1057+2303 (Figure~\ref{fig:1057mod}) having a VHE peak consistent with the CTAO detectability threshold.

We cross-checked these results with the VHE detectability tool on the Firmamento platform. The tool computes CTAO detectability for a given blazar provided a redshift based on extrapolation of predictive power of infrared data of blazars in comparison with the population of current TeV detected blazars as reported in the TeVCAT. In all cases we found an agreement. The sources 5BZBJ1253+3826 and 5BZBJ1251$-$2958 show a probability of detection of 71\%, while 5BZBJ0040$-$2719 has a lower detection probability of 33\%. In contrast, 5BZBJ1636$-$1248, 5BZBJ1057+2303 and 5BZBJ1258+0134 are expected to be detected with probabilities exceeding 95\% with CTAO.

\begin{table*}
\caption{\label{tab:ctao_det} Observations needed for best chances of detectability of the sources according to the modelled broadband SED.}
\centering
\scalebox{0.8}{
\renewcommand*{\arraystretch}{1.2}
\begin{tabular}{p{3cm} 
>{\centering\arraybackslash}p{3.6cm}
>{\centering\arraybackslash}p{3.6cm}
>{\centering\arraybackslash}p{3cm}}
\toprule
& \multicolumn{3}{c}{CTAO detectability}\\
Source & log parabola model &  broken power law model  & array visibility\\
\midrule
5BZBJ0333$-$3619   & no & no & CTAO-S\\
5BZBJ1253+3826     & $\ge$100 h  & no & CTAO-N\\
5BZBJ1636$-$1248     &  50h & $\ge$100 h & CTAO-N\\
5BZBJ1251$-$2958   & 50 h & $\ge$100 h & both \\
5BZBJ0040$-$2719   & 50 h & no & both \\
5BZBJ1302+5056     & no & no & CTAO-N\\
5BZBJ1057+2303     & 50 h & $\ge$100 h & both\\
5BZBJ2217$-$3106   & no & no & CTAO-S\\
5BZBJ0124+0918     & no & no & both \\
5BZBJ1258+0134     & $\ge$100 h & no & both \\
\bottomrule
\end{tabular}}
\tablefoot{ The detectability was assessed visually by comparing the fluxes reached by the model with the CTAO-N or CTAO-S sensitivity. We assumed that a target is visible by an array if during the year it reaches an altitude higher than 30$^\circ$ above the horizon at its site declination.}
\end{table*}

\begin{table*}
\caption{\label{tab:detectable} Properties of the sources with chances of observation as reported in Table~\ref{tab:ctao_det}. }
\centering
\scalebox{0.8}{
\renewcommand*{\arraystretch}{1.2}
\begin{tabular}{cccccc}
\toprule
Source & RA, DEC & redshift & $ \rm \nu F_{\nu,\  2-10 \ keV}$ & $ \rm \nu F_{\nu,\ 100 \ GeV, \ LP \ model}$ & $ \rm \nu F_{\nu,\ 100 \ GeV , \ BPL \ model}$\\
& & &   [10$^{-12}$ erg s$^{-1}$ cm$^{-2}$] & [10$^{-12}$ erg s$^{-1}$ cm$^{-2}$] & [10$^{-12}$ erg s$^{-1}$ cm$^{-2}$]\\
\midrule
5BZBJ1253+3826     & 12:53:00.90, +38:26:26.01 & 0.371 & $1.46^{+0.22}_{-0.19}$ & 0.17 - 0.63 & 0.13 - 0.15\\
5BZBJ1636$-$1248   & 16:36:58.42,  \ $-$12:48:36.5 & 0.246 & $0.61^{+0.40}_{-0.26}$  & 0.30 - 1.52 & 0.21 - 0.51\\
5BZBJ1251$-$2958   & 12:51:34.88,  \ $-$29:58:42.88 & 0.382 & $0.73^{+0.12}_{-0.10}$  & 0.37 - 1.46 & 0.29 - 0.64\\
5BZBJ0040$-$2719   & 00:40:16.40,  \ $-$27:19:11.60 & 0.172 &  $0.59^{+0.08}_{-0.08}$ & 0.19 - 1.00 & 0.12 - 0.18 \\
5BZBJ1057+2303     & 10:57:23.09, +23:03:18.79 & 0.379 & $1.22^{+0.33}_{-0.26}$ & 0.51 -1.38 & 0.58 - 1.16\\
5BZBJ1258+0134     & 12:58:54.59, +01:34:41.41 & -99 & $0.29^{+0.56}_{-0.17}$ & 0.16 - 1.10 & 0.09 - 0.37\\
\bottomrule
\end{tabular}}
\tablefoot{ The coordinates (RA, DEC) and redshift are reported from the 5BZCAT catalogue, while the fluxes come from this work. The X-ray flux in the band 2-10 keV comes from the X-ray analysis results, as reported in Table~\ref{tab:result_xanalysis}, while the VHE flux at 100 GeV corresponds to the range of fluxes established by the models in Section~\ref{sec:SEDmod}. As mentioned in the text, we caution that the lack of a spectroscopic redshift can affect the reliability of the VHE predictions for 5BZBJ1258+0134.}
\end{table*}

We report a recap of the observable sources according to the two models and from the two CTAO sites in Table~\ref{tab:ctao_det}: the source is considered observable from a given observatory if it reaches an altitude above 30$^\circ$ in the sky during the year at the array site. 
In Table~\ref{tab:detectable} we instead report the specific properties of the sources that can be expected to be detectable by CTAO. Here, one can find evidence of the X-ray to VHE correlation typically observed in EHSPs, since from the models we derive a VHE flux with the same order of magnitude or just one order of magnitude below the X-ray flux.

\section{Conclusions and future developments}\label{sec:conclusions}

We analysed a sample of 78 X-ray bright, non-4FGL-detected HSP and EHSP blazars with the goal of identifying sources with clear evidence of a spectral curvature proving the presence of the synchrotron peak in the X-ray and obtained an estimate of their VHE flux through modelling. Our sample is selected from a larger population of $\sim 1000$ X-ray emitting blazar, not reported in the 4FGL-DR4 catalogue by \lat, compiled by \citet{marchesi25} from the 5BZCAT catalogue: we picked those 78 HSP/EHSPs (log($\rm \nu_{peak}) > 16$) with a high X-ray flux ($\rm log(F_{0.2-12 \ keV})>-12.5$).

EHSPs are a population of extreme emitters that have only recently been studied, and is therefore still poorly characterised. Only a few of them have complete, broadband SEDs, due to their faintness with respect to other blazar classes, which also leads to limited detections in the VHE band with current Cherenkov telescopes. However, the upcoming CTAO, with highly improved sensitivity in the 20 GeV–300 TeV range, will significantly expand this view. Therefore, it is important to make predictions on the population of blazars which will be detectable by CTAO, developing new strategies to complement predictions based on extrapolations from $\gamma$-ray data, which in principle could overlook sources whose IC peak is still rising, and therefore not detectable, at \lat\ energies. 

The main results of this work can be summarised as follows:
\begin{enumerate}
    \item We performed an X-ray analysis using calibrated data retrieved by \cha, \xrt, \xmm\ and/or eROSITA. Comparing fits between log parabola and power law models, 17 sources showed significant curvature ($\rm \Delta Cstat>2.7$, and reasonable values of $\beta$) showing evidence of synchrotron peaks in the X-ray band. Due to such properties, these sources were considered viable candidates for TeV emission and selected for further analysis.
    \item Out of the 17 sources with X-ray spectrum best fitted by the log parabola model, we further selected the 10 objects with SED dominated by the jet emission (i.e., those sources classified as BZB in the 5BZCAT).  We then used the \texttt{easyfermi} software to compute the 100\,MeV -- 300\,GeV upper limits, taking advantage of over 16 years of \lat\ data
    \item We modelled the broadband SEDs of these 10 sources using the physically motivated SED modelling tool JetSeT, assuming both log parabola and broken power law electron distributions, with physical parameters guided by the modelling of the EHSP 1ES 0229+200 performed by \citet{hota24}. EBL attenuation in the VHE band was also considered and included in the modelling, causing strong attenuation at energies $>$ 100 GeV. Through visual inspection, all sources showed a good agreement between the models and the data (detections and upper limits). With this analysis, we validated the adopted JetSeT models, showing how calibrating them on 1ES 0229+200 provides a reliable description of the observed SEDs. 
    \item By comparing the predicted IC component attenuated by the EBL with the CTAO Northern and Southern array sensitivity curves, we identified 6 sources as promising candidates for detection in the VHE band: 5BZBJ1636$-$1248, 5BZBJ1251$-$2958, 5BZBJ0040$-$2719, 5BZBJ1057+2303, 5BZBJ1253+3826  and 5BZBJ1258+0134. Assuming a log parabolic distribution for electrons, the first 4 sources are bright enough to be detected by CTAO in a $\sim$50-hour, survey--like observation, while the latter 2 may require deeper pointed campaigns. For 5BZBJ1258+0134, the lack of information on its redshift reduces the reliability of the modelled IC peak, making it a lower-priority CTAO target.
\end{enumerate}

Clearly, a higher number of candidate TeV emitting HSP/EHSP blazars may exist beyond the small subset identified here. For many targets in our sample, the X-ray data quality from archival observations  was often insufficient to perform a statistically meaningful comparison between the broken power law model and the log parabola one. To address this problem, dedicated follow-up campaigns with \xmm\ would allow one to work with higher-quality X-ray spectra, potentially leading to the observation of the spectral curvature expected for TeV candidates even in some of the sources excluded in this work.

Furthermore, for the most promising objects among those selected, additional simultaneous observations with \textit{NuSTAR} (which would extend the X-ray spectral coverage up to $\sim$ 50 \ keV) and one of the other X-ray telescopes covering the $<$3\,keV band missed by \textit{NuSTAR} would provide valuable coverage of the hard X-ray band, helping to better constrain the entire shape of the synchrotron peak. This would reduce uncertainties in the modelling of the one-zone emission region and of the particle population, and in turn obtain even stronger constraints on the location and intensity of the IC peak. 

At higher energies, in the MeV to TeV regime, additional data and analysis of some selected sources would also be highly valuable to constrain their IC peak. Additional \lat\ data could be collected and analysed exploiting long exposures and targeted likelihood fits to uncover faint HSPs below the detection threshold of other catalogues \citep[a similar approach was explored in the 2BIGB catalogue of blazars candidates,][]{arsioli20}, and is currently being done to obtain the 4FHL catalogue \citep[hard-spectrum sources above 50 GeV,][]{rico25}. Alternatively, one could use a likelihood profile stacking technique to measure the emission from the selected EHSP population as a whole, expanding the work by \citealt{paliya19} on another large population of non-$\gamma$-ray detected HSP. Using these techniques, as well as attempting observations with current IACTs \citep[mainly MAGIC, in combination with the LSTs already operating; respectively][]{aleksic16, cortina19} for the most promising sources, may help confirm or re-evaluate their chances of detection by CTAO.

\section*{Data Availability}
Table A.1 is only available in electronic form at the CDS via anonymous ftp to cdsarc.u-strasbg.fr (130.79.128.5) or via \url{http://cdsweb.u-strasbg.fr/cgi-bin/qcat?J/A+A/}.

\section*{Acknowledgments} 
We thank the referee for their useful suggestions.
The research activities described in this paper were carried out with contribution of the Next Generation EU funds within the National Recovery and Resilience Plan (PNRR), Mission 4 - Education and Research, Component 2 - From Research to Business (M4C2), Investment Line 3.1 - Strengthening and creation of Research Infrastructures, Project IR0000012 – ``CTA+ - Cherenkov Telescope Array Plus''.
This work made use of data supplied by the UK Swift Science Data Centre at the University of Leicester. This work made use of Gammapy \citep{gammapy:2023}, a community-developed Python package. The Gammapy team acknowledges all Gammapy past and current contributors, as well as all contributors of the main Gammapy dependency libraries: \href{https://numpy.org/}{NumPy} \citep{harris20}, \href{https://scipy.org/}{SciPy}, \href{http://www.astropy.org}{Astropy} \citep{astropy22}, \href{https://astropy-regions.readthedocs.io/}{Astropy Regions}, \href{https://scikit-hep.org/iminuit/}{iminuit}, \href{https://matplotlib.org/}{Matplotlib} \citep{hunter07}.
We thank Andrea Tramacere for helpful clarifications on the use of the JetSet software.

\bibliographystyle{aa}
\bibliography{Xray_blazars_biblio}

@ARTICLE{abdo10,
       author = {{Abdo}, A.~A. and {Ackermann}, M. and {Agudo}, I. and {Ajello}, M. and {Aller}, H.~D. and {Aller}, M.~F. and {Angelakis}, E. and {Arkharov}, A.~A. and {Axelsson}, M. and {Bach}, U. and {Baldini}, L. and {Ballet}, J. and {Barbiellini}, G. and {Bastieri}, D. and {Baughman}, B.~M. and {Bechtol}, K. and {Bellazzini}, R. and {Benitez}, E. and {Berdyugin}, A. and {Berenji}, B. and {Blandford}, R.~D. and {Bloom}, E.~D. and {Boettcher}, M. and {Bonamente}, E. and {Borgland}, A.~W. and {Bregeon}, J. and {Brez}, A. and {Brigida}, M. and {Bruel}, P. and {Burnett}, T.~H. and {Burrows}, D. and {Buson}, S. and {Caliandro}, G.~A. and {Calzoletti}, L. and {Cameron}, R.~A. and {Capalbi}, M. and {Caraveo}, P.~A. and {Carosati}, D. and {Casandjian}, J.~M. and {Cavazzuti}, E. and {Cecchi}, C. and {{\c{C}}elik}, {\"O}. and {Charles}, E. and {Chaty}, S. and {Chekhtman}, A. and {Chen}, W.~P. and {Chiang}, J. and {Chincarini}, G. and {Ciprini}, S. and {Claus}, R. and {Cohen-Tanugi}, J. and {Colafrancesco}, S. and {Cominsky}, L.~R. and {Conrad}, J. and {Costamante}, L. and {Cutini}, S. and {D'ammando}, F. and {Deitrick}, R. and {D'Elia}, V. and {Dermer}, C.~D. and {de Angelis}, A. and {de Palma}, F. and {Digel}, S.~W. and {Donnarumma}, I. and {Silva}, E. do Couto e. and {Drell}, P.~S. and {Dubois}, R. and {Dultzin}, D. and {Dumora}, D. and {Falcone}, A. and {Farnier}, C. and {Favuzzi}, C. and {Fegan}, S.~J. and {Focke}, W.~B. and {Forn{\'e}}, E. and {Fortin}, P. and {Frailis}, M. and {Fuhrmann}, L. and {Fukazawa}, Y. and {Funk}, S. and {Fusco}, P. and {G{\'o}mez}, J.~L. and {Gargano}, F. and {Gasparrini}, D. and {Gehrels}, N. and {Germani}, S. and {Giebels}, B. and {Giglietto}, N. and {Giommi}, P. and {Giordano}, F. and {Giuliani}, A. and {Glanzman}, T. and {Godfrey}, G. and {Grenier}, I.~A. and {Gronwall}, C. and {Grove}, J.~E. and {Guillemot}, L. and {Guiriec}, S. and {Gurwell}, M.~A. and {Hadasch}, D. and {Hanabata}, Y. and {Harding}, A.~K. and {Hayashida}, M. and {Hays}, E. and {Healey}, S.~E. and {Heidt}, J. and {Hiriart}, D. and {Horan}, D. and {Hoversten}, E.~A. and {Hughes}, R.~E. and {Itoh}, R. and {Jackson}, M.~S. and {J{\'o}hannesson}, G. and {Johnson}, A.~S. and {Johnson}, W.~N. and {Jorstad}, S.~G. and {Kadler}, M. and {Kamae}, T. and {Katagiri}, H. and {Kataoka}, J. and {Kawai}, N. and {Kennea}, J. and {Kerr}, M. and {Kimeridze}, G. and {Kn{\"o}dlseder}, J. and {Kocian}, M.~L. and {Kopatskaya}, E.~N. and {Koptelova}, E. and {Konstantinova}, T.~S. and {Kovalev}, Y.~Y. and {Kovalev}, Yu. A. and {Kurtanidze}, O.~M. and {Kuss}, M. and {Lande}, J. and {Larionov}, V.~M. and {Latronico}, L. and {Leto}, P. and {Lindfors}, E. and {Longo}, F. and {Loparco}, F. and {Lott}, B. and {Lovellette}, M.~N. and {Lubrano}, P. and {Madejski}, G.~M. and {Makeev}, A. and {Marchegiani}, P. and {Marscher}, A.~P. and {Marshall}, F. and {Max-Moerbeck}, W. and {Mazziotta}, M.~N. and {McConville}, W. and {McEnery}, J.~E. and {Meurer}, C. and {Michelson}, P.~F. and {Mitthumsiri}, W. and {Mizuno}, T. and {Moiseev}, A.~A. and {Monte}, C. and {Monzani}, M.~E. and {Morselli}, A. and {Moskalenko}, I.~V. and {Murgia}, S. and {Nestoras}, I. and {Nilsson}, K. and {Nizhelsky}, N.~A. and {Nolan}, P.~L. and {Norris}, J.~P. and {Nuss}, E. and {Ohsugi}, T. and {Ojha}, R. and {Omodei}, N. and {Orlando}, E. and {Ormes}, J.~F. and {Osborne}, J. and {Ozaki}, M. and {Pacciani}, L. and {Padovani}, P. and {Pagani}, C. and {Page}, K. and {Paneque}, D. and {Panetta}, J.~H. and {Parent}, D. and {Pasanen}, M. and {Pavlidou}, V. and {Pelassa}, V. and {Pepe}, M. and {Perri}, M. and {Pesce-Rollins}, M. and {Piranomonte}, S. and {Piron}, F. and {Pittori}, C. and {Porter}, T.~A. and {Puccetti}, S. and {Rahoui}, F. and {Rain{\`o}}, S. and {Raiteri}, C. and {Rando}, R. and {Razzano}, M. and {Reimer}, A. and {Reimer}, O. and {Reposeur}, T. and {Richards}, J.~L. and {Ritz}, S. and {Rochester}, L.~S. and {Rodriguez}, A.~Y. and {Romani}, R.~W. and {Ros}, J.~A. and {Roth}, M. and {Roustazadeh}, P. and {Ryde}, F. and {Sadrozinski}, H.~F. -W. and {Sadun}, A. and {Sanchez}, D. and {Sander}, A. and {Saz Parkinson}, P.~M. and {Scargle}, J.~D. and {Sellerholm}, A. and {Sgr{\`o}}, C. and {Shaw}, M.~S. and {Sigua}, L.~A. and {Siskind}, E.~J. and {Smith}, D.~A. and {Smith}, P.~D. and {Spandre}, G. and {Spinelli}, P. and {Starck}, J. -L. and {Stevenson}, M. and {Stratta}, G. and {Strickman}, M.~S. and {Suson}, D.~J. and {Tajima}, H. and {Takahashi}, H. and {Takahashi}, T. and {Takalo}, L.~O. and {Tanaka}, T. and {Thayer}, J.~B. and {Thayer}, J.~G. and {Thompson}, D.~J. and {Tibaldo}, L. and {Torres}, D.~F. and {Tosti}, G. and {Tramacere}, A. and {Uchiyama}, Y. and {Usher}, T.~L. and {Vasileiou}, V. and {Verrecchia}, F. and {Vilchez}, N. and {Villata}, M. and {Vitale}, V. and {Waite}, A.~P. and {Wang}, P. and {Winer}, B.~L. and {Wood}, K.~S. and {Ylinen}, T. and {Zensus}, J.~A. and {Zhekanis}, G.~V. and {Ziegler}, M.},
        title = "{The Spectral Energy Distribution of Fermi Bright Blazars}",
      journal = {\apj},
         year = 2010,
        month = jun,
       volume = {716},
       number = {1},
        pages = {30-70},
          doi = {10.1088/0004-637X/716/1/30},
archivePrefix = {arXiv},
       eprint = {0912.2040},
 primaryClass = {astro-ph.CO},
       adsurl = {https://ui.adsabs.harvard.edu/abs/2010ApJ...716...30A}
}

@ARTICLE{abdollahi22,
       author = {{Abdollahi}, S. and {Acero}, F. and {Baldini}, L. and {Ballet}, J. and {Bastieri}, D. and {Bellazzini}, R. and {Berenji}, B. and {Berretta}, A. and {Bissaldi}, E. and {Blandford}, R.~D. and {Bloom}, E. and {Bonino}, R. and {Brill}, A. and {Britto}, R.~J. and {Bruel}, P. and {Burnett}, T.~H. and {Buson}, S. and {Cameron}, R.~A. and {Caputo}, R. and {Caraveo}, P.~A. and {Castro}, D. and {Chaty}, S. and {Cheung}, C.~C. and {Chiaro}, G. and {Cibrario}, N. and {Ciprini}, S. and {Coronado-Bl{\'a}zquez}, J. and {Crnogorcevic}, M. and {Cutini}, S. and {D'Ammando}, F. and {De Gaetano}, S. and {Digel}, S.~W. and {Di Lalla}, N. and {Dirirsa}, F. and {Di Venere}, L. and {Dom{\'\i}nguez}, A. and {Fallah Ramazani}, V. and {Fegan}, S.~J. and {Ferrara}, E.~C. and {Fiori}, A. and {Fleischhack}, H. and {Franckowiak}, A. and {Fukazawa}, Y. and {Funk}, S. and {Fusco}, P. and {Galanti}, G. and {Gammaldi}, V. and {Gargano}, F. and {Garrappa}, S. and {Gasparrini}, D. and {Giacchino}, F. and {Giglietto}, N. and {Giordano}, F. and {Giroletti}, M. and {Glanzman}, T. and {Green}, D. and {Grenier}, I.~A. and {Grondin}, M. -H. and {Guillemot}, L. and {Guiriec}, S. and {Gustafsson}, M. and {Harding}, A.~K. and {Hays}, E. and {Hewitt}, J.~W. and {Horan}, D. and {Hou}, X. and {J{\'o}hannesson}, G. and {Karwin}, C. and {Kayanoki}, T. and {Kerr}, M. and {Kuss}, M. and {Landriu}, D. and {Larsson}, S. and {Latronico}, L. and {Lemoine-Goumard}, M. and {Li}, J. and {Liodakis}, I. and {Longo}, F. and {Loparco}, F. and {Lott}, B. and {Lubrano}, P. and {Maldera}, S. and {Malyshev}, D. and {Manfreda}, A. and {Mart{\'\i}-Devesa}, G. and {Mazziotta}, M.~N. and {Mereu}, I. and {Meyer}, M. and {Michelson}, P.~F. and {Mirabal}, N. and {Mitthumsiri}, W. and {Mizuno}, T. and {Moiseev}, A.~A. and {Monzani}, M.~E. and {Morselli}, A. and {Moskalenko}, I.~V. and {Negro}, M. and {Nuss}, E. and {Omodei}, N. and {Orienti}, M. and {Orlando}, E. and {Paneque}, D. and {Pei}, Z. and {Perkins}, J.~S. and {Persic}, M. and {Pesce-Rollins}, M. and {Petrosian}, V. and {Pillera}, R. and {Poon}, H. and {Porter}, T.~A. and {Principe}, G. and {Rain{\`o}}, S. and {Rando}, R. and {Rani}, B. and {Razzano}, M. and {Razzaque}, S. and {Reimer}, A. and {Reimer}, O. and {Reposeur}, T. and {S{\'a}nchez-Conde}, M. and {Saz Parkinson}, P.~M. and {Scotton}, L. and {Serini}, D. and {Sgr{\`o}}, C. and {Siskind}, E.~J. and {Smith}, D.~A. and {Spandre}, G. and {Spinelli}, P. and {Sueoka}, K. and {Suson}, D.~J. and {Tajima}, H. and {Tak}, D. and {Thayer}, J.~B. and {Thompson}, D.~J. and {Torres}, D.~F. and {Troja}, E. and {Valverde}, J. and {Wood}, K. and {Zaharijas}, G.},
        title = "{Incremental Fermi Large Area Telescope Fourth Source Catalog}",
      journal = {\apjs},
         year = 2022,
        month = jun,
       volume = {260},
       number = {2},
          eid = {53},
        pages = {53},
          doi = {10.3847/1538-4365/ac6751},
archivePrefix = {arXiv},
       eprint = {2201.11184},
 primaryClass = {astro-ph.HE},
       adsurl = {https://ui.adsabs.harvard.edu/abs/2022ApJS..260...53A}
}

@ARTICLE{arsioli15,
       author = {{Arsioli}, B. and {Fraga}, B. and {Giommi}, P. and {Padovani}, P. and {Marrese}, P.~M.},
        title = "{1WHSP: An IR-based sample of \raisebox{-0.5ex}\textasciitilde1000 VHE {\ensuremath{\gamma}}-ray blazar candidates}",
      journal = {\aap},
         year = 2015,
        month = jul,
       volume = {579},
          eid = {A34},
        pages = {A34},
          doi = {10.1051/0004-6361/201424148},
archivePrefix = {arXiv},
       eprint = {1504.02801},
 primaryClass = {astro-ph.HE},
       adsurl = {https://ui.adsabs.harvard.edu/abs/2015A&A...579A..34A}
}

@ARTICLE{ballet23,
       author = {{Ballet}, J. and {Bruel}, P. and {Burnett}, T.~H. and {Lott}, B. and {The Fermi-LAT collaboration}},
        title = "{Fermi Large Area Telescope Fourth Source Catalog Data Release 4 (4FGL-DR4)}",
      journal = {arXiv e-prints},
         year = 2023,
        month = jul,
          eid = {arXiv:2307.12546},
        pages = {arXiv:2307.12546},
          doi = {10.48550/arXiv.2307.12546},
archivePrefix = {arXiv},
       eprint = {2307.12546},
 primaryClass = {astro-ph.HE},
       adsurl = {https://ui.adsabs.harvard.edu/abs/2023arXiv230712546B}
}

@ARTICLE{bennett14,
       author = {{Bennett}, C.~L. and {Larson}, D. and {Weiland}, J.~L. and {Hinshaw}, G.},
        title = "{The 1\% Concordance Hubble Constant}",
      journal = {\apj},
         year = 2014,
        month = oct,
       volume = {794},
       number = {2},
          eid = {135},
        pages = {135},
          doi = {10.1088/0004-637X/794/2/135},
archivePrefix = {arXiv},
       eprint = {1406.1718},
 primaryClass = {astro-ph.CO},
       adsurl = {https://ui.adsabs.harvard.edu/abs/2014ApJ...794..135B}
}

@ARTICLE{bonnoli15,
       author = {{Bonnoli}, G. and {Tavecchio}, F. and {Ghisellini}, G. and {Sbarrato}, T.},
        title = "{An emerging population of BL Lacs with extreme properties: towards a class of EBL and cosmic magnetic field probes?}",
      journal = {\mnras},
         year = 2015,
        month = jul,
       volume = {451},
       number = {1},
        pages = {611-621},
          doi = {10.1093/mnras/stv953},
archivePrefix = {arXiv},
       eprint = {1501.01974},
 primaryClass = {astro-ph.HE},
       adsurl = {https://ui.adsabs.harvard.edu/abs/2015MNRAS.451..611B}
}

@ARTICLE{chang17,
       author = {{Chang}, Y. -L. and {Arsioli}, B. and {Giommi}, P. and {Padovani}, P.},
        title = "{2WHSP: A multi-frequency selected catalogue of high energy and very high energy {\ensuremath{\gamma}}-ray blazars and blazar candidates}",
      journal = {\aap},
         year = 2017,
        month = feb,
       volume = {598},
          eid = {A17},
        pages = {A17},
          doi = {10.1051/0004-6361/201629487},
archivePrefix = {arXiv},
       eprint = {1609.05808},
 primaryClass = {astro-ph.HE},
       adsurl = {https://ui.adsabs.harvard.edu/abs/2017A&A...598A..17C}
}

@ARTICLE{chang19,
       author = {{Chang}, Y. -L. and {Arsioli}, B. and {Giommi}, P. and {Padovani}, P. and {Brandt}, C.~H.},
        title = "{The 3HSP catalogue of extreme and high-synchrotron peaked blazars}",
      journal = {\aap},
         year = 2019,
        month = dec,
       volume = {632},
          eid = {A77},
        pages = {A77},
          doi = {10.1051/0004-6361/201834526},
archivePrefix = {arXiv},
       eprint = {1909.08279},
 primaryClass = {astro-ph.HE},
       adsurl = {https://ui.adsabs.harvard.edu/abs/2019A&A...632A..77C}
}

@ARTICLE{chang20,
       author = {{Chang}, Y. -L. and {Brandt}, C.~H. and {Giommi}, P.},
        title = "{The Open Universe VOU-Blazars tool}",
      journal = {Astronomy and Computing},
         year = 2020,
        month = jan,
       volume = {30},
          eid = {100350},
        pages = {100350},
          doi = {10.1016/j.ascom.2019.100350},
archivePrefix = {arXiv},
       eprint = {1909.11455},
 primaryClass = {astro-ph.HE},
       adsurl = {https://ui.adsabs.harvard.edu/abs/2020A&C....3000350C}
}

@ARTICLE{chen09,
       author = {{Chen}, Zhaoyu and {Gu}, Minfeng and {Cao}, Xinwu},
        title = "{Flat-spectrum radio quasars from the SDSS DR3 quasar catalogue}",
      journal = {\mnras},
         year = 2009,
        month = aug,
       volume = {397},
       number = {4},
        pages = {1713-1727},
          doi = {10.1111/j.1365-2966.2009.14875.x},
archivePrefix = {arXiv},
       eprint = {0904.1452},
 primaryClass = {astro-ph.GA},
       adsurl = {https://ui.adsabs.harvard.edu/abs/2009MNRAS.397.1713C}
}

@ARTICLE{costamante18,
       author = {{Costamante}, L. and {Bonnoli}, G. and {Tavecchio}, F. and {Ghisellini}, G. and {Tagliaferri}, G. and {Khangulyan}, D.},
        title = "{The NuSTAR view on hard-TeV BL Lacs}",
      journal = {\mnras},
         year = 2018,
        month = jul,
       volume = {477},
       number = {3},
        pages = {4257-4268},
          doi = {10.1093/mnras/sty857},
archivePrefix = {arXiv},
       eprint = {1711.06282},
 primaryClass = {astro-ph.HE},
       adsurl = {https://ui.adsabs.harvard.edu/abs/2018MNRAS.477.4257C}
}

@ARTICLE{costamante20,
       author = {{Costamante}, L.},
        title = "{TeV-peaked candidate BL Lac objects}",
      journal = {\mnras},
         year = 2020,
        month = jan,
       volume = {491},
       number = {2},
        pages = {2771-2778},
          doi = {10.1093/mnras/stz3018},
archivePrefix = {arXiv},
       eprint = {1911.05027},
 primaryClass = {astro-ph.HE},
       adsurl = {https://ui.adsabs.harvard.edu/abs/2020MNRAS.491.2771C}
}

@ARTICLE{evans24,
       author = {{Evans}, Ian N. and {Evans}, Janet D. and {Mart{\'\i}nez-Galarza}, J. Rafael and {Miller}, Joseph B. and {Primini}, Francis A. and {Azadi}, Mojegan and {Burke}, Douglas J. and {Civano}, Francesca M. and {D'Abrusco}, Raffaele and {Fabbiano}, Giuseppina and {Graessle}, Dale E. and {Grier}, John D. and {Houck}, John C. and {Lauer}, Jennifer and {McCollough}, Michael L. and {Nowak}, Michael A. and {Plummer}, David A. and {Rots}, Arnold H. and {Siemiginowska}, Aneta and {Tibbetts}, Michael S.},
        title = "{The Chandra Source Catalog Release 2 Series}",
      journal = {\apjs},
         year = 2024,
        month = oct,
       volume = {274},
       number = {2},
          eid = {22},
        pages = {22},
          doi = {10.3847/1538-4365/ad6319},
archivePrefix = {arXiv},
       eprint = {2407.10799},
 primaryClass = {astro-ph.HE},
       adsurl = {https://ui.adsabs.harvard.edu/abs/2024ApJS..274...22E}
}

@ARTICLE{evansP20_swi,
       author = {{Evans}, P.~A. and {Page}, K.~L. and {Osborne}, J.~P. and {Beardmore}, A.~P. and {Willingale}, R. and {Burrows}, D.~N. and {Kennea}, J.~A. and {Perri}, M. and {Capalbi}, M. and {Tagliaferri}, G. and {Cenko}, S.~B.},
        title = "{2SXPS: An Improved and Expanded Swift X-Ray Telescope Point-source Catalog}",
      journal = {\apjs},
         year = 2020,
        month = apr,
       volume = {247},
       number = {2},
          eid = {54},
        pages = {54},
          doi = {10.3847/1538-4365/ab7db9},
archivePrefix = {arXiv},
       eprint = {1911.11710},
 primaryClass = {astro-ph.IM},
       adsurl = {https://ui.adsabs.harvard.edu/abs/2020ApJS..247...54E}
}

@ARTICLE{foffano19,
       author = {{Foffano}, L. and {Prandini}, E. and {Franceschini}, A. and {Paiano}, S.},
        title = "{A new hard X-ray-selected sample of extreme high-energy peaked BL Lac objects and their TeV gamma-ray properties}",
      journal = {\mnras},
         year = 2019,
        month = jun,
       volume = {486},
       number = {2},
        pages = {1741-1762},
          doi = {10.1093/mnras/stz812},
archivePrefix = {arXiv},
       eprint = {1903.07972},
 primaryClass = {astro-ph.HE},
       adsurl = {https://ui.adsabs.harvard.edu/abs/2019MNRAS.486.1741F}
}

@ARTICLE{ghisellini98,
       author = {{Ghisellini}, G. and {Celotti}, A. and {Fossati}, G. and {Maraschi}, L. and {Comastri}, A.},
        title = "{A theoretical unifying scheme for gamma-ray bright blazars}",
      journal = {\mnras},
         year = 1998,
        month = dec,
       volume = {301},
       number = {2},
        pages = {451-468},
          doi = {10.1046/j.1365-8711.1998.02032.x},
archivePrefix = {arXiv},
       eprint = {astro-ph/9807317},
 primaryClass = {astro-ph},
       adsurl = {https://ui.adsabs.harvard.edu/abs/1998MNRAS.301..451G}
}

@ARTICLE{giommi99,
       author = {{Giommi}, P. and {Menna}, M.~T. and {Padovani}, P.},
        title = "{The sedentary multifrequency survey - I. Statistical identification and cosmological properties of high-energy peaked BL Lacs}",
      journal = {\mnras},
         year = 1999,
        month = dec,
       volume = {310},
       number = {2},
        pages = {465-475},
          doi = {10.1046/j.1365-8711.1999.02942.x},
archivePrefix = {arXiv},
       eprint = {astro-ph/9907014},
 primaryClass = {astro-ph},
       adsurl = {https://ui.adsabs.harvard.edu/abs/1999MNRAS.310..465G}
}

@ARTICLE{giommi24,
       author = {{Giommi}, P. and {Sahakyan}, N. and {Israyelyan}, D. and {Manvelyan}, M.},
        title = "{The Remarkable Predictive Power of Infrared Data of Blazars}",
      journal = {\apj},
         year = 2024,
        month = mar,
       volume = {963},
       number = {1},
          eid = {48},
        pages = {48},
          doi = {10.3847/1538-4357/ad20cb},
archivePrefix = {arXiv},
       eprint = {2401.10548},
 primaryClass = {astro-ph.HE},
       adsurl = {https://ui.adsabs.harvard.edu/abs/2024ApJ...963...48G}
}

@ARTICLE{glauch22,
       author = {{Glauch}, T. and {Kerscher}, T. and {Giommi}, P.},
        title = "{BlaST - A machine-learning estimator for the synchrotron peak of blazars}",
      journal = {Astronomy and Computing},
         year = 2022,
        month = oct,
       volume = {41},
          eid = {100646},
        pages = {100646},
          doi = {10.1016/j.ascom.2022.100646},
archivePrefix = {arXiv},
       eprint = {2207.03813},
 primaryClass = {astro-ph.HE},
       adsurl = {https://ui.adsabs.harvard.edu/abs/2022A&C....4100646G}
}

@ARTICLE{green17,
       author = {{Green}, T.~S. and {Edge}, A.~C. and {Ebeling}, H. and {Burgett}, W.~S. and {Draper}, P.~W. and {Kaiser}, N. and {Kudritzki}, R. -P. and {Magnier}, E.~A. and {Metcalfe}, N. and {Wainscoat}, R.~J. and {Waters}, C.},
        title = "{Hiding in plain sight - recovering clusters of galaxies with the strongest AGN in their cores}",
      journal = {\mnras},
         year = 2017,
        month = mar,
       volume = {465},
       number = {4},
        pages = {4872-4885},
          doi = {10.1093/mnras/stw3059},
archivePrefix = {arXiv},
       eprint = {1611.07996},
 primaryClass = {astro-ph.GA},
       adsurl = {https://ui.adsabs.harvard.edu/abs/2017MNRAS.465.4872G}
}

@ARTICLE{harris20,
       author = {{Harris}, Charles R. and {Millman}, K. Jarrod and {van der Walt}, St{\'e}fan J. and {Gommers}, Ralf and {Virtanen}, Pauli and {Cournapeau}, David and {Wieser}, Eric and {Taylor}, Julian and {Berg}, Sebastian and {Smith}, Nathaniel J. and {Kern}, Robert and {Picus}, Matti and {Hoyer}, Stephan and {van Kerkwijk}, Marten H. and {Brett}, Matthew and {Haldane}, Allan and {del R{\'\i}o}, Jaime Fern{\'a}ndez and {Wiebe}, Mark and {Peterson}, Pearu and {G{\'e}rard-Marchant}, Pierre and {Sheppard}, Kevin and {Reddy}, Tyler and {Weckesser}, Warren and {Abbasi}, Hameer and {Gohlke}, Christoph and {Oliphant}, Travis E.},
        title = "{Array programming with NumPy}",
      journal = {\nat},
         year = 2020,
        month = sep,
       volume = {585},
       number = {7825},
        pages = {357-362},
          doi = {10.1038/s41586-020-2649-2},
archivePrefix = {arXiv},
       eprint = {2006.10256},
 primaryClass = {cs.MS},
       adsurl = {https://ui.adsabs.harvard.edu/abs/2020Natur.585..357H}
}

@ARTICLE{hofmann23,
       author = {{Hofmann}, Werner and {Zanin}, Roberta},
        title = "{The Cherenkov Telescope Array}",
      journal = {arXiv e-prints},
         year = 2023,
        month = may,
          eid = {arXiv:2305.12888},
        pages = {arXiv:2305.12888},
          doi = {10.48550/arXiv.2305.12888},
archivePrefix = {arXiv},
       eprint = {2305.12888},
 primaryClass = {astro-ph.IM},
       adsurl = {https://ui.adsabs.harvard.edu/abs/2023arXiv230512888H}
}

@ARTICLE{hunter07,
       author = {{Hunter}, John D.},
        title = "{Matplotlib: A 2D Graphics Environment}",
      journal = {Computing in Science and Engineering},
         year = 2007,
        month = may,
       volume = {9},
       number = {3},
        pages = {90-95},
          doi = {10.1109/MCSE.2007.55},
       adsurl = {https://ui.adsabs.harvard.edu/abs/2007CSE.....9...90H}
}

@ARTICLE{kirk98,
       author = {{Kirk}, J.~G. and {Rieger}, F.~M. and {Mastichiadis}, A.},
        title = "{Particle acceleration and synchrotron emission in blazar jets}",
      journal = {\aap},
         year = 1998,
        month = may,
       volume = {333},
        pages = {452-458},
          doi = {10.48550/arXiv.astro-ph/9801265},
archivePrefix = {arXiv},
       eprint = {astro-ph/9801265},
 primaryClass = {astro-ph},
       adsurl = {https://ui.adsabs.harvard.edu/abs/1998A&A...333..452K}
}

@ARTICLE{lakshminarayanan16,
       author = {{Lakshminarayanan}, Balaji and {Pritzel}, Alexander and {Blundell}, Charles},
        title = "{Simple and Scalable Predictive Uncertainty Estimation using Deep Ensembles}",
      journal = {arXiv e-prints},
         year = 2016,
        month = dec,
          eid = {arXiv:1612.01474},
        pages = {arXiv:1612.01474},
          doi = {10.48550/arXiv.1612.01474},
archivePrefix = {arXiv},
       eprint = {1612.01474},
 primaryClass = {stat.ML},
       adsurl = {https://ui.adsabs.harvard.edu/abs/2016arXiv161201474L}
}

@ARTICLE{massaro11b,
       author = {{Massaro}, F. and {Paggi}, A. and {Cavaliere}, A.},
        title = "{X-Ray and TeV Emissions from High-frequency-peaked BL Lac Objects}",
      journal = {\apjl},
         year = 2011,
        month = dec,
       volume = {742},
       number = {2},
          eid = {L32},
        pages = {L32},
          doi = {10.1088/2041-8205/742/2/L32},
archivePrefix = {arXiv},
       eprint = {1203.1924},
 primaryClass = {astro-ph.HE},
       adsurl = {https://ui.adsabs.harvard.edu/abs/2011ApJ...742L..32M}
}

@ARTICLE{massaro13,
       author = {{Massaro}, F. and {Paggi}, A. and {Errando}, M. and {D'Abrusco}, R. and {Masetti}, N. and {Tosti}, G. and {Funk}, S.},
        title = "{BL Lac Candidates for TeV Observations}",
      journal = {\apjs},
         year = 2013,
        month = jul,
       volume = {207},
       number = {1},
          eid = {16},
        pages = {16},
          doi = {10.1088/0067-0049/207/1/16},
archivePrefix = {arXiv},
       eprint = {1307.8113},
 primaryClass = {astro-ph.HE},
       adsurl = {https://ui.adsabs.harvard.edu/abs/2013ApJS..207...16M}
}

@ARTICLE{massaro15,
       author = {{Massaro}, E. and {Maselli}, A. and {Leto}, C. and {Marchegiani}, P. and {Perri}, M. and {Giommi}, P. and {Piranomonte}, S.},
        title = "{The 5th edition of the Roma-BZCAT. A short presentation}",
      journal = {\apss},
         year = 2015,
        month = may,
       volume = {357},
       number = {1},
          eid = {75},
        pages = {75},
          doi = {10.1007/s10509-015-2254-2},
archivePrefix = {arXiv},
       eprint = {1502.07755},
 primaryClass = {astro-ph.HE},
       adsurl = {https://ui.adsabs.harvard.edu/abs/2015Ap&SS.357...75M}
}

@ARTICLE{merloni24,
       author = {{Merloni}, A. and {Lamer}, G. and {Liu}, T. and {Ramos-Ceja}, M.~E. and {Brunner}, H. and {Bulbul}, E. and {Dennerl}, K. and {Doroshenko}, V. and {Freyberg}, M.~J. and {Friedrich}, S. and {Gatuzz}, E. and {Georgakakis}, A. and {Haberl}, F. and {Igo}, Z. and {Kreykenbohm}, I. and {Liu}, A. and {Maitra}, C. and {Malyali}, A. and {Mayer}, M.~G.~F. and {Nandra}, K. and {Predehl}, P. and {Robrade}, J. and {Salvato}, M. and {Sanders}, J.~S. and {Stewart}, I. and {Tub{\'\i}n-Arenas}, D. and {Weber}, P. and {Wilms}, J. and {Arcodia}, R. and {Artis}, E. and {Aschersleben}, J. and {Avakyan}, A. and {Aydar}, C. and {Bahar}, Y.~E. and {Balzer}, F. and {Becker}, W. and {Berger}, K. and {Boller}, T. and {Bornemann}, W. and {Br{\"u}ggen}, M. and {Brusa}, M. and {Buchner}, J. and {Burwitz}, V. and {Camilloni}, F. and {Clerc}, N. and {Comparat}, J. and {Coutinho}, D. and {Czesla}, S. and {Dannhauer}, S.~M. and {Dauner}, L. and {Dauser}, T. and {Dietl}, J. and {Dolag}, K. and {Dwelly}, T. and {Egg}, K. and {Ehl}, E. and {Freund}, S. and {Friedrich}, P. and {Gaida}, R. and {Garrel}, C. and {Ghirardini}, V. and {Gokus}, A. and {Gr{\"u}nwald}, G. and {Grandis}, S. and {Grotova}, I. and {Gruen}, D. and {Gueguen}, A. and {H{\"a}mmerich}, S. and {Hamaus}, N. and {Hasinger}, G. and {Haubner}, K. and {Homan}, D. and {Ider Chitham}, J. and {Joseph}, W.~M. and {Joyce}, A. and {K{\"o}nig}, O. and {Kaltenbrunner}, D.~M. and {Khokhriakova}, A. and {Kink}, W. and {Kirsch}, C. and {Kluge}, M. and {Knies}, J. and {Krippendorf}, S. and {Krumpe}, M. and {Kurpas}, J. and {Li}, P. and {Liu}, Z. and {Locatelli}, N. and {Lorenz}, M. and {M{\"u}ller}, S. and {Magaudda}, E. and {Mannes}, C. and {McCall}, H. and {Meidinger}, N. and {Michailidis}, M. and {Migkas}, K. and {Mu{\~n}oz-Giraldo}, D. and {Musiimenta}, B. and {Nguyen-Dang}, N.~T. and {Ni}, Q. and {Olechowska}, A. and {Ota}, N. and {Pacaud}, F. and {Pasini}, T. and {Perinati}, E. and {Pires}, A.~M. and {Pommranz}, C. and {Ponti}, G. and {Poppenhaeger}, K. and {P{\"u}hlhofer}, G. and {Rau}, A. and {Reh}, M. and {Reiprich}, T.~H. and {Roster}, W. and {Saeedi}, S. and {Santangelo}, A. and {Sasaki}, M. and {Schmitt}, J. and {Schneider}, P.~C. and {Schrabback}, T. and {Schuster}, N. and {Schwope}, A. and {Seppi}, R. and {Serim}, M.~M. and {Shreeram}, S. and {Sokolova-Lapa}, E. and {Starck}, H. and {Stelzer}, B. and {Stierhof}, J. and {Suleimanov}, V. and {Tenzer}, C. and {Traulsen}, I. and {Tr{\"u}mper}, J. and {Tsuge}, K. and {Urrutia}, T. and {Veronica}, A. and {Waddell}, S.~G.~H. and {Willer}, R. and {Wolf}, J. and {Yeung}, M.~C.~H. and {Zainab}, A. and {Zangrandi}, F. and {Zhang}, X. and {Zhang}, Y. and {Zheng}, X.},
        title = "{The SRG/eROSITA all-sky survey. First X-ray catalogues and data release of the western Galactic hemisphere}",
      journal = {\aap},
         year = 2024,
        month = feb,
       volume = {682},
          eid = {A34},
        pages = {A34},
          doi = {10.1051/0004-6361/202347165},
archivePrefix = {arXiv},
       eprint = {2401.17274},
 primaryClass = {astro-ph.HE},
       adsurl = {https://ui.adsabs.harvard.edu/abs/2024A&A...682A..34M}
}

@ARTICLE{middei22,
       author = {{Middei}, R. and {Giommi}, P. and {Perri}, M. and {Turriziani}, S. and {Sahakyan}, N. and {Chang}, Y.~L. and {Leto}, C. and {Verrecchia}, F.},
        title = "{The first hard X-ray spectral catalogue of Blazars observed by NuSTAR}",
      journal = {\mnras},
         year = 2022,
        month = aug,
       volume = {514},
       number = {3},
        pages = {3179-3190},
          doi = {10.1093/mnras/stac1185},
archivePrefix = {arXiv},
       eprint = {2205.05089},
 primaryClass = {astro-ph.HE},
       adsurl = {https://ui.adsabs.harvard.edu/abs/2022MNRAS.514.3179M}
}

@ARTICLE{padovani95,
       author = {{Padovani}, Paolo and {Giommi}, Paolo},
        title = "{The Connection between X-Ray-- and Radio-selected BL Lacertae Objects}",
      journal = {\apj},
         year = 1995,
        month = may,
       volume = {444},
        pages = {567},
          doi = {10.1086/175631},
archivePrefix = {arXiv},
       eprint = {astro-ph/9412073},
 primaryClass = {astro-ph},
       adsurl = {https://ui.adsabs.harvard.edu/abs/1995ApJ...444..567P}
}

@ARTICLE{prandini22,
       author = {{Prandini}, Elisa and {Ghisellini}, Gabriele},
        title = "{The Blazar Sequence and Its Physical Understanding}",
      journal = {Galaxies},
         year = 2022,
        month = feb,
       volume = {10},
       number = {1},
          eid = {35},
        pages = {35},
          doi = {10.3390/galaxies10010035},
archivePrefix = {arXiv},
       eprint = {2202.07490},
 primaryClass = {astro-ph.HE},
       adsurl = {https://ui.adsabs.harvard.edu/abs/2022Galax..10...35P}
}

@ARTICLE{predehl21,
       author = {{Predehl}, P. and {Andritschke}, R. and {Arefiev}, V. and {Babyshkin}, V. and {Batanov}, O. and {Becker}, W. and {B{\"o}hringer}, H. and {Bogomolov}, A. and {Boller}, T. and {Borm}, K. and {Bornemann}, W. and {Br{\"a}uninger}, H. and {Br{\"u}ggen}, M. and {Brunner}, H. and {Brusa}, M. and {Bulbul}, E. and {Buntov}, M. and {Burwitz}, V. and {Burkert}, W. and {Clerc}, N. and {Churazov}, E. and {Coutinho}, D. and {Dauser}, T. and {Dennerl}, K. and {Doroshenko}, V. and {Eder}, J. and {Emberger}, V. and {Eraerds}, T. and {Finoguenov}, A. and {Freyberg}, M. and {Friedrich}, P. and {Friedrich}, S. and {F{\"u}rmetz}, M. and {Georgakakis}, A. and {Gilfanov}, M. and {Granato}, S. and {Grossberger}, C. and {Gueguen}, A. and {Gureev}, P. and {Haberl}, F. and {H{\"a}lker}, O. and {Hartner}, G. and {Hasinger}, G. and {Huber}, H. and {Ji}, L. and {Kienlin}, A. v. and {Kink}, W. and {Korotkov}, F. and {Kreykenbohm}, I. and {Lamer}, G. and {Lomakin}, I. and {Lapshov}, I. and {Liu}, T. and {Maitra}, C. and {Meidinger}, N. and {Menz}, B. and {Merloni}, A. and {Mernik}, T. and {Mican}, B. and {Mohr}, J. and {M{\"u}ller}, S. and {Nandra}, K. and {Nazarov}, V. and {Pacaud}, F. and {Pavlinsky}, M. and {Perinati}, E. and {Pfeffermann}, E. and {Pietschner}, D. and {Ramos-Ceja}, M.~E. and {Rau}, A. and {Reiffers}, J. and {Reiprich}, T.~H. and {Robrade}, J. and {Salvato}, M. and {Sanders}, J. and {Santangelo}, A. and {Sasaki}, M. and {Scheuerle}, H. and {Schmid}, C. and {Schmitt}, J. and {Schwope}, A. and {Shirshakov}, A. and {Steinmetz}, M. and {Stewart}, I. and {Str{\"u}der}, L. and {Sunyaev}, R. and {Tenzer}, C. and {Tiedemann}, L. and {Tr{\"u}mper}, J. and {Voron}, V. and {Weber}, P. and {Wilms}, J. and {Yaroshenko}, V.},
        title = "{The eROSITA X-ray telescope on SRG}",
      journal = {\aap},
         year = 2021,
        month = mar,
       volume = {647},
          eid = {A1},
        pages = {A1},
          doi = {10.1051/0004-6361/202039313},
archivePrefix = {arXiv},
       eprint = {2010.03477},
 primaryClass = {astro-ph.HE},
       adsurl = {https://ui.adsabs.harvard.edu/abs/2021A&A...647A...1P}
}

@ARTICLE{saldana21,
       author = {{Saldana-Lopez}, Alberto and {Dom{\'\i}nguez}, Alberto and {P{\'e}rez-Gonz{\'a}lez}, Pablo G. and {Finke}, Justin and {Ajello}, Marco and {Primack}, Joel R. and {Paliya}, Vaidehi S. and {Desai}, Abhishek},
        title = "{An observational determination of the evolving extragalactic background light from the multiwavelength HST/CANDELS survey in the Fermi and CTA era}",
      journal = {\mnras},
         year = 2021,
        month = nov,
       volume = {507},
       number = {4},
        pages = {5144-5160},
          doi = {10.1093/mnras/stab2393},
archivePrefix = {arXiv},
       eprint = {2012.03035},
 primaryClass = {astro-ph.CO},
       adsurl = {https://ui.adsabs.harvard.edu/abs/2021MNRAS.507.5144S}
}

@ARTICLE{sunyaev21,
       author = {{Sunyaev}, R. and {Arefiev}, V. and {Babyshkin}, V. and {Bogomolov}, A. and {Borisov}, K. and {Buntov}, M. and {Brunner}, H. and {Burenin}, R. and {Churazov}, E. and {Coutinho}, D. and {Eder}, J. and {Eismont}, N. and {Freyberg}, M. and {Gilfanov}, M. and {Gureyev}, P. and {Hasinger}, G. and {Khabibullin}, I. and {Kolmykov}, V. and {Komovkin}, S. and {Krivonos}, R. and {Lapshov}, I. and {Levin}, V. and {Lomakin}, I. and {Lutovinov}, A. and {Medvedev}, P. and {Merloni}, A. and {Mernik}, T. and {Mikhailov}, E. and {Molodtsov}, V. and {Mzhelsky}, P. and {M{\"u}ller}, S. and {Nandra}, K. and {Nazarov}, V. and {Pavlinsky}, M. and {Poghodin}, A. and {Predehl}, P. and {Robrade}, J. and {Sazonov}, S. and {Scheuerle}, H. and {Shirshakov}, A. and {Tkachenko}, A. and {Voron}, V.},
        title = "{SRG X-ray orbital observatory. Its telescopes and first scientific results}",
      journal = {\aap},
         year = 2021,
        month = dec,
       volume = {656},
          eid = {A132},
        pages = {A132},
          doi = {10.1051/0004-6361/202141179},
archivePrefix = {arXiv},
       eprint = {2104.13267},
 primaryClass = {astro-ph.HE},
       adsurl = {https://ui.adsabs.harvard.edu/abs/2021A&A...656A.132S}
}

@ARTICLE{tramacere11,
       author = {{Tramacere}, A. and {Massaro}, E. and {Taylor}, A.~M.},
        title = "{Stochastic Acceleration and the Evolution of Spectral Distributions in Synchro-Self-Compton Sources: A Self-consistent Modeling of Blazars' Flares}",
      journal = {\apj},
         year = 2011,
        month = oct,
       volume = {739},
       number = {2},
          eid = {66},
        pages = {66},
          doi = {10.1088/0004-637X/739/2/66},
archivePrefix = {arXiv},
       eprint = {1107.1879},
 primaryClass = {astro-ph.HE},
       adsurl = {https://ui.adsabs.harvard.edu/abs/2011ApJ...739...66T}
}

@INPROCEEDINGS{wakely18,
       author = {{Wakely}, S.~P. and {Horan}, D.},
        title = "{TeVCat: An online catalog for Very High Energy Gamma-Ray Astronomy}",
    booktitle = {International Cosmic Ray Conference},
         year = 2008,
       series = {International Cosmic Ray Conference},
       volume = {3},
        month = jan,
        pages = {1341-1344},
       adsurl = {https://ui.adsabs.harvard.edu/abs/2008ICRC....3.1341W}
}

@ARTICLE{webb20,
       author = {{Webb}, N.~A. and {Coriat}, M. and {Traulsen}, I. and {Ballet}, J. and {Motch}, C. and {Carrera}, F.~J. and {Koliopanos}, F. and {Authier}, J. and {de la Calle}, I. and {Ceballos}, M.~T. and {Colomo}, E. and {Chuard}, D. and {Freyberg}, M. and {Garcia}, T. and {Kolehmainen}, M. and {Lamer}, G. and {Lin}, D. and {Maggi}, P. and {Michel}, L. and {Page}, C.~G. and {Page}, M.~J. and {Perea-Calderon}, J.~V. and {Pineau}, F. -X. and {Rodriguez}, P. and {Rosen}, S.~R. and {Santos Lleo}, M. and {Saxton}, R.~D. and {Schwope}, A. and {Tom{\'a}s}, L. and {Watson}, M.~G. and {Zakardjian}, A.},
        title = "{The XMM-Newton serendipitous survey. IX. The fourth XMM-Newton serendipitous source catalogue}",
      journal = {\aap},
         year = 2020,
        month = sep,
       volume = {641},
          eid = {A136},
        pages = {A136},
          doi = {10.1051/0004-6361/201937353},
archivePrefix = {arXiv},
       eprint = {2007.02899},
 primaryClass = {astro-ph.HE},
       adsurl = {https://ui.adsabs.harvard.edu/abs/2020A&A...641A.136W}
}

@ARTICLE{marchesi25,
       author = {{Marchesi}, Stefano and {Iuliano}, Antonio and {Prandini}, Elisa and {Da Vela}, Paolo and {Doro}, Michele and {Loporchio}, Serena and {Miceli}, Davide and {Righi}, Chiara and {Zanin}, Roberta and {Bronzini}, Ettore and {Vignali}, Cristian},
        title = "{A new look at the extragalactic very high energy sky: Searching for TeV-emitting candidates among the X-ray-bright, non-Fermi-detected blazar population}",
      journal = {\aap},
         year = 2025,
        month = jan,
       volume = {693},
          eid = {A142},
        pages = {A142},
          doi = {10.1051/0004-6361/202451924},
archivePrefix = {arXiv},
       eprint = {2410.18278},
 primaryClass = {astro-ph.HE},
       adsurl = {https://ui.adsabs.harvard.edu/abs/2025A&A...693A.142M}
}

@ARTICLE{kaastra17,
       author = {{Kaastra}, J.~S.},
        title = "{On the use of C-stat in testing models for X-ray spectra}",
      journal = {\aap},
         year = 2017,
        month = sep,
       volume = {605},
          eid = {A51},
        pages = {A51},
          doi = {10.1051/0004-6361/201629319},
archivePrefix = {arXiv},
       eprint = {1707.09202},
 primaryClass = {astro-ph.HE},
       adsurl = {https://ui.adsabs.harvard.edu/abs/2017A&A...605A..51K}
}

@ARTICLE{cash79,
       author = {{Cash}, W.},
        title = "{Parameter estimation in astronomy through application of the likelihood ratio.}",
      journal = {\apj},
         year = 1979,
        month = mar,
       volume = {228},
        pages = {939-947},
          doi = {10.1086/156922},
       adsurl = {https://ui.adsabs.harvard.edu/abs/1979ApJ...228..939C}
}

@ARTICLE{deMenezes22,
       author = {{de Menezes}, R.},
        title = "{easyFermi: A graphical interface for performing Fermi-LAT data analyses}",
      journal = {Astronomy and Computing},
         year = 2022,
        month = jul,
       volume = {40},
          eid = {100609},
        pages = {100609},
          doi = {10.1016/j.ascom.2022.100609},
archivePrefix = {arXiv},
       eprint = {2206.11272},
 primaryClass = {astro-ph.HE},
       adsurl = {https://ui.adsabs.harvard.edu/abs/2022A&C....4000609D}
}

@ARTICLE{giommi25,
       author = {{Giommi}, Paolo},
        title = "{The Firmamento platform}",
      journal = {arXiv e-prints},
         year = 2025,
        month = mar,
          eid = {arXiv:2503.04434},
        pages = {arXiv:2503.04434},
          doi = {10.48550/arXiv.2503.04434},
archivePrefix = {arXiv},
       eprint = {2503.04434},
 primaryClass = {astro-ph.IM},
       adsurl = {https://ui.adsabs.harvard.edu/abs/2025arXiv250304434G}
}

@ARTICLE{hota24,
       author = {{Hota}, Jyotishree and {Khatoon}, Rukaiya and {Misra}, Ranjeev and {Pradhan}, Ananta C.},
        title = "{Multiwavelength Study of Extreme High-energy Peaked BL Lac Source 1ES 0229+200 Using Ultraviolet, X-Ray, and {\ensuremath{\gamma}}-Ray Observations}",
      journal = {\apj},
         year = 2024,
        month = nov,
       volume = {976},
       number = {1},
          eid = {69},
        pages = {69},
          doi = {10.3847/1538-4357/ad8085},
archivePrefix = {arXiv},
       eprint = {2409.12827},
 primaryClass = {astro-ph.HE},
       adsurl = {https://ui.adsabs.harvard.edu/abs/2024ApJ...976...69H}
}

@ARTICLE{dominguez24,
       author = {{Dom{\'\i}nguez}, A. and {{\O}stergaard Kirkeberg}, P. and {Wojtak}, R. and {Saldana-Lopez}, A. and {Desai}, A. and {Primack}, J.~R. and {Finke}, J. and {Ajello}, M. and {P{\'e}rez-Gonz{\'a}lez}, P.~G. and {Paliya}, V.~S. and {Hartmann}, D.},
        title = "{A new derivation of the Hubble constant from {\ensuremath{\gamma}}-ray attenuation using improved optical depths for the Fermi and CTA era}",
      journal = {\mnras},
         year = 2024,
        month = jan,
       volume = {527},
       number = {3},
        pages = {4632-4642},
          doi = {10.1093/mnras/stad3425},
archivePrefix = {arXiv},
       eprint = {2306.09878},
 primaryClass = {astro-ph.HE},
       adsurl = {https://ui.adsabs.harvard.edu/abs/2024MNRAS.527.4632D}
}

@BOOK{ghisellini13,
       author = {{Ghisellini}, Gabriele},
        title = "{Radiative Processes in High Energy Astrophysics}",
         year = 2013,
       volume = {873},
          doi = {10.1007/978-3-319-00612-3},
        publisher = {Springer},
       adsurl = {https://ui.adsabs.harvard.edu/abs/2013LNP...873.....G}
}

@ARTICLE{bottcher07,
       author = {{B{\"o}ttcher}, Markus},
        title = "{Modeling the emission processes in blazars}",
      journal = {\apss},
         year = 2007,
        month = jun,
       volume = {309},
       number = {1-4},
        pages = {95-104},
          doi = {10.1007/s10509-007-9404-0},
archivePrefix = {arXiv},
       eprint = {astro-ph/0608713},
 primaryClass = {astro-ph},
       adsurl = {https://ui.adsabs.harvard.edu/abs/2007Ap&SS.309...95B}
}

@ARTICLE{maraschi92,
       author = {{Maraschi}, L. and {Ghisellini}, G. and {Celotti}, A.},
        title = "{A Jet Model for the Gamma-Ray--emitting Blazar 3C 279}",
      journal = {\apjl},
         year = 1992,
        month = sep,
       volume = {397},
        pages = {L5},
          doi = {10.1086/186531},
       adsurl = {https://ui.adsabs.harvard.edu/abs/1992ApJ...397L...5M}
}

@ARTICLE{arsioli20,
       author = {{Arsioli}, B. and {Chang}, Y. -L. and {Musiimenta}, B.},
        title = "{Extreme and high synchrotron peak blazars beyond 4FGL: The 2BIGB {\ensuremath{\gamma}}-ray catalogue}",
      journal = {\mnras},
         year = 2020,
        month = apr,
       volume = {493},
       number = {2},
        pages = {2438-2451},
          doi = {10.1093/mnras/staa368},
archivePrefix = {arXiv},
       eprint = {1911.08912},
 primaryClass = {astro-ph.HE},
       adsurl = {https://ui.adsabs.harvard.edu/abs/2020MNRAS.493.2438A}
}

@ARTICLE{tramacere09,
       author = {{Tramacere}, A. and {Giommi}, P. and {Perri}, M. and {Verrecchia}, F. and {Tosti}, G.},
        title = "{Swift observations of the very intense flaring activity of Mrk 421 during 2006. I. Phenomenological picture of electron acceleration and predictions for MeV/GeV emission}",
      journal = {\aap},
         year = 2009,
        month = jul,
       volume = {501},
       number = {3},
        pages = {879-898},
          doi = {10.1051/0004-6361/200810865},
archivePrefix = {arXiv},
       eprint = {0901.4124},
 primaryClass = {astro-ph.HE},
       adsurl = {https://ui.adsabs.harvard.edu/abs/2009A&A...501..879T}
}

@software{tramacere20,
       author = {{Tramacere}, Andrea},
        title = "{JetSeT: Numerical modeling and SED fitting tool for relativistic jets}",
 howpublished = {Astrophysics Source Code Library, record ascl:2009.001},
         year = 2020,
        month = sep,
          eid = {ascl:2009.001},
archivePrefix = {ascl},
       eprint = {2009.001},
       adsurl = {https://ui.adsabs.harvard.edu/abs/2020ascl.soft09001T}
}

@INPROCEEDINGS{rico25,
       author = {{Rico}, Alba and {Ajello}, Marco and {Dominguez}, Alberto},
        title = "{4FHL: The Deepest Fermi-LAT Catalog of Hard Gamma-Ray Sources Above 50 GeV}",
    booktitle = {246th Meeting of the American Astronomical Society},
         year = 2025,
       series = {American Astronomical Society Meeting Abstracts},
       volume = {246},
        month = jun,
          eid = {213.02},
        pages = {213.02},
       adsurl = {https://ui.adsabs.harvard.edu/abs/2025AAS...24621302R}
}

@ARTICLE{paliya19,
       author = {{Paliya}, Vaidehi S. and {Dom{\'\i}nguez}, A. and {Ajello}, M. and {Franckowiak}, A. and {Hartmann}, D.},
        title = "{Fermi-LAT Stacking Analysis Technique: An Application to Extreme Blazars and Prospects for their CTA Detection}",
      journal = {\apjl},
         year = 2019,
        month = sep,
       volume = {882},
       number = {1},
          eid = {L3},
        pages = {L3},
          doi = {10.3847/2041-8213/ab398a},
archivePrefix = {arXiv},
       eprint = {1908.02496},
 primaryClass = {astro-ph.HE},
       adsurl = {https://ui.adsabs.harvard.edu/abs/2019ApJ...882L...3P}
}

@ARTICLE{massaro04,
       author = {{Massaro}, E. and {Perri}, M. and {Giommi}, P. and {Nesci}, R. and {Verrecchia}, F.},
        title = "{Log-parabolic spectra and particle acceleration in blazars.  II. The BeppoSAX wide band X-ray spectra of Mkn 501}",
      journal = {\aap},
         year = 2004,
        month = jul,
       volume = {422},
        pages = {103-111},
          doi = {10.1051/0004-6361:20047148},
archivePrefix = {arXiv},
       eprint = {astro-ph/0405152},
 primaryClass = {astro-ph},
       adsurl = {https://ui.adsabs.harvard.edu/abs/2004A&A...422..103M}
}

@ARTICLE{Metzger25,
       author = {{Metzger}, Cassidy and {Gokus}, Andrea and {Errando}, Manel},
        title = "{New TeV-emitting BL Lac Candidates from the eROSITA X-Ray Survey}",
      journal = {\apj},
         year = 2025,
        month = oct,
       volume = {992},
       number = {2},
          eid = {184},
        pages = {184},
          doi = {10.3847/1538-4357/adff87},
archivePrefix = {arXiv},
       eprint = {2501.12520},
 primaryClass = {astro-ph.HE},
       adsurl = {https://ui.adsabs.harvard.edu/abs/2025ApJ...992..184M}
}

@ARTICLE{mattox96,
       author = {{Mattox}, J.~R. and {Bertsch}, D.~L. and {Chiang}, J. and {Dingus}, B.~L. and {Digel}, S.~W. and {Esposito}, J.~A. and {Fierro}, J.~M. and {Hartman}, R.~C. and {Hunter}, S.~D. and {Kanbach}, G. and {Kniffen}, D.~A. and {Lin}, Y.~C. and {Macomb}, D.~J. and {Mayer-Hasselwander}, H.~A. and {Michelson}, P.~F. and {von Montigny}, C. and {Mukherjee}, R. and {Nolan}, P.~L. and {Ramanamurthy}, P.~V. and {Schneid}, E. and {Sreekumar}, P. and {Thompson}, D.~J. and {Willis}, T.~D.},
        title = "{The Likelihood Analysis of EGRET Data}",
      journal = {\apj},
         year = 1996,
        month = apr,
       volume = {461},
        pages = {396},
          doi = {10.1086/177068},
       adsurl = {https://ui.adsabs.harvard.edu/abs/1996ApJ...461..396M}
}

@ARTICLE{biteau20,
       author = {{Biteau}, J. and {Prandini}, E. and {Costamante}, L. and {Lemoine}, M. and {Padovani}, P. and {Pueschel}, E. and {Resconi}, E. and {Tavecchio}, F. and {Taylor}, A. and {Zech}, A.},
        title = "{Progress in unveiling extreme particle acceleration in persistent astrophysical jets}",
      journal = {Nature Astronomy},
         year = 2020,
        month = feb,
       volume = {4},
        pages = {124-131},
          doi = {10.1038/s41550-019-0988-4},
archivePrefix = {arXiv},
       eprint = {2001.09222},
 primaryClass = {astro-ph.HE},
       adsurl = {https://ui.adsabs.harvard.edu/abs/2020NatAs...4..124B}
}

@ARTICLE{astropy22,
       author = {{Astropy Collaboration} and {Price-Whelan}, Adrian M. and {Lim}, Pey Lian and {Earl}, Nicholas and {Starkman}, Nathaniel and {Bradley}, Larry and {Shupe}, David L. and {Patil}, Aarya A. and {Corrales}, Lia and {Brasseur}, C.~E. and {N{\"o}the}, Maximilian and {Donath}, Axel and {Tollerud}, Erik and {Morris}, Brett M. and {Ginsburg}, Adam and {Vaher}, Eero and {Weaver}, Benjamin A. and {Tocknell}, James and {Jamieson}, William and {van Kerkwijk}, Marten H. and {Robitaille}, Thomas P. and {Merry}, Bruce and {Bachetti}, Matteo and {G{\"u}nther}, H. Moritz and {Aldcroft}, Thomas L. and {Alvarado-Montes}, Jaime A. and {Archibald}, Anne M. and {B{\'o}di}, Attila and {Bapat}, Shreyas and {Barentsen}, Geert and {Baz{\'a}n}, Juanjo and {Biswas}, Manish and {Boquien}, M{\'e}d{\'e}ric and {Burke}, D.~J. and {Cara}, Daria and {Cara}, Mihai and {Conroy}, Kyle E. and {Conseil}, Simon and {Craig}, Matthew W. and {Cross}, Robert M. and {Cruz}, Kelle L. and {D'Eugenio}, Francesco and {Dencheva}, Nadia and {Devillepoix}, Hadrien A.~R. and {Dietrich}, J{\"o}rg P. and {Eigenbrot}, Arthur Davis and {Erben}, Thomas and {Ferreira}, Leonardo and {Foreman-Mackey}, Daniel and {Fox}, Ryan and {Freij}, Nabil and {Garg}, Suyog and {Geda}, Robel and {Glattly}, Lauren and {Gondhalekar}, Yash and {Gordon}, Karl D. and {Grant}, David and {Greenfield}, Perry and {Groener}, Austen M. and {Guest}, Steve and {Gurovich}, Sebastian and {Handberg}, Rasmus and {Hart}, Akeem and {Hatfield-Dodds}, Zac and {Homeier}, Derek and {Hosseinzadeh}, Griffin and {Jenness}, Tim and {Jones}, Craig K. and {Joseph}, Prajwel and {Kalmbach}, J. Bryce and {Karamehmetoglu}, Emir and {Ka{\l}uszy{\'n}ski}, Miko{\l}aj and {Kelley}, Michael S.~P. and {Kern}, Nicholas and {Kerzendorf}, Wolfgang E. and {Koch}, Eric W. and {Kulumani}, Shankar and {Lee}, Antony and {Ly}, Chun and {Ma}, Zhiyuan and {MacBride}, Conor and {Maljaars}, Jakob M. and {Muna}, Demitri and {Murphy}, N.~A. and {Norman}, Henrik and {O'Steen}, Richard and {Oman}, Kyle A. and {Pacifici}, Camilla and {Pascual}, Sergio and {Pascual-Granado}, J. and {Patil}, Rohit R. and {Perren}, Gabriel I. and {Pickering}, Timothy E. and {Rastogi}, Tanuj and {Roulston}, Benjamin R. and {Ryan}, Daniel F. and {Rykoff}, Eli S. and {Sabater}, Jose and {Sakurikar}, Parikshit and {Salgado}, Jes{\'u}s and {Sanghi}, Aniket and {Saunders}, Nicholas and {Savchenko}, Volodymyr and {Schwardt}, Ludwig and {Seifert-Eckert}, Michael and {Shih}, Albert Y. and {Jain}, Anany Shrey and {Shukla}, Gyanendra and {Sick}, Jonathan and {Simpson}, Chris and {Singanamalla}, Sudheesh and {Singer}, Leo P. and {Singhal}, Jaladh and {Sinha}, Manodeep and {Sip{\H{o}}cz}, Brigitta M. and {Spitler}, Lee R. and {Stansby}, David and {Streicher}, Ole and {{\v{S}}umak}, Jani and {Swinbank}, John D. and {Taranu}, Dan S. and {Tewary}, Nikita and {Tremblay}, Grant R. and {de Val-Borro}, Miguel and {Van Kooten}, Samuel J. and {Vasovi{\'c}}, Zlatan and {Verma}, Shresth and {de Miranda Cardoso}, Jos{\'e} Vin{\'\i}cius and {Williams}, Peter K.~G. and {Wilson}, Tom J. and {Winkel}, Benjamin and {Wood-Vasey}, W.~M. and {Xue}, Rui and {Yoachim}, Peter and {Zhang}, Chen and {Zonca}, Andrea and {Astropy Project Contributors}},
        title = "{The Astropy Project: Sustaining and Growing a Community-oriented Open-source Project and the Latest Major Release (v5.0) of the Core Package}",
      journal = {\apj},
         year = 2022,
        month = aug,
       volume = {935},
       number = {2},
          eid = {167},
        pages = {167},
          doi = {10.3847/1538-4357/ac7c74},
archivePrefix = {arXiv},
       eprint = {2206.14220},
 primaryClass = {astro-ph.IM},
       adsurl = {https://ui.adsabs.harvard.edu/abs/2022ApJ...935..167A}
}

@book{CTAscience18,
   title={Science with the Cherenkov Telescope Array},
   ISBN={9789813270091},
   url={http://dx.doi.org/10.1142/10986},
   DOI={10.1142/10986},
   publisher={WORLD SCIENTIFIC},
   year={2018},
   month=feb }

@INPROCEEDINGS{Arnaud96,
       author = {{Arnaud}, K.~A.},
        title = "{XSPEC: The First Ten Years}",
    booktitle = {Astronomical Data Analysis Software and Systems V},
         year = 1996,
       editor = {{Jacoby}, George H. and {Barnes}, Jeannette},
       series = {Astronomical Society of the Pacific Conference Series},
       volume = {101},
        month = jan,
        pages = {17},
       adsurl = {https://ui.adsabs.harvard.edu/abs/1996ASPC..101...17A}
}

@ARTICLE{Evans07,
       author = {{Evans}, P.~A. and {Beardmore}, A.~P. and {Page}, K.~L. and {Tyler}, L.~G. and {Osborne}, J.~P. and {Goad}, M.~R. and {O'Brien}, P.~T. and {Vetere}, L. and {Racusin}, J. and {Morris}, D. and {Burrows}, D.~N. and {Capalbi}, M. and {Perri}, M. and {Gehrels}, N. and {Romano}, P.},
        title = "{An online repository of Swift/XRT light curves of {\ensuremath{\gamma}}-ray bursts}",
      journal = {\aap},
         year = 2007,
        month = jul,
       volume = {469},
       number = {1},
        pages = {379-385},
          doi = {10.1051/0004-6361:20077530},
archivePrefix = {arXiv},
       eprint = {0704.0128},
 primaryClass = {astro-ph},
       adsurl = {https://ui.adsabs.harvard.edu/abs/2007A&A...469..379E}
}

@ARTICLE{Evans09,
       author = {{Evans}, P.~A. and {Beardmore}, A.~P. and {Page}, K.~L. and {Osborne}, J.~P. and {O'Brien}, P.~T. and {Willingale}, R. and {Starling}, R.~L.~C. and {Burrows}, D.~N. and {Godet}, O. and {Vetere}, L. and {Racusin}, J. and {Goad}, M.~R. and {Wiersema}, K. and {Angelini}, L. and {Capalbi}, M. and {Chincarini}, G. and {Gehrels}, N. and {Kennea}, J.~A. and {Margutti}, R. and {Morris}, D.~C. and {Mountford}, C.~J. and {Pagani}, C. and {Perri}, M. and {Romano}, P. and {Tanvir}, N.},
        title = "{Methods and results of an automatic analysis of a complete sample of Swift-XRT observations of GRBs}",
      journal = {\mnras},
         year = 2009,
        month = aug,
       volume = {397},
       number = {3},
        pages = {1177-1201},
          doi = {10.1111/j.1365-2966.2009.14913.x},
archivePrefix = {arXiv},
       eprint = {0812.3662},
 primaryClass = {astro-ph},
       adsurl = {https://ui.adsabs.harvard.edu/abs/2009MNRAS.397.1177E}
}

@dataset{cherenkov2021,
  author       = {Cherenkov Telescope Array         
                  Observatory and                 Cherenkov Telescope Array Consortium},
  title        = {CTAO Instrument Response 
                   Functions - prod5 version
                   v0.1},
  month        = sep,
  year         = 2021,
  publisher    = {Zenodo},
  version      = {v0.1},
  doi          = {10.5281/zenodo.5499840},
  url          = {https://doi.org/10.5281/zenodo.5499840},
}

@article{gammapy:2023,
 author = {{Donath}, Axel and {Terrier}, R\'egis and {Remy}, Quentin and {Sinha}, Atreyee and {Nigro}, Cosimo and
 {Pintore}, Fabio and {Kh\'elifi}, Bruno and {Olivera-Nieto}, Laura and {Ruiz}, Jose Enrique and
 {Br\"ugge}, Kai and {Linhoff}, Maximilian and {Contreras}, Jose Luis and {Acero}, Fabio and
 {Aguasca-Cabot}, Arnau and {Berge}, David and {Bhattacharjee}, Pooja and {Buchner}, Johannes and
 {Boisson}, Catherine and {Carreto Fidalgo}, David and {Chen}, Andrew and {de Bony de Lavergne}, Mathieu and
 {de Miranda Cardoso}, Jos\'e Vinicius and {Deil}, Christoph and {F\"u\ss{}ling}, Matthias and
 {Funk}, Stefan and {Giunti}, Luca and {Hinton}, Jim and {Jouvin}, L\'ea and {King}, Johannes and
 {Lefaucheur}, Julien and {Lemoine-Goumard}, Marianne and {Lenain}, Jean-Philippe and {L\'opez-Coto}, Rub\'en
 and {Mohrmann}, Lars and {Morcuende}, Daniel and {Panny}, Sebastian and {Regeard}, Maxime and {Saha}, Lab
 and {Siejkowski}, Hubert and {Siemiginowska}, Aneta and {Sip"ocz}, Brigitta M. and {Unbehaun}, Tim
 and {van Eldik}, Christopher and {Vuillaume}, Thomas and {Zanin}, Roberta},
 title = {Gammapy: A Python package for gamma-ray astronomy},
 DOI= "10.1051/0004-6361/202346488",
 url= "https://doi.org/10.1051/0004-6361/202346488",
 journal = {A\&A},
 year = 2023,
 volume = 678,
 pages = "A157",
 }

@software{gammapy:zenodo-1.2,
  author       = {Acero, Fabio and
                  Bernete, Juan and
                  Biederbeck, Noah and
                  Djuvsland, Julia and
                  Donath, Axel and
                  Feijen, Kirsty and
                  Fröse, Stefan and
                  Galelli, Claudio and
                  Khélifi, Bruno and
                  Konrad, Jana and
                  Kornecki, Paula and
                  Linhoff, Maximilian and
                  McKee, Kurt and
                  Mender, Simone and
                  Morcuende, Daniel and
                  Olivera-Nieto, Laura and
                  Pintore, Fabio and
                  Punch, Michael and
                  Regeard, Maxime and
                  Remy, Quentin and
                  Sinha, Atreyee and
                  Stapel, Hanna and
                  Streil, Katrin and
                  Terrier, Régis and
                  Unbehaun, Tim},
  title        = {Gammapy v1.2: Python toolbox for gamma-ray astronomy},
  month        = 02,
  year         = 2024,
  publisher    = {Zenodo},
  version      = {v1.2},
  doi          = {10.5281/zenodo.10726484},
  url          = {https://doi.org/10.5281/zenodo.10726484}
}

@ARTICLE{kalberla05,
       author = {{Kalberla}, P.~M.~W. and {Burton}, W.~B. and {Hartmann}, Dap and {Arnal}, E.~M. and {Bajaja}, E. and {Morras}, R. and {P{\"o}ppel}, W.~G.~L.},
        title = "{The Leiden/Argentine/Bonn (LAB) Survey of Galactic HI. Final data release of the combined LDS and IAR surveys with improved stray-radiation corrections}",
      journal = {\aap},
         year = 2005,
        month = sep,
       volume = {440},
       number = {2},
        pages = {775-782},
          doi = {10.1051/0004-6361:20041864},
archivePrefix = {arXiv},
       eprint = {astro-ph/0504140},
 primaryClass = {astro-ph},
       adsurl = {https://ui.adsabs.harvard.edu/abs/2005A&A...440..775K}
}

@ARTICLE{celotti08,
       author = {{Celotti}, Annalisa and {Ghisellini}, Gabriele},
        title = "{The power of blazar jets}",
      journal = {\mnras},
         year = 2008,
        month = mar,
       volume = {385},
       number = {1},
        pages = {283-300},
          doi = {10.1111/j.1365-2966.2007.12758.x},
archivePrefix = {arXiv},
       eprint = {0711.4112},
 primaryClass = {astro-ph},
       adsurl = {https://ui.adsabs.harvard.edu/abs/2008MNRAS.385..283C}
}

@ARTICLE{ghisellini09,
       author = {{Ghisellini}, G. and {Tavecchio}, F.},
        title = "{Canonical high-power blazars}",
      journal = {\mnras},
         year = 2009,
        month = aug,
       volume = {397},
       number = {2},
        pages = {985-1002},
          doi = {10.1111/j.1365-2966.2009.15007.x},
archivePrefix = {arXiv},
       eprint = {0902.0793},
 primaryClass = {astro-ph.CO},
       adsurl = {https://ui.adsabs.harvard.edu/abs/2009MNRAS.397..985G}
}

@ARTICLE{klinger24,
       author = {{Klinger}, Marc and {Rudolph}, Annika and {Rodrigues}, Xavier and {Yuan}, Chengchao and {Fichet de Clairfontaine}, Ga{\"e}tan and {Fedynitch}, Anatoli and {Winter}, Walter and {Pohl}, Martin and {Gao}, Shan},
        title = "{AM$^{3}$: An Open-source Tool for Time-dependent Lepto-hadronic Modeling of Astrophysical Sources}",
      journal = {\apjs},
         year = 2024,
        month = nov,
       volume = {275},
       number = {1},
          eid = {4},
        pages = {4},
          doi = {10.3847/1538-4365/ad725c},
archivePrefix = {arXiv},
       eprint = {2312.13371},
 primaryClass = {astro-ph.HE},
       adsurl = {https://ui.adsabs.harvard.edu/abs/2024ApJS..275....4K}
}

@ARTICLE{stathopoulos24,
       author = {{Stathopoulos}, S.~I. and {Petropoulou}, M. and {Vasilopoulos}, G. and {Mastichiadis}, A.},
        title = "{LeHaMoC: A versatile time-dependent lepto-hadronic modeling code for high-energy astrophysical sources}",
      journal = {\aap},
         year = 2024,
        month = mar,
       volume = {683},
          eid = {A225},
        pages = {A225},
          doi = {10.1051/0004-6361/202347277},
archivePrefix = {arXiv},
       eprint = {2308.06174},
 primaryClass = {astro-ph.HE},
       adsurl = {https://ui.adsabs.harvard.edu/abs/2024A&A...683A.225S}
}

@INPROCEEDINGS{wood17,
       author = {{Wood}, M. and {Caputo}, R. and {Charles}, E. and {Di Mauro}, M. and {Magill}, J. and {Perkins}, J.~S. and {Fermi-LAT Collaboration}},
        title = "{Fermipy: An open-source Python package for analysis of Fermi-LAT Data}",
    booktitle = {35th International Cosmic Ray Conference (ICRC2017)},
         year = 2017,
       series = {International Cosmic Ray Conference},
       volume = {301},
        month = jul,
          eid = {824},
        pages = {824},
          doi = {10.22323/1.301.0824},
archivePrefix = {arXiv},
       eprint = {1707.09551},
 primaryClass = {astro-ph.IM},
       adsurl = {https://ui.adsabs.harvard.edu/abs/2017ICRC...35..824W}
}

@misc{cooray16,
      title={Extragalactic Background Light: Measurements and Applications}, 
      author={Asantha Cooray},
      year={2016},
      eprint={1602.03512},
      archivePrefix={arXiv},
      primaryClass={astro-ph.CO},
      url={https://arxiv.org/abs/1602.03512}, 
}

@INPROCEEDINGS{cortina19,
       author = {{Cortina}, J. and {Project}, C. l.},
        title = "{Status of the Large Size Telescopes of the Cherenkov Telescope Array}",
    booktitle = {36th International Cosmic Ray Conference (ICRC2019)},
         year = 2019,
       series = {International Cosmic Ray Conference},
       volume = {36},
        month = jul,
          eid = {653},
        pages = {653},
          doi = {10.22323/1.358.0653},
archivePrefix = {arXiv},
       eprint = {1907.10146},
 primaryClass = {astro-ph.IM},
       adsurl = {https://ui.adsabs.harvard.edu/abs/2019ICRC...36..653C}
}

@ARTICLE{aleksic16,
       author = {{Aleksi{\'c}}, J. and {Ansoldi}, S. and {Antonelli}, L.~A. and {Antoranz}, P. and {Babic}, A. and {Bangale}, P. and {Barcel{\'o}}, M. and {Barrio}, J.~A. and {Becerra Gonz{\'a}lez}, J. and {Bednarek}, W. and {Bernardini}, E. and {Biasuzzi}, B. and {Biland}, A. and {Bitossi}, M. and {Blanch}, O. and {Bonnefoy}, S. and {Bonnoli}, G. and {Borracci}, F. and {Bretz}, T. and {Carmona}, E. and {Carosi}, A. and {Cecchi}, R. and {Colin}, P. and {Colombo}, E. and {Contreras}, J.~L. and {Corti}, D. and {Cortina}, J. and {Covino}, S. and {Da Vela}, P. and {Dazzi}, F. and {De Angelis}, A. and {De Caneva}, G. and {De Lotto}, B. and {de O{\~n}a Wilhelmi}, E. and {Delgado Mendez}, C. and {Dettlaff}, A. and {Dominis Prester}, D. and {Dorner}, D. and {Doro}, M. and {Einecke}, S. and {Eisenacher}, D. and {Elsaesser}, D. and {Fidalgo}, D. and {Fink}, D. and {Fonseca}, M.~V. and {Font}, L. and {Frantzen}, K. and {Fruck}, C. and {Galindo}, D. and {Garc{\'\i}a L{\'o}pez}, R.~J. and {Garczarczyk}, M. and {Garrido Terrats}, D. and {Gaug}, M. and {Giavitto}, G. and {Godinovi{\'c}}, N. and {Gonz{\'a}lez Mu{\~n}oz}, A. and {Gozzini}, S.~R. and {Haberer}, W. and {Hadasch}, D. and {Hanabata}, Y. and {Hayashida}, M. and {Herrera}, J. and {Hildebrand}, D. and {Hose}, J. and {Hrupec}, D. and {Idec}, W. and {Illa}, J.~M. and {Kadenius}, V. and {Kellermann}, H. and {Knoetig}, M.~L. and {Kodani}, K. and {Konno}, Y. and {Krause}, J. and {Kubo}, H. and {Kushida}, J. and {La Barbera}, A. and {Lelas}, D. and {Lemus}, J.~L. and {Lewandowska}, N. and {Lindfors}, E. and {Lombardi}, S. and {Longo}, F. and {L{\'o}pez}, M. and {L{\'o}pez-Coto}, R. and {L{\'o}pez-Oramas}, A. and {Lorca}, A. and {Lorenz}, E. and {Lozano}, I. and {Makariev}, M. and {Mallot}, K. and {Maneva}, G. and {Mankuzhiyil}, N. and {Mannheim}, K. and {Maraschi}, L. and {Marcote}, B. and {Mariotti}, M. and {Mart{\'\i}nez}, M. and {Mazin}, D. and {Menzel}, U. and {Miranda}, J.~M. and {Mirzoyan}, R. and {Moralejo}, A. and {Munar-Adrover}, P. and {Nakajima}, D. and {Negrello}, M. and {Neustroev}, V. and {Niedzwiecki}, A. and {Nilsson}, K. and {Nishijima}, K. and {Noda}, K. and {Orito}, R. and {Overkemping}, A. and {Paiano}, S. and {Palatiello}, M. and {Paneque}, D. and {Paoletti}, R. and {Paredes}, J.~M. and {Paredes-Fortuny}, X. and {Persic}, M. and {Poutanen}, J. and {Prada Moroni}, P.~G. and {Prandini}, E. and {Puljak}, I. and {Reinthal}, R. and {Rhode}, W. and {Rib{\'o}}, M. and {Rico}, J. and {Rodriguez Garcia}, J. and {R{\"u}gamer}, S. and {Saito}, T. and {Saito}, K. and {Satalecka}, K. and {Scalzotto}, V. and {Scapin}, V. and {Schultz}, C. and {Schlammer}, J. and {Schmidl}, S. and {Schweizer}, T. and {Shore}, S.~N. and {Sillanp{\"a}{\"a}}, A. and {Sitarek}, J. and {Snidaric}, I. and {Sobczynska}, D. and {Spanier}, F. and {Stamerra}, A. and {Steinbring}, T. and {Storz}, J. and {Strzys}, M. and {Takalo}, L. and {Takami}, H. and {Tavecchio}, F. and {Tejedor}, L.~A. and {Temnikov}, P. and {Terzi{\'c}}, T. and {Tescaro}, D. and {Teshima}, M. and {Thaele}, J. and {Tibolla}, O. and {Torres}, D.~F. and {Toyama}, T. and {Treves}, A. and {Vogler}, P. and {Wetteskind}, H. and {Will}, M. and {Zanin}, R.},
        title = "{The major upgrade of the MAGIC telescopes, Part II: A performance study using observations of the Crab Nebula}",
      journal = {Astroparticle Physics},
         year = 2016,
        month = jan,
       volume = {72},
        pages = {76-94},
          doi = {10.1016/j.astropartphys.2015.02.005},
archivePrefix = {arXiv},
       eprint = {1409.5594},
 primaryClass = {astro-ph.IM},
       adsurl = {https://ui.adsabs.harvard.edu/abs/2016APh....72...76A}
}

@ARTICLE{jansen01,
       author = {{Jansen}, F. and {Lumb}, D. and {Altieri}, B. and {Clavel}, J. and {Ehle}, M. and {Erd}, C. and {Gabriel}, C. and {Guainazzi}, M. and {Gondoin}, P. and {Much}, R. and {Munoz}, R. and {Santos}, M. and {Schartel}, N. and {Texier}, D. and {Vacanti}, G.},
        title = "{XMM-Newton observatory. I. The spacecraft and operations}",
      journal = {\aap},
         year = 2001,
        month = jan,
       volume = {365},
        pages = {L1-L6},
          doi = {10.1051/0004-6361:20000036},
       adsurl = {https://ui.adsabs.harvard.edu/abs/2001A&A...365L...1J}
}

@ARTICLE{weisskopf02,
       author = {{Weisskopf}, M.~C. and {Brinkman}, B. and {Canizares}, C. and {Garmire}, G. and {Murray}, S. and {Van Speybroeck}, L.~P.},
        title = "{An Overview of the Performance and Scientific Results from the Chandra X-Ray Observatory}",
      journal = {\pasp},
         year = 2002,
        month = jan,
       volume = {114},
       number = {791},
        pages = {1-24},
          doi = {10.1086/338108},
archivePrefix = {arXiv},
       eprint = {astro-ph/0110308},
 primaryClass = {astro-ph},
       adsurl = {https://ui.adsabs.harvard.edu/abs/2002PASP..114....1W}
}

@article{Gehrels04,
doi = {10.1086/422091},
url = {https://doi.org/10.1086/422091},
year = {2004},
month = {aug},
publisher = {},
volume = {611},
number = {2},
pages = {1005},
author = {Gehrels, N. and Chincarini, G. and Giommi, P. and Mason, K. O. and Nousek, J. A. and Wells, A. A. and White, N. E. and Barthelmy, S. D. and Burrows, D. N. and Cominsky, L. R. and Hurley, K. C. and Marshall, F. E. and Mészáros, P. and Roming, P. W. A. and Angelini, L. and Barbier, L. M. and Belloni, T. and Campana, S. and Caraveo, P. A. and Chester, M. M. and Citterio, O. and Cline, T. L. and Cropper, M. S. and Cummings, J. R. and Dean, A. J. and Feigelson, E. D. and Fenimore, E. E. and Frail, D. A. and Fruchter, A. S. and Garmire, G. P. and Gendreau, K. and Ghisellini, G. and Greiner, J. and Hill, J. E. and Hunsberger, S. D. and Krimm, H. A. and Kulkarni, S. R. and Kumar, P. and Lebrun, F. and Lloyd-Ronning, N. M. and Markwardt, C. B. and Mattson, B. J. and Mushotzky, R. F. and Norris, J. P. and Osborne, J. and Paczynski, B. and Palmer, D. M. and Park, H.-S. and Parsons, A. M. and Paul, J. and Rees, M. J. and Reynolds, C. S. and Rhoads, J. E. and Sasseen, T. P. and Schaefer, B. E. and Short, A. T. and Smale, A. P. and Smith, I. A. and Stella, L. and Tagliaferri, G. and Takahashi, T. and Tashiro, M. and Townsley, L. K. and Tueller, J. and Turner, M. J. L. and Vietri, M. and Voges, W. and Ward, M. J. and Willingale, R. and Zerbi, F. M. and Zhang, W. W.},
title = {The Swift Gamma-Ray Burst Mission},
journal = {The Astrophysical Journal}
}

@ARTICLE{cao23,
       author = {{Cao}, Zhen and {Chen}, Songzhan and {Liu}, Ruoyu and {Yang}, Ruizhi},
        title = "{Ultra-High-Energy Gamma-Ray Astronomy}",
      journal = {Annual Review of Nuclear and Particle Science},
         year = 2023,
        month = sep,
       volume = {73},
        pages = {341-363},
          doi = {10.1146/annurev-nucl-112822-025357},
archivePrefix = {arXiv},
       eprint = {2310.01744},
 primaryClass = {astro-ph.HE},
       adsurl = {https://ui.adsabs.harvard.edu/abs/2023ARNPS..73..341C}
}

@ARTICLE{tavecchio98,
       author = {{Tavecchio}, Fabrizio and {Maraschi}, Laura and {Ghisellini}, Gabriele},
        title = "{Constraints on the Physical Parameters of TeV Blazars}",
      journal = {\apj},
         year = 1998,
        month = dec,
       volume = {509},
       number = {2},
        pages = {608-619},
          doi = {10.1086/306526},
archivePrefix = {arXiv},
       eprint = {astro-ph/9809051},
 primaryClass = {astro-ph},
       adsurl = {https://ui.adsabs.harvard.edu/abs/1998ApJ...509..608T}
}

\appendix
\onecolumn
\begin{landscape}
\section{X-ray analysis results}\label{app:xresults}
In this Appendix, we report the results of the fits performed on the 73 sources of our sample with data of sufficient quality. In Table~\ref{tab:result_xanalysis} we report the parameters resulting from the fits, and the fit quality.

{\fontsize{9}{10}\selectfont
\begin{longtable}{
>{\raggedright\arraybackslash}p{3.5cm} 
>{\centering\arraybackslash}p{0.7cm}
>{\centering\arraybackslash}p{0.6cm}
>{\centering\arraybackslash}p{1.1cm}
>{\centering\arraybackslash}p{1.4cm}
>{\centering\arraybackslash}p{1.2cm}
>{\centering\arraybackslash}p{1.4cm}
>{\centering\arraybackslash}p{1.5cm}
>{\centering\arraybackslash}p{1.2cm}
>{\centering\arraybackslash}p{2cm}
>{\centering\arraybackslash}p{0.8cm}
>{\raggedright\arraybackslash}p{2.5cm}
}
\toprule
           source &  z &  $\nu_{\rm{peak}}$ &      $\Gamma$ & Cstat/DoF &             $\alpha$ &        $\beta$ & Cstat/DoF &                     $ \rm \nu F_{\nu ,0.3-8 keV}$  &                  log(L$ \rm _{2-10 \ keV})$ &  $\rm{\Delta Cstat}$ &                    catalogue \\
           & & & & (powerlaw) & & &(logpar) &   [10$^{-12}$ erg s$^{-1}$ cm$^{-2}$] & [log(erg s$^{-1})]$ & &  \\
         
\midrule
   5BZBJ0333$-$3619(1) &     0.308 &                   17.1 & $1.65^{+0.03}_{-0.03}$ &      565.9/530 &        $1.46^{+0.05}_{-0.05}$ &   $0.64^{+0.11}_{-0.10}$ &      463.8/529 &    $2.04^{+0.07}_{-0.06}$ &  $44.66^{+0.02}_{-0.02}$ &                    102.1 &                      2SXPS \\
5BZGJ1510+3335(1) &     0.114 &                   17.1 & $1.94^{+0.03}_{-0.03}$ &      459.1/413 &        $1.74^{+0.06}_{-0.06}$ &    $0.41^{+0.10}_{-0.10}$ &      410.4/412 &    $2.46^{+0.06}_{-0.05}$ & $43.69^{+0.01}_{-0.01}$ &                     48.7 &                       2CSC \\
5BZGJ1510+3335(2) &     0.114 &                   17.1 & $1.82^{+0.02}_{-0.02}$ &      489.7/439 &        $1.67^{+0.05}_{-0.05}$ &  $0.23^{+0.08}_{-0.08}$ &      458.9/438 &    $2.55^{+0.05}_{-0.05}$ & $43.73^{+0.01}_{-0.01}$ &                     30.8 &                       2CSC \\
   5BZBJ0333$-$3619(2) &     0.308 &                   17.1 &  $2.00^{+0.02}_{-0.02}$ &      590.3/536 &        $2.01^{+0.02}_{-0.02}$ &  $0.24^{+0.09}_{-0.09}$ &      571.9/535 &    $1.75^{+0.06}_{-0.06}$ & $44.48^{+0.01}_{-0.01}$ &                     18.4 &                  4XMM-DR14 \\
5BZBJ1253+3826(1) &     0.371 &                   17.1 & $1.82^{+0.07}_{-0.06}$ &      391.1/397 &        $1.69^{+0.10}_{-0.09}$ &  $0.57^{+0.24}_{-0.26}$ &      373.3/396 &    $1.91^{+0.13}_{-0.12}$ &  $44.77^{+0.03}_{-0.04}$ &                     17.8 &                      2SXPS \\
5BZGJ1510+3335(3) &     0.114 &                   17.1 & $1.93^{+0.04}_{-0.04}$ &      722.6/728 &        $1.93^{+0.04}_{-0.04}$ &  $0.35^{+0.15}_{-0.15}$ &      706.6/727 &    $2.35^{+0.14}_{-0.11}$ & $43.69^{+0.02}_{-0.03}$ &                     16.0 &                  4XMM-DR14 \\
5BZBJ1636$-$1248 $\ddagger$ &     0.246 &                   16.7 &  $1.20^{+0.26}_{-0.28}$ &       105.9/90 &        $0.19^{+0.61}_{-0.81}$ &  $2.64^{+0.95}_{-1.25}$ &        91.0/89 &   $1.06^{+0.36}_{-6.66}$ &  $44.40^{+0.14}_{-0.14}$ &                     14.9 &                      2SXPS \\
   5BZBJ1251$-$2958(1) &     0.382 &                   17.0 & $2.22^{+0.04}_{-0.04}$ &      550.3/524 &        $2.21^{+0.04}_{-0.04}$ &   $0.40^{+0.18}_{-0.18}$ &      536.4/523 &    $1.99^{+0.11}_{-0.09}$ & $43.94^{+0.02}_{-0.02}$ &                     13.9 &                  4XMM-DR14 \\
5BZBJ1253+3826(2) &     0.371 &                   17.1 & $1.73^{+0.12}_{-0.13}$ &      200.3/211 &        $1.99^{+0.10}_{-0.18}$ &  $0.87^{+0.41}_{-0.44}$ &      186.7/210 &    $2.56^{+0.32}_{-0.27}$ &  $44.95^{+0.06}_{-0.06}$ &                     13.6 &                      2SXPS \\
5BZGJ1510+3335(4) &     0.114 &                   17.1 & $1.88^{+0.03}_{-0.03}$ &      532.2/536 &         $1.90^{+0.03}_{-0.03}$ &  $0.29^{+0.14}_{-0.14}$ &      520.2/535 &    $3.46^{+0.21}_{-0.17}$ & $43.87^{+0.02}_{-0.02}$ &                     12.0 &                  4XMM-DR14 \\
   5BZBJ0040$-$2719(1) &     0.172 &                   16.8 &  $2.60^{+0.03}_{-0.03}$ &      550.6/533 &        $2.66^{+0.04}_{-0.04}$ &  $0.45^{+0.15}_{-0.15}$ &      524.0/532 &     $3.25^{+0.10}_{-0.09}$ &  $43.90^{+0.02}_{-0.02}$ &                      9.0 &                  4XMM-DR14 \\
   5BZGJ1510+3335(5) &     0.114 &                   17.1 & $1.81^{+0.09}_{-0.09}$ &      259.4/274 &         $1.46^{+0.22}_{-0.22}$ &  $0.63^{+0.36}_{-0.35}$ &      250.6/273 &    $2.77^{+0.28}_{-0.21}$ & $43.74^{+0.05}_{-0.05}$ &                     8.8 &                  2CSC \\
   5BZGJ1201$-$0011 &     0.164 &                   16.9 & $1.52^{+0.18}_{-0.19}$ &      125.6/147 &         $1.22^{+0.27}_{-0.30}$ &  $1.14^{+0.73}_{-0.67}$ &      117.4/146 &    $0.98^{+0.21}_{-0.15}$ &   $43.84^{+0.10}_{-0.11}$ &                      8.2 &                      2SXPS \\
   5BZBJ1302+5056 &   0.688* &                   17.2 & $1.92^{+0.09}_{-0.09}$ &      243.6/268 &        $1.84^{+0.11}_{-0.11}$ &   $0.49^{+0.31}_{-0.30}$ &      236.0/267 &     $2.10^{+0.23}_{-0.16}$ &  $45.42^{+0.04}_{-0.04}$ &                      7.6 &                      2SXPS \\
   5BZBJ1057+2303 &     0.379 &                   17.1 &    $2.10^{+0.10}_{-0.10}$ &      205.4/233 &        $2.02^{+0.12}_{-0.12}$ &  $0.51^{+0.34}_{-0.33}$ &      198.6/232 &    $3.25^{+0.34}_{-0.27}$ &  $44.93^{+0.06}_{-0.06}$ &                      6.8 &                      2SXPS \\
   5BZGJ1544+0458(1) &     0.326 &                   16.6 & $2.51^{+0.13}_{-0.13}$ &      160.6/213 &        $2.21^{+0.26}_{-0.27}$ &  $0.69^{+0.54}_{-0.51}$ &      155.6/212 &    $0.75^{+0.11}_{-0.07}$ & $44.07^{+0.05}_{-0.05}$ &                      5.0 &                       2CSC \\
   5BZGJ1444+6336 &     0.298 &                   17.2 & $1.76^{+0.11}_{-0.11}$ &      209.2/232 &        $1.66^{+0.14}_{-0.14}$ &  $0.47^{+0.39}_{-0.37}$ &      204.8/231 &    $3.22^{+0.48}_{-0.34}$ &  $44.80^{+0.06}_{-0.06}$ &                      4.4 &                      2SXPS \\
   5BZGJ2310$-$4347(1) &     0.089 &                   16.8 &   $1.78^{+0.20}_{-0.20}$ &      103.5/125 &        $1.68^{+0.23}_{-0.25}$ &  $0.93^{+0.81}_{-0.75}$ &       99.2/124 &    $2.38^{+0.67}_{-0.35}$ &  $43.57^{+0.13}_{-0.13}$ &                      4.3 &                      2SXPS \\
5BZGJ1616+3756(1) &     0.202 &                   17.2 & $1.94^{+0.08}_{-0.08}$ &      284.7/294 &        $1.89^{+0.09}_{-0.09}$ &  $0.32^{+0.27}_{-0.26}$ &      280.6/293 &     $4.58^{+0.40}_{-0.35}$ &  $44.50^{+0.05}_{-0.05}$ &                      4.1 &                      2SXPS \\
   5BZGJ1552+3159 &     0.584 &                   16.6 &  $1.80^{+0.16}_{-0.16}$ &      158.9/163 &        $1.46^{+0.35}_{-0.37}$ &  $0.67^{+0.63}_{-0.59}$ &      155.4/162 &     $1.30^{+0.23}_{-0.16}$ & $45.09^{+0.06}_{-0.06}$ &                      3.5 &                       2CSC \\
5BZGJ1616+3756(2) &     0.202 &                   17.2 & $1.93^{+0.08}_{-0.08}$ &      286.3/296 &        $1.88^{+0.09}_{-0.09}$ &  $0.29^{+0.26}_{-0.26}$ &      282.9/295 &    $4.59^{+0.45}_{-0.35}$ &  $44.50^{+0.05}_{-0.05}$ &                      3.4 &                      2SXPS \\
   5BZBJ2217$-$3106 &     0.460* &                   17.1 & $1.85^{+0.32}_{-0.31}$ &        66.6/69 &        $1.46^{+0.49}_{-0.55}$ &  $1.34^{+1.37}_{-1.22}$ &        63.2/68 &    $2.09^{+0.79}_{-0.47}$ &  $45.08^{+0.14}_{-0.15}$ &                      3.4 &                      2SXPS \\
   5BZBJ0124+0918(1) &     0.338 &                   17.1 &  $1.90^{+0.08}_{-0.08}$ &      467.5/467 &        $1.59^{+0.08}_{-0.08}$ &  $1.09^{+0.39}_{-0.38}$ &      463.6/466 &     $0.99^{+0.17}_{-0.10}$ & $44.46^{+0.05}_{-0.05}$ &                      3.4 &                  4XMM-DR14 \\
   5BZBJ1258+0134 &   -99 &                   16.2 & $2.41^{+0.16}_{-0.16}$ &      326.1/396 &        $2.29^{+0.19}_{-0.17}$ &  $-0.73^{+0.70}_{-0.63}$ &      323.2/395 &    $0.48^{+0.29}_{-0.12}$ &      NaN &                      3.1 &                  4XMM-DR14 \\
5BZGJ1510+3335(6) &     0.114 &                   17.1 & $1.92^{+0.06}_{-0.06}$ &     738.6/2516 &         $1.90^{+0.07}_{-0.07}$ &    $0.20^{+0.20}_{-0.19}$ &     735.8/2515 &    $3.47^{+0.24}_{-0.21}$ & $43.83^{+0.04}_{-0.04}$ &                      2.8 &                  4XMM-DR14 \\ \\
\hline \\
   
   5BZBJ2136$-$4443 &   -99 &                   16.9 & $2.25^{+0.39}_{-0.39}$ &        66.7/69 &        $1.69^{+0.59}_{-0.55}$ &  $1.85^{+1.53}_{-1.42}$ &        62.8/68 &  $8.87^{+173.13}_{-7.22}$ &      NaN &                      3.9 &                     eRASS1 \\
   5BZGJ2238$-$3940(1) &     0.251 &                   17.0 & $1.89^{+0.67}_{-0.68}$ &        22.8/26 &        $1.99^{+0.52}_{-0.56}$ & $-2.34^{+2.02}_{-6.57}$ &        19.3/25 &   $9.33^{+53.37}_{-7.24}$ &  $44.33^{+0.39}_{-0.41}$ &                      3.5 &                      2SXPS \\
   5BZBJ1027+3526 &   0.470* &                   16.7 & $1.65^{+0.69}_{-0.71}$ &        24.9/29 &        $2.21^{+0.65}_{-0.81}$ & $-3.34^{+2.96}_{-2.61}$ &        21.5/28 &   $21.30^{+41.90}_{-19.54}$ &  $44.91^{+0.33}_{-0.32}$ &                      3.4 &                      2SXPS \\
   5BZGJ1322+1344 &     0.377 &                   16.2 & $2.14^{+0.44}_{-0.39}$ &        44.6/69 &         $1.26^{+0.95}_{-1.10}$ &  $2.49^{+4.16}_{-2.36}$ &        41.6/68 &     $0.20^{+0.15}_{-0.04}$ &  $43.79^{+0.17}_{-0.23}$ &                      3.0 &                      2SXPS \\
   5BZBJ0124+0918(2) &     0.338 &                   17.1 & $0.97^{+0.63}_{-0.71}$ &        21.6/24 &         $1.59^{+0.60}_{-0.84}$ & $-2.09^{+2.01}_{-1.82}$ &        18.7/23 &   $6.77^{+21.33}_{-4.81}$ &  $44.78^{+0.32}_{-0.31}$ &                      2.9 &                      2SXPS \\
   5BZBJ0755+3726 &     0.606 &                   17.0 & $1.77^{+0.34}_{-0.35}$ &        54.3/69 &         $1.70^{+0.38}_{-0.41}$ &  $1.44^{+1.69}_{-1.51}$ &        51.9/68 &     $0.69^{+0.50}_{-0.15}$ &  $45.00^{+0.16}_{-0.16}$ &                      2.4 &                      2SXPS \\
   5BZGJ2310$-$4347(2) &     0.089 &                   16.8 & $1.86^{+0.03}_{-0.03}$ &      454.1/409 &        $1.83^{+0.05}_{-0.05}$ &  $0.08^{+0.10}_{-0.10}$ &      452.0/408 &    $2.72^{+0.09}_{-0.07}$ & $43.59^{+0.02}_{-0.02}$ &                      2.1 &                       2CSC \\
   5BZBJ0403$-$2429 &     0.357 &                   17.4 & $1.84^{+0.14}_{-0.14}$ &      150.3/177 &         $1.74^{+0.19}_{-0.20}$ &  $0.37^{+0.47}_{-0.44}$ &      148.5/176 &    $3.34^{+0.56}_{-0.42}$ &  $44.97^{+0.07}_{-0.08}$ &                      1.8 &                      2SXPS \\
5BZGJ0643+4214(1) &     0.089 &                   17.2 & $1.88^{+0.34}_{-0.33}$ &        51.3/77 &         $2.28^{+0.55}_{-0.60}$ & $-1.01^{+1.26}_{-1.18}$ &        49.6/76 &     $3.57^{+2.60}_{-1.16}$ &  $43.57^{+0.17}_{-0.17}$ &                      1.7 &                      2SXPS \\
   5BZBJ0737+2846 &     0.272 &                   16.2 & $2.61^{+0.27}_{-0.27}$ &        75.3/83 &         $2.60^{+0.29}_{-0.29}$ &  $0.81^{+1.12}_{-0.99}$ &        73.6/82 &      $0.60^{+0.17}_{-0.10}$ &  $43.73^{+0.15}_{-0.16}$ &                      1.7 &                      2SXPS \\
   5BZBJ1729+5255 &     0.349 &                   16.4 &   $2.17^{+0.20}_{-0.20}$ &      122.5/122 &        $2.14^{+0.22}_{-0.22}$ &   $0.55^{+0.80}_{-0.75}$ &      121.0/121 &     $1.68^{+0.50}_{-0.25}$ &  $44.59^{+0.11}_{-0.11}$ &                      1.5 &                      2SXPS \\
   5BZBJ0951+0102 &     0.502 &                   16.3 &  $2.30^{+0.48}_{-0.48}$ &        44.6/47 &        $2.74^{+0.92}_{-0.77}$ &    $1.70^{+2.80}_{-2.29}$ &        43.1/46 &    $0.59^{+0.77}_{-0.15}$ &  $44.60^{+0.29}_{-0.29}$ &                      1.5 &                     eRASS1 \\
   5BZGJ2116$-$0628 &     0.292 &                   16.1 & $2.13^{+0.21}_{-0.21}$ &      130.6/121 &        $2.05^{+0.25}_{-0.27}$ &   $0.47^{+0.77}_{-0.70}$ &      129.4/120 &     $0.50^{+0.13}_{-0.08}$ &  $43.92^{+0.11}_{-0.11}$ &                      1.2 &                      2SXPS \\
   5BZBJ1401+5209 &     0.482 &                   16.9 & $1.75^{+0.33}_{-0.33}$ &        49.8/64 &        $1.84^{+0.32}_{-0.34}$ & $-0.73^{+1.12}_{-1.04}$ &        48.6/63 &    $2.58^{+2.85}_{-0.95}$ &  $44.99^{+0.17}_{-0.17}$ &                      1.2 &                      2SXPS \\
   5BZBJ1251$-$2958(2) &     0.382 &                   17.0 &  $2.00^{+0.18}_{-0.18}$ &      136.8/154 &        $2.06^{+0.19}_{-0.21}$ & $-0.32^{+0.55}_{-0.52}$ &      135.8/153 &     $3.03^{+0.72}_{-0.50}$ &  $44.91^{+0.09}_{-0.09}$ &                      1.0 &                      2SXPS \\
   5BZBJ1411+3404 &     0.421 &                   16.2 & $2.37^{+0.29}_{-0.28}$ &        57.9/76 &        $2.39^{+0.27}_{-0.27}$ &  $-0.49^{+0.90}_{-0.81}$ &        57.0/75 &    $0.54^{+0.25}_{-0.12}$ &  $44.09^{+0.15}_{-0.16}$ &                      0.9 &                      2SXPS \\
   5BZBJ0624$-$3230 &     0.252 &                   17.2 & $2.05^{+0.22}_{-0.22}$ &      131.5/148 &        $2.18^{+0.33}_{-0.31}$ &  $0.59^{+1.08}_{-0.99}$ &      130.6/147 &    $2.28^{+1.58}_{-0.48}$ &  $44.48^{+0.15}_{-0.15}$ &                      0.9 &                     eRASS1 \\
5BZBJ1025+0402(1) &     0.208 &                   17.1 & $1.57^{+0.62}_{-0.65}$ &        28.4/32 &        $1.41^{+0.73}_{-1.03}$ &  $1.47^{+3.51}_{-2.53}$ &        27.6/31 &     $1.18^{+2.90}_{-0.42}$ &  $44.22^{+0.37}_{-0.37}$ &                      0.8 &                      2SXPS \\
   
   5BZGJ1324+5739 &     0.115 &                   16.1 & $2.33^{+0.14}_{-0.14}$ &      117.1/154 &          $2.26^{+0.20}_{-0.20}$ &  $0.23^{+0.51}_{-0.49}$ &      116.5/153 &    $1.17^{+0.14}_{-0.13}$ &  $43.20^{+0.08}_{-0.08}$ &                      0.6 &                       2CSC \\
   5BZGJ2319$-$0116 &     0.284 &                   16.8 & $2.12^{+0.19}_{-0.19}$ &      116.8/138 &          $2.10^{+0.20}_{-0.21}$ &   $0.30^{+0.73}_{-0.69}$ &      116.3/137 &    $1.68^{+0.44}_{-0.27}$ &  $44.37^{+0.11}_{-0.11}$ &                      0.5 &                      2SXPS \\
   5BZGJ2308$-$2219(1) &     0.137 &                   16.1 &   $2.26^{+0.10}_{-0.10}$ &      211.4/243 &        $2.32^{+0.17}_{-0.18}$ & $-0.15^{+0.36}_{-0.34}$ &      210.9/242 &    $4.66^{+0.51}_{-0.36}$ & $44.09^{+0.07}_{-0.06}$ &                      0.5 &                       2CSC \\
   5BZBJ0943$-$0709 &     0.433 &                   16.5 & $1.66^{+0.49}_{-0.51}$ &        37.4/45 &        $1.83^{+0.73}_{-0.65}$ &  $1.03^{+2.91}_{-2.35}$ &        36.9/44 &     $0.80^{+3.87}_{-0.27}$ &   $44.75^{+0.32}_{-0.30}$ &                      0.5 & eRASS1 \\
   5BZGJ1604+3345 &     0.177 &                   16.4 & $1.47^{+0.67}_{-0.72}$ &        29.0/20 &         $1.42^{+0.74}_{-0.90}$ &  $1.25^{+3.88}_{-2.99}$ &        28.6/19 &   $0.94^{+19.16}_{-0.41}$ &  $44.00^{+0.43}_{-0.42}$ &                      0.4 &                      2SXPS \\
   5BZGJ1449+2746 &     0.227 &                   16.8 &  $1.90^{+0.28}_{-0.29}$ &        64.3/84 &        $1.85^{+0.31}_{-0.34}$ &  $0.39^{+1.08}_{-0.99}$ &        63.9/83 &    $1.73^{+0.86}_{-0.39}$ &  $44.24^{+0.16}_{-0.17}$ &                      0.4 &                      2SXPS \\
   5BZGJ1221+0821(1) &     0.132 &                   16.2 & $2.22^{+0.33}_{-0.33}$ &        77.0/74 &          $2.40^{+0.60}_{-0.54}$ & $0.56^{+1.52}_{-0.14}$ &        76.6/73 &    $0.99^{+1.23}_{-0.22}$ &  $43.37^{+0.24}_{-0.25}$ &                      0.4 &                     eRASS1 \\
   5BZGJ0823+1524 &     0.167 &                   16.1 &  $2.27^{+0.70}_{-0.71}$ &        21.6/28 &        $2.59^{+1.35}_{-1.05}$ &  $1.36^{+4.55}_{-3.38}$ &        21.2/27 &    $0.37^{+7.68}_{-0.13}$ &   $43.26^{+0.49}_{-0.50}$ &                      0.4 &                     eRASS1 \\
5BZBJ2240+1326(1) &     0.660 &                   16.6 &  $1.66^{+0.50}_{-0.52}$ &        24.2/39 &        $1.61^{+0.53}_{-0.61}$ &  $0.84^{+2.44}_{-2.07}$ &        23.8/38 &     $1.70^{+4.25}_{-0.59}$ &  $45.44^{+0.23}_{-0.23}$ &                      0.4 &                      2SXPS \\
   5BZBJ1004+3752(1) &     0.440 &                   16.5 & $2.31^{+0.51}_{-0.51}$ &        36.9/44 &         $2.03^{+1.00}_{-0.90}$ &  $-0.81^{+2.47}_{-2.20}$ &        36.5/43 &   $1.54^{+74.06}_{-0.96}$ &  $44.43^{+0.33}_{-0.35}$ &                      0.4 &                     eRASS1 \\
   5BZUJ1129$-$4435 &     0.317 &                   16.4 & $1.81^{+0.34}_{-0.33}$ &        56.7/61 &        $1.63^{+0.61}_{-0.67}$ &    $0.40^{+1.34}_{-1.20}$ &        56.4/60 &    $0.73^{+0.32}_{-0.16}$ &  $44.50^{+0.13}_{-0.15}$ &                      0.3 &                      2SXPS \\
5BZGJ0643+4214(2) &     0.089 &                   17.2 & $2.03^{+0.23}_{-0.23}$ &      102.9/135 &         $1.92^{+0.40}_{-0.43}$ &  $0.31^{+0.91}_{-0.85}$ &      102.6/134 &    $4.06^{+1.23}_{-0.66}$ &  $43.70^{+0.11}_{-0.11}$ &                      0.3 &                      2SXPS \\
   5BZBJ1311+0853 &     0.469 &                   17.2 & $1.97^{+0.12}_{-0.11}$ &      225.6/257 &        $2.02^{+0.18}_{-0.19}$ & $-0.14^{+0.42}_{-0.41}$ &      225.3/256 &    $2.61^{+0.41}_{-0.28}$ &  $45.05^{+0.05}_{-0.05}$ &                      0.3 &                      2SXPS \\
   5BZBJ0851+0549(2) $\dagger$ &     0.486 &                   16.1 & $2.26^{+0.69}_{-0.72}$ &        28.1/22 &        $2.48^{+1.08}_{-0.86}$ &  $1.32^{+4.49}_{-3.82}$ &        27.8/21 &    $0.38^{+7.16}_{-0.13}$ &  $44.32^{+3.42}_{-0.39}$ &                      0.3 &                      2SXPS \\
   5BZGJ1226+2604 &     0.176 &                   17.0 & $2.12^{+0.19}_{-0.19}$ &      128.6/166 &        $2.18^{+0.32}_{-0.30}$ &   $0.22^{+0.85}_{-0.80}$ &      128.4/165 &    $3.48^{+2.21}_{-0.74}$ &  $44.19^{+0.15}_{-0.14}$ &                      0.2 &                     eRASS1 \\
   5BZGJ0903+4055 &     0.188 &                   16.2 & $2.33^{+0.24}_{-0.24}$ &       88.1/131 &        $2.32^{+0.27}_{-0.26}$ &  $0.17^{+0.96}_{-0.74}$ &       87.9/130 &     $0.31^{+0.10}_{-0.05}$ &  $43.08^{+0.14}_{-0.16}$ &                      0.2 &                      2SXPS \\
   5BZBJ2123$-$1036 &   0.230* &                   16.4 & $2.23^{+0.47}_{-0.48}$ &        45.3/36 &        $2.25^{+0.51}_{-0.51}$ &  $0.55^{+2.25}_{-1.94}$ &        45.1/35 &    $1.46^{+2.43}_{-0.41}$ &  $41.96^{+0.32}_{-0.32}$ &                      0.2 &                      2SXPS \\
   5BZBJ2038$-$2636 &     0.437 &                   16.5 &   $2.18^{+0.70}_{-0.70}$ &      532.7/522 &        $2.19^{+0.07}_{-0.07}$ &   $0.10^{+0.33}_{-0.31}$ &      532.5/521 &    $0.86^{+0.11}_{-0.08}$ &  $44.51^{+0.04}_{-0.40}$ &                      0.2 &                  4XMM-DR14 \\
   5BZBJ1256+0609 &     0.423 &                   16.9 &  $1.80^{+0.27}_{-0.26}$ &       73.4/102 &        $1.89^{+0.43}_{-0.46}$ &  $-0.25^{+1.00}_{-0.94}$ &       73.2/101 &    $1.03^{+0.47}_{-0.25}$ &  $44.57^{+0.11}_{-0.12}$ &                      0.2 &                      2SXPS \\
   5BZBJ1209+0210 &   0.291* &                   16.6 & $2.22^{+0.41}_{-0.41}$ &        41.4/55 &        $2.11^{+0.67}_{-0.60}$ & $-0.37^{+1.77}_{-1.59}$ &        41.2/54 &    $1.19^{+8.47}_{-0.51}$ &  $44.07^{+0.28}_{-0.21}$ &                      0.2 &                     eRASS1 \\
   5BZBJ1208$-$2937(1) &     0.249 &                   16.2 &  $1.92^{+0.70}_{-0.73}$ &        13.5/18 &        $1.93^{+0.67}_{-0.71}$ & $-0.79^{+3.04}_{-2.61}$ &        13.3/18 &     $2.30^{+32.90}_{-1.35}$ &  $44.23^{+5.81}_{-0.41}$ &                      0.2 &                      2SXPS \\
   5BZBJ1033$-$1436 &     0.367 &                   16.5 & $2.72^{+0.24}_{-0.23}$ &      101.2/110 &        $2.67^{+0.31}_{-0.32}$ &  $0.26^{+1.04}_{-0.88}$ &      101.0/109 &    $0.69^{+0.19}_{-0.11}$ &  $44.06^{+0.10}_{-0.11}$ &                      0.2 &                       2CSC \\
5BZBJ1025+0402(2) &     0.208 &                   17.1 & $1.83^{+0.16}_{-0.16}$ &      149.7/177 &        $1.81^{+0.17}_{-0.18}$ &  $0.16^{+0.58}_{-0.55}$ &      149.5/176 &    $3.46^{+0.92}_{-0.57}$ &    $44.45^{+0.10}_{-0.10}$ &                      0.2 &                      2SXPS \\
   5BZBJ0956+0156 &   0.650* &                   16.6 &  $1.60^{+0.42}_{-0.44}$ &        33.5/53 &        $1.62^{+0.41}_{-0.45}$ & $-0.39^{+1.81}_{-1.58}$ &        33.3/52 &     $0.58^{+1.99}_{-0.30}$ &   $44.73^{+0.20}_{-0.21}$ &                      0.2 &                      2SXPS \\
   5BZGJ2308$-$2219(2) &     0.137 &                   16.1 & $1.85^{+0.17}_{-0.17}$ &      133.8/165 &        $1.84^{+0.18}_{-0.18}$ &  $0.07^{+0.58}_{-0.56}$ &      133.7/164 &    $1.24^{+0.36}_{-0.21}$ &    $43.57^{+0.10}_{-0.10}$ &                      0.1 &                      2SXPS \\
   5BZGJ2238$-$3940(2) &     0.251 &                   17.0 & $2.14^{+0.35}_{-0.35}$ &        77.4/75 &        $2.24^{+0.66}_{-0.60}$ &  $0.29^{+1.65}_{-1.49}$ &        77.3/74 &    $1.22^{+3.88}_{-0.37}$ &  $44.09^{+0.25}_{-0.25}$ &                      0.1 &                     eRASS1 \\
   5BZGJ1221+0821(2) &     0.132 &                   16.2 & $2.04^{+0.59}_{-0.59}$ &        16.4/23 &        $2.03^{+0.62}_{-0.65}$ &  $0.37^{+2.56}_{-2.09}$ &        16.3/22 &    $1.25^{+3.61}_{-0.49}$ &   $43.51^{+nan}_{-0.38}$ &                      0.1 &                      2SXPS \\
   5BZGJ1157+2822 &     0.300 &                   16.7 & $1.89^{+0.34}_{-0.35}$ &        57.3/73 &        $1.81^{+0.53}_{-0.49}$ &  $-0.31^{+1.50}_{-1.36}$ &        57.2/72 &    $2.36^{+9.84}_{-1.16}$ &  $44.50^{+0.24}_{-0.24}$ &                      0.1 &                     eRASS1 \\
   5BZBJ0855+4237 &     0.517 &                   16.7 &  $1.90^{+0.09}_{-0.09}$ &      234.0/235 &        $1.89^{+0.15}_{-0.15}$ &  $0.04^{+0.33}_{-0.32}$ &      233.9/234 &    $1.67^{+0.17}_{-0.16}$ &  $45.00^{+0.04}_{-0.04}$ &                      0.1 &                       2CSC \\
   5BZBJ0823+1125 &     0.441 &                   16.3 &  $2.50^{+1.04}_{-1.05}$ &         8.9/15 &        $2.78^{+2.62}_{-1.94}$ &   $0.82^{+6.70}_{-2.95}$ &         8.8/14 & $0.23^{+1969.77}_{-0.12}$ &  $43.90^{+0.64}_{-0.72}$ &                      0.1 &                     eRASS1 \\
   5BZBJ0301+3441 &     0.246 &                   17.3 &   $2.00^{+1.00}_{-0.99}$ &         9.3/19 &        $1.76^{+1.84}_{-1.76}$ &  $0.82^{+0.81}_{-0.81}$ &         9.2/18 &     $2.30^{+34.00}_{-0.95}$ &  $44.52^{+3.07}_{-0.44}$ &                      0.1 &                      2SXPS \\
   5BZBJ0040$-$2719(2) &     0.172 &                   16.8 &  $2.60^{+0.11}_{-0.11}$ &      201.7/199 &         $2.60^{+0.11}_{-0.11}$ &   $0.08^{+0.40}_{-0.38}$ &      201.6/198 &    $0.78^{+0.08}_{-0.06}$ &  $43.26^{+0.07}_{-0.07}$ &                      0.1 &                      2SXPS \\
   5BZGJ2248$-$0036 &     0.212 &                   16.3 & $2.31^{+0.35}_{-0.35}$ &        69.3/72 &        $2.31^{+0.36}_{-0.36}$ &  $0.18^{+1.59}_{-1.42}$ &        69.3/71 &    $0.35^{+0.39}_{-0.09}$ &   $43.36^{+0.20}_{-0.21}$ &                      0.0 &                      2SXPS \\
   5BZGJ1544+0458(2) &     0.326 &                   16.6 &  $2.26^{+0.28}_{-0.28}$ &        86.8/98 &        $2.25^{+0.30}_{-0.29}$ & $-0.11^{+1.21}_{-1.11}$ &        86.8/97 &    $1.46^{+2.04}_{-0.40}$ &  $44.36^{+0.16}_{-0.16}$ &                      0.0 &                      2SXPS \\
   5BZGJ1510+3335(7) &     0.114 &                   17.1 &  $2.00^{+0.58}_{-0.57}$ &        15.3/26 &         $2.00^{+0.58}_{-0.61}$ &   $0.06^{+2.40}_{-1.91}$ &        15.3/23 &    $1.46^{+2.53}_{-0.56}$ &  $43.41^{+0.35}_{-0.38}$ &                      0.0 &                      2SXPS \\
   5BZGJ1234$-$2432 &     0.182 &                   16.3 & $1.54^{+0.52}_{-0.56}$ &        48.1/62 &         $1.54^{+0.60}_{-0.57}$ & $-0.03^{+3.01}_{-2.48}$ &        48.1/61 &   $1.35^{+21.95}_{-0.82}$ &  $44.00^{+0.35}_{-0.34}$ &                      0.0 &                     eRASS1 \\
   5BZGJ1201$-$0007 &     0.165 &                   16.3 &   $1.98^{+0.20}_{-0.20}$ &       93.2/122 &        $1.97^{+0.22}_{-0.24}$ &  $0.06^{+0.66}_{-0.61}$ &       93.2/121 &    $0.91^{+0.21}_{-0.14}$ &  $43.57^{+0.12}_{-0.12}$ &                      0.0 &                      2SXPS \\
   5BZGJ0932+3630 &     0.154 &                   16.4 & $2.05^{+0.21}_{-0.21}$ &       92.7/103 &        $2.04^{+0.24}_{-0.25}$ &  $0.06^{+0.65}_{-0.59}$ &       92.7/102 &    $1.35^{+0.34}_{-0.22}$ &  $43.64^{+0.13}_{-0.13}$ &                      0.0 &                      2SXPS \\
5BZBJ2240+1326(2) &     0.660 &                   16.6 &   $2.30^{+0.30}_{-0.21}$ &       96.9/132 &         $2.30^{+0.22}_{-0.21}$ & $-0.03^{+0.92}_{-0.87}$ &       96.9/131 &    $2.21^{+1.16}_{-0.48}$ &    $45.31^{+0.10}_{-0.10}$ &                      0.0 &                      2SXPS \\
   5BZBJ2146$-$4748(1) &   0.461* &                   17.3 & $2.23^{+0.32}_{-0.32}$ &        65.3/91 &        $2.22^{+0.53}_{-0.49}$ &  $-0.01^{+1.40}_{-1.26}$ &        65.3/90 &     $1.67^{+4.00}_{-0.51}$ &    $44.78^{+0.20}_{-0.20}$ &                      0.0 &                     eRASS1 \\
   5BZBJ1536+0138 &     0.311 &                   17.2 & $1.68^{+0.31}_{-0.31}$ &        62.0/66 &         $1.70^{+0.38}_{-0.44}$ & $-0.06^{+1.02}_{-0.92}$ &        62.0/65 &     $3.98^{+2.10}_{-1.01}$ &  $44.92^{+0.15}_{-0.09}$ &                      0.0 &                      2SXPS \\
   5BZBJ1406+1236 &   0.450* &                   16.3 & $2.03^{+0.39}_{-0.39}$ &        57.9/67 &        $2.04^{+0.67}_{-0.62}$ &  $0.03^{+1.81}_{-1.63}$ &        57.9/66 &    $0.86^{+6.29}_{-0.35}$ &  $44.53^{+0.25}_{-0.25}$ &                      0.0 &                     eRASS1 \\
   5BZBJ1228$-$0221 &     0.323 &                   17.0 & $2.24^{+0.27}_{-0.27}$ &      101.0/122 &        $2.18^{+0.42}_{-0.39}$ & $-0.21^{+1.17}_{-1.09}$ &      101.0/121 &    $2.07^{+3.45}_{-0.68}$ &  $44.45^{+0.18}_{-0.18}$ &                      0.0 &                     eRASS1 \\
   5BZBJ1208$-$2937(2) &     0.249 &                   16.2 & $1.94^{+0.44}_{-0.45}$ &        47.7/55 &         $1.9^{+0.62}_{.0.56}$ &  $-0.10^{+2.04}_{-1.88}$ &        47.7/54 &   $0.99^{+11.81}_{-0.48}$ &   $44.00^{+0.30}_{-0.29}$ &                      0.0 &                     eRASS1 \\
   5BZBJ1155+0926 &   0.470* &                   16.1 &  $2.30^{+0.68}_{-0.71}$ &        14.3/28 &        $2.24^{+1.59}_{-1.39}$ & $-0.16^{+3.62}_{-2.98}$ &        14.3/27 &  $0.48^{+150.52}_{-0.25}$ &  $44.10^{+0.47}_{-0.46}$ &                      0.0 &                     eRASS1 \\
   5BZBJ0323$-$0108 &     0.392 &                   17.1 & $1.75^{+0.12}_{-0.12}$ &      265.8/256 &        $1.75^{+0.24}_{-0.25}$ & $-0.01^{+0.44}_{-0.42}$ &      265.8/255 &    $2.73^{+0.36}_{-0.28}$ &  $45.00^{+0.05}_{-0.06}$ &                      0.0 &                      2SXPS \\
   5BZGJ1154+0238 &     0.211 &                   16.2 & $2.42^{+0.59}_{-0.59}$ &        29.9/36 &            NaN &      NaN &          NaN &       NaN &  $43.38^{+0.43}_{-0.43}$ &                      NaN &                     eRASS1 \\
   5BZUJ2129+0035 &     0.426 &                   16.9 & $1.42^{+1.04}_{-1.23}$ &         3.9/10 &            NaN &      NaN &          NaN &       NaN &      NaN &                      NaN &                      2SXPS \\
   5BZBJ2146$-$4748(2) &   0.461* &                   17.3 & $1.37^{+0.98}_{-1.16}$ &        15.2/14 &            NaN &      NaN &          NaN &       NaN &      NaN &                      NaN &                      2SXPS \\
   5BZBJ1410+0515 $\dagger$ &     0.544 &                   16.2 &     NaN &          NaN &         $2.31^{+0.80}_{-0.63}$ &  $0.67^{+3.13}_{-2.82}$ &        46.2/34 &    $0.24^{+1.88}_{-0.07}$ &      NaN &                      NaN &                      2SXPS \\
   5BZBJ1004+3752(2) &     0.440 &                   17.1 &     NaN &          NaN &         $2.23^{+0.81}_{-0.80}$ &  $-0.96^{+2.90}_{-2.72}$ &        57.1/54 &   $0.93^{+53.67}_{-0.54}$ &      NaN &                      NaN &                      2SXPS \\
\bottomrule
\\
\caption{Sample of analysed sources and results of the fit with a power law and with a log parabola model, ordered by $\Delta$Cstat. The sources before the first splitting lines are those which have been selected as best fitted with a log parabola model. Sources with $\Delta$Cstat$>$2.7 reported below the splitting line have unphysical $\beta$ best-fit values.\\ For sources for which the entire analysis was possible, the table reports first the redshift and the synchrotron peak frequency; then one can find the results from both fit in terms of parameters ($\Gamma$, $\alpha$ and $\beta$ all reported with the relative errors) and quality of the fit (Cstat and degrees of freedom);  finally are reported the observed flux and the luminosity of the X-ray emission as well as their relative errors and the final $\Delta$Cstat and the catalog from which their data were retrieved.
For some sources only some columns are compiled due to the low quality of the data, which did not allow us to properly complete both fits.
Sources where more than one analysis is present, which are either due to multiple observations by the same telescope or by different telescopes, are reported with a progressive number indicating the number of appearances. \\
 * indicates redshift coming from the 3HSP catalog for sources where it was not reported in the 5BZCAT.\\
 $\dagger$ indicates the sources for which no information concerning the light curve was provided so data from different observations was analysed as a single dataset.\\ 
 $\ddagger$ indicates the source with $\beta$ outside the range $-1<\beta<$1.5 \citep{middei22} included in the analysis due to the significantly hard photon index ($\Gamma \sim 1.2$).
\label{tab:result_xanalysis} }
\end{longtable}
}

\end{landscape}
\twocolumn
\FloatBarrier

\section{Targets SEDs}\label{app:modelling}

\subsection{1ES0229200}
\begin{figure*} 
\begin{minipage}{0.5\textwidth} 
 \centering 
 \includegraphics[width=1\textwidth]{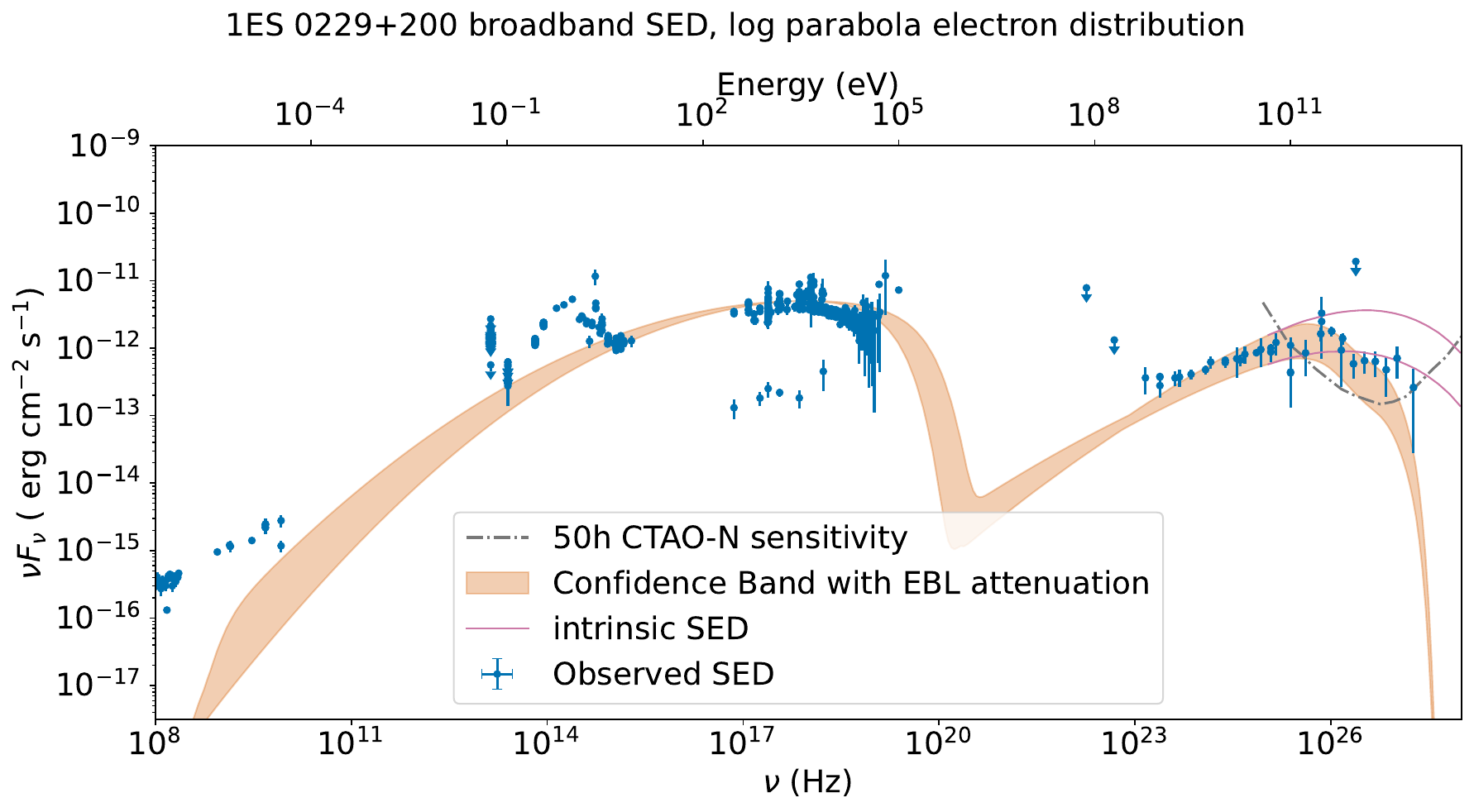} 
\end{minipage}
\begin{minipage}{0.49\textwidth} 
 \centering 
 \includegraphics[width=1\textwidth]{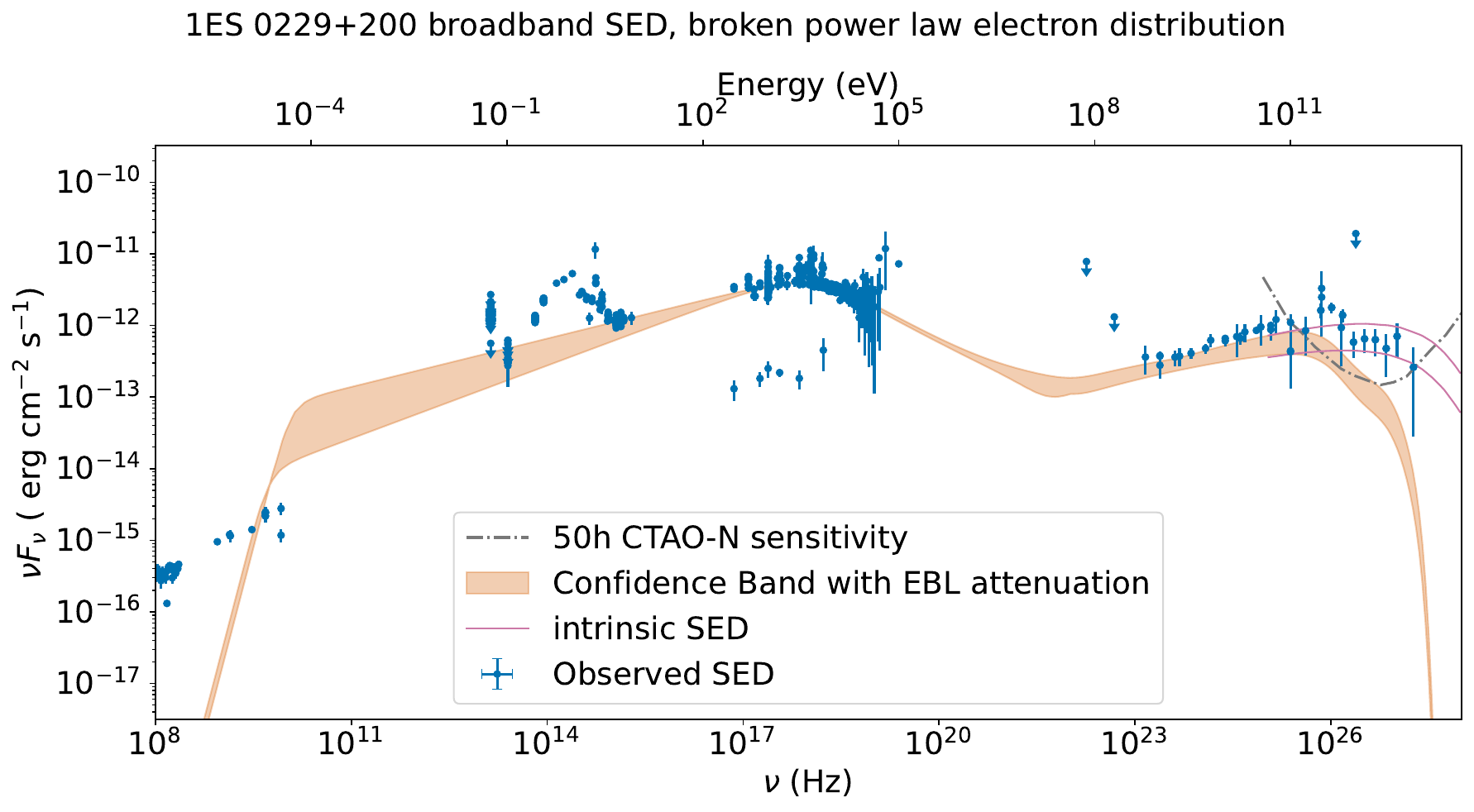} 
\end{minipage}
    \caption{ Broadband spectral energy distribution of 1ES 0229+200 ($z$=0.139, $\gamma_0 \sim 10^5$, $\gamma_{break} \sim 10^6$). The data points in blue are those available from past observations, while the upper limits in green are derived from Fermi data. The regions in orange are the one-zone models assuming on the left a log parabola distribution and on the right a broken power law distribution for the electron population and EBL attenuation according to \citet{dominguez24}. Additionally, pink lines show how the intrinsic models would behave without EBL attenuation, while in grey we show the sensitivity of the CTAO-N array in survey mode for a 50-hour observation at zenith 20$^\circ$.}
    \label{fig:1ESmod}
\end{figure*}

The models for the model source 1ES0229200 are shown in Figure \ref{fig:1ESmod} assuming a log parabola or a broken power law electron distribution.\\ Their behaviour is described in Section \ref{sec:ModDescr}.

\subsection{5BZBJ0333-3619}
\begin{figure*} 
\begin{minipage}{0.5\textwidth} 
 \centering 
 \includegraphics[width=1\textwidth]{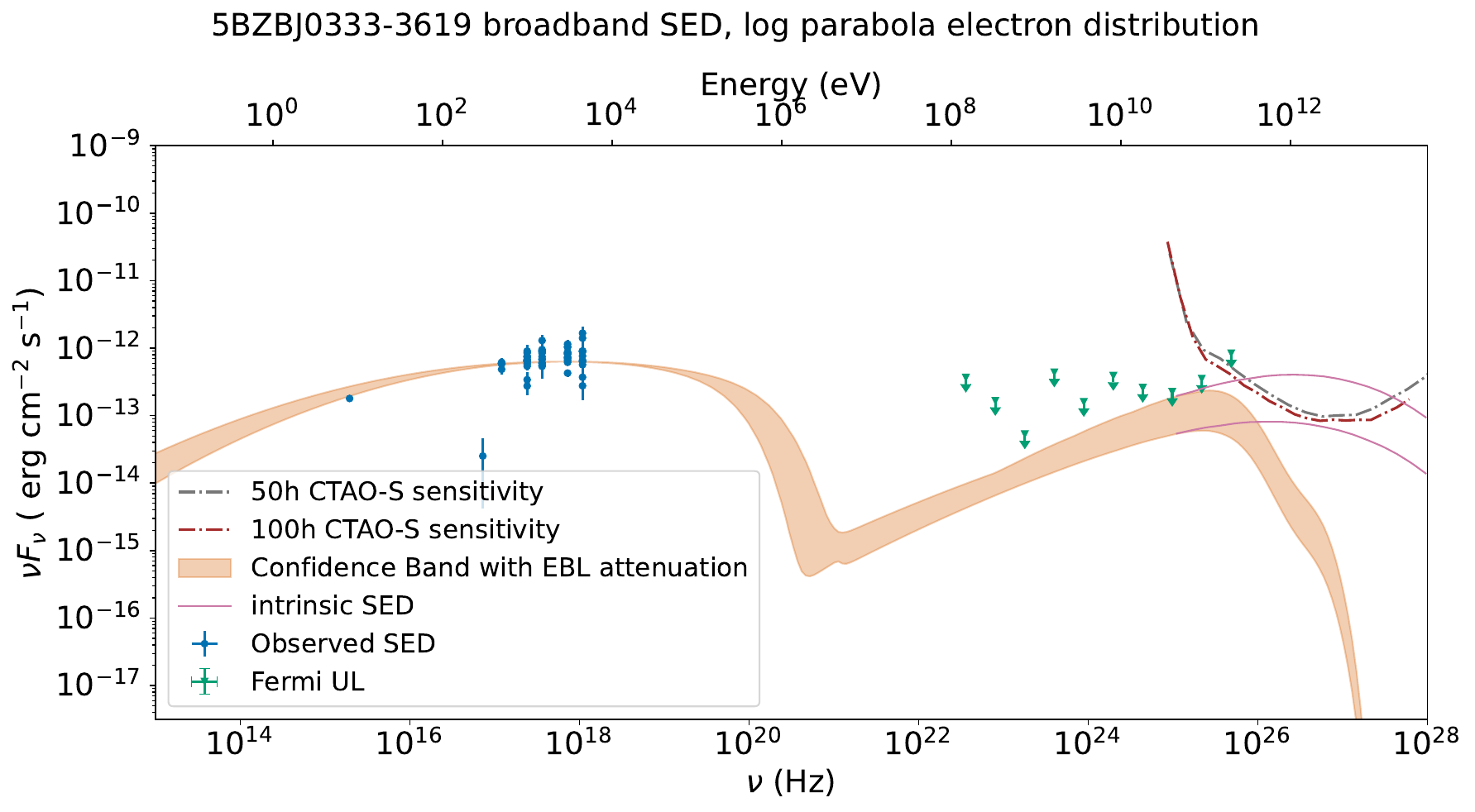} 
\end{minipage}
\begin{minipage}{0.49\textwidth} 
 \centering 
 \includegraphics[width=1\textwidth]{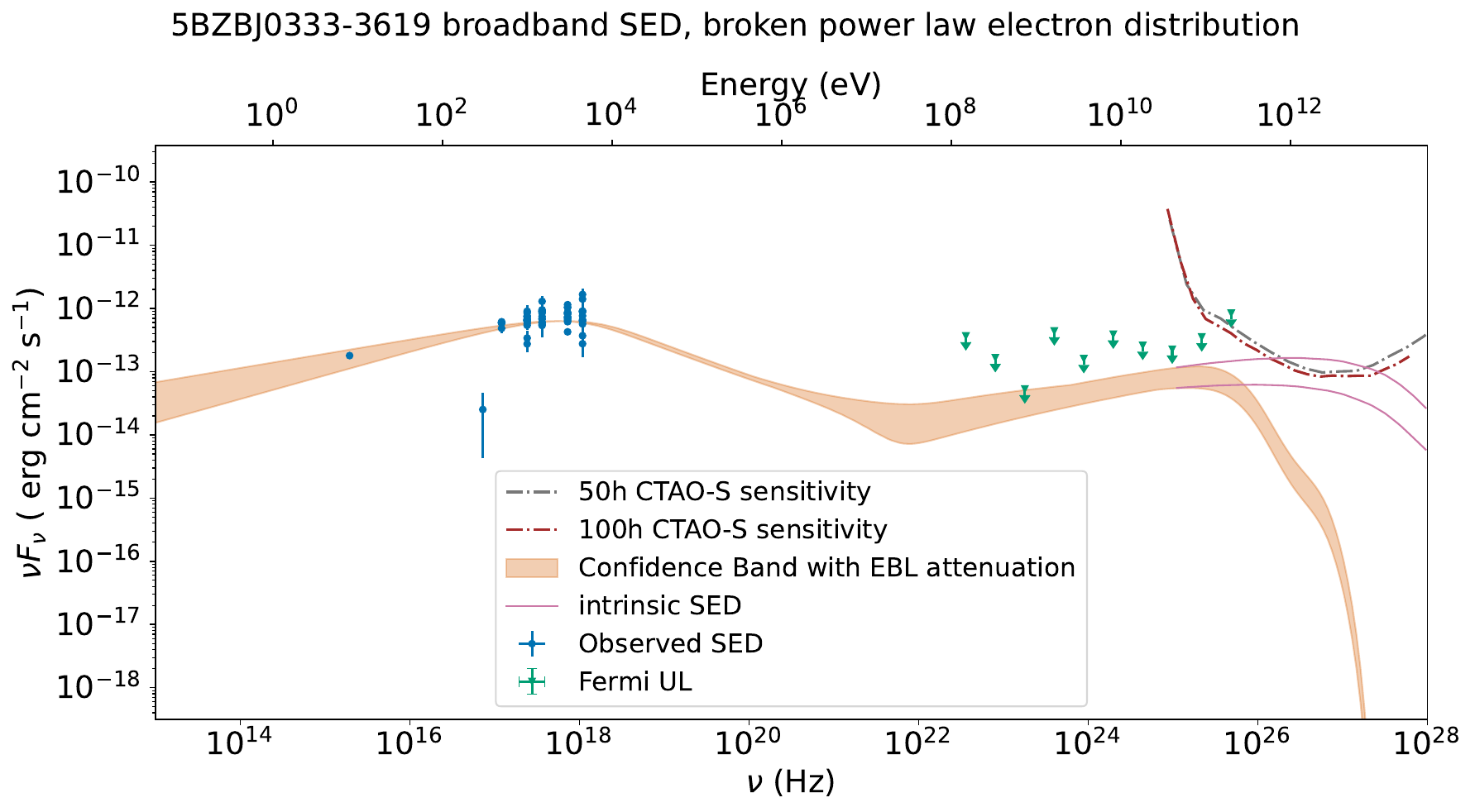} 
\end{minipage}
    \caption{ Broadband spectral energy distribution of 5BZBJ0333-3619 ($z$=0.308, $\gamma_0 \sim 10^5$, $\gamma_{break} \sim 10^6$). The data points in blue are those available from past observations, while the upper limits in green are derived from Fermi data. The regions in orange are the one-zone models assuming on the left a log parabola distribution and on the right a broken power law distribution for the electron population and EBL attenuation according to \citet{dominguez24}. Additionally, pink lines show how the intrinsic models would behave without EBL attenuation, while in grey and red we show respectively the sensitivity of the CTAO-S array in survey mode for a 50-hour and a 100-hour observation at zenith 20$^\circ$.}
    \label{fig:0333mod}
\end{figure*}

The models for the source 5BZBJ0333-3619 are shown in Figure \ref{fig:0333mod} assuming a log parabola or a broken power law electron distribution. 

In both models, the synchrotron peak is well constrained in frequency, the range in observed fluxes being very likely due to variability, since X-ray data points are obtained by multiple \xrt\ observations. The whole region of IC peak modelling is also consistent with the upper limits.  However, EBL attenuation seems to prevent chances of detection by CTAO-S in survey mode, especially if the power law distribution is assumed to be correct. 

\subsection{5BZBJ1253+3826}
\begin{figure*} 
\begin{minipage}{0.5\textwidth} 
 \centering 
 \includegraphics[width=1\textwidth]{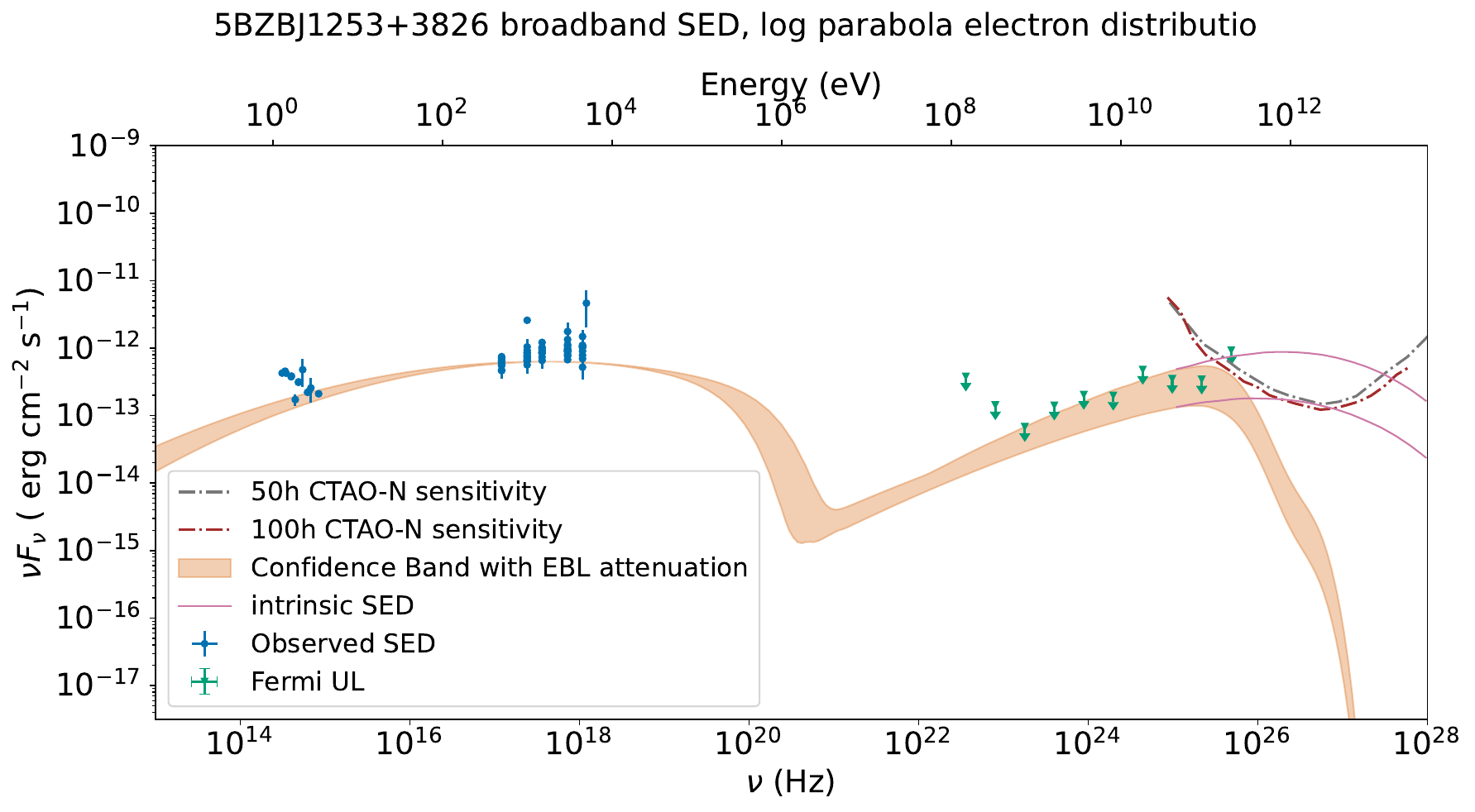} 
\end{minipage}
\begin{minipage}{0.49\textwidth} 
 \centering 
 \includegraphics[width=1\textwidth]{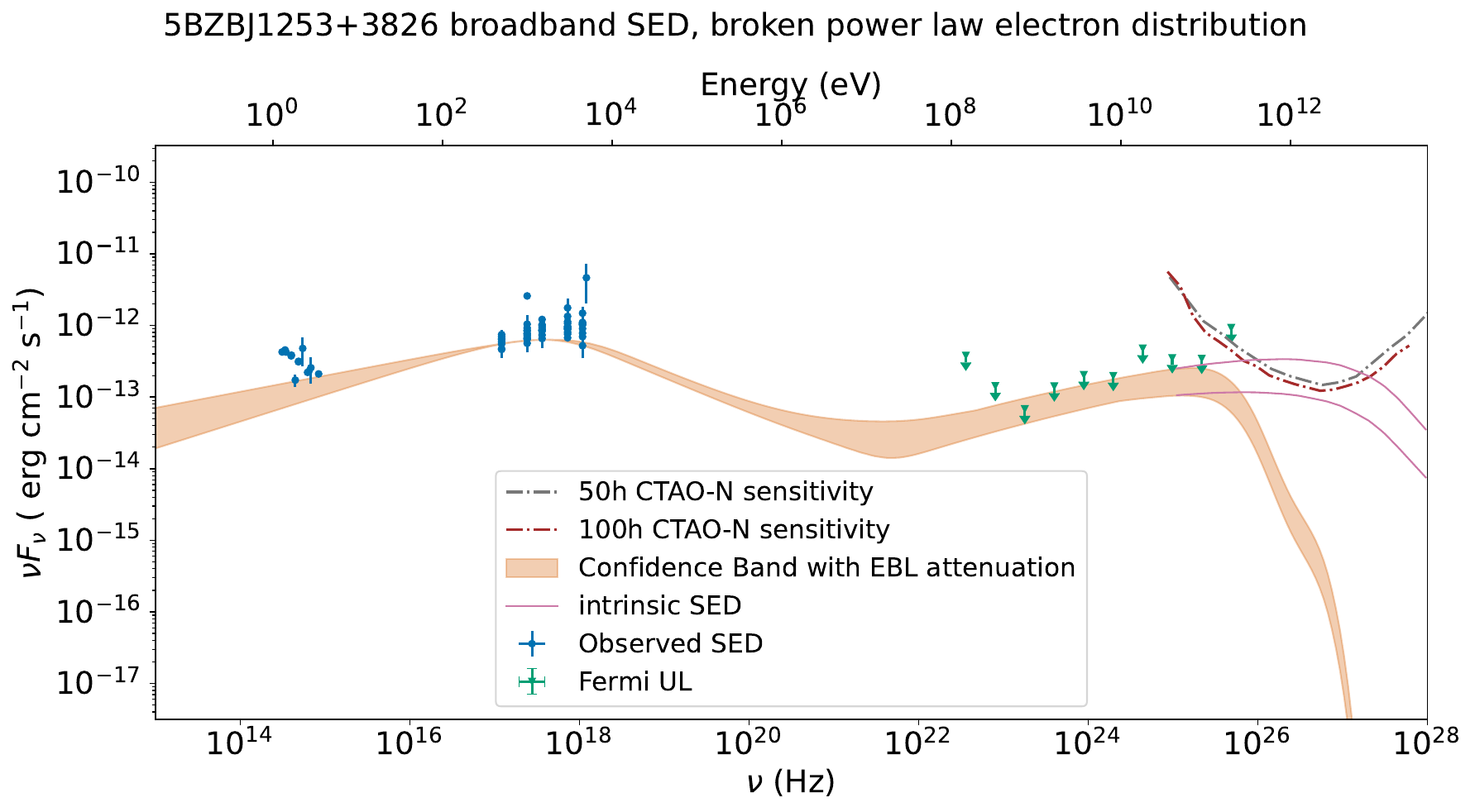} 
\end{minipage}
    \caption{ Broadband spectral energy distribution of 5BZBJ1253+3826 ($z$=0.371, $\gamma_0 \sim 10^5$, $\gamma_{break} \sim 10^6$). The data points in blue are those available from past observations, while the upper limits in green are derived from Fermi data. The regions in orange are the one-zone models assuming on the left a log parabola distribution and on the right a broken power law distribution for the electron population and EBL attenuation according to \citet{dominguez24}. Additionally, pink lines show how the intrinsic models would behave without EBL attenuation, while in grey and red we show respectively the sensitivity of the CTAO-S array in survey mode for a 50-hour and a 100-hour observation at zenith 20$^\circ$.}
    \label{fig:1253mod}
\end{figure*}

The models for the source 5BZBJ1253+3826 are shown in Figure \ref{fig:1253mod} assuming a log parabola or a broken power law electron distribution.

In both models, the synchrotron peak is well constrained in frequency, the range in observed fluxes being very likely due to variability, since X-ray data points are obtained by multiple observations from different instruments. The modelling of the IC peak is also reasonable with respect to the upper limits. In particular, the models that seem the most consistent with the data are those resulting in the lowest values of fluxes in
the range sampled by the different models used in this work. As discussed in Section \ref{sec:SEDparams}, this might suggest that the magnetic field in the jet tends to have the highest values amongst those considered. Based on this modelling, EBL attenuation likely will prevent chances of detection by CTAO-N in survey mode if the power law distribution is assumed to be correct. If the log parabola distribution model is correct, however, the source might be detectable with some specific, long-lasting, pointed observations.

\subsection{5BZBJ1636--1248}
\begin{figure*} 
\begin{minipage}{0.5\textwidth} 
 \centering 
 \includegraphics[width=1\textwidth]{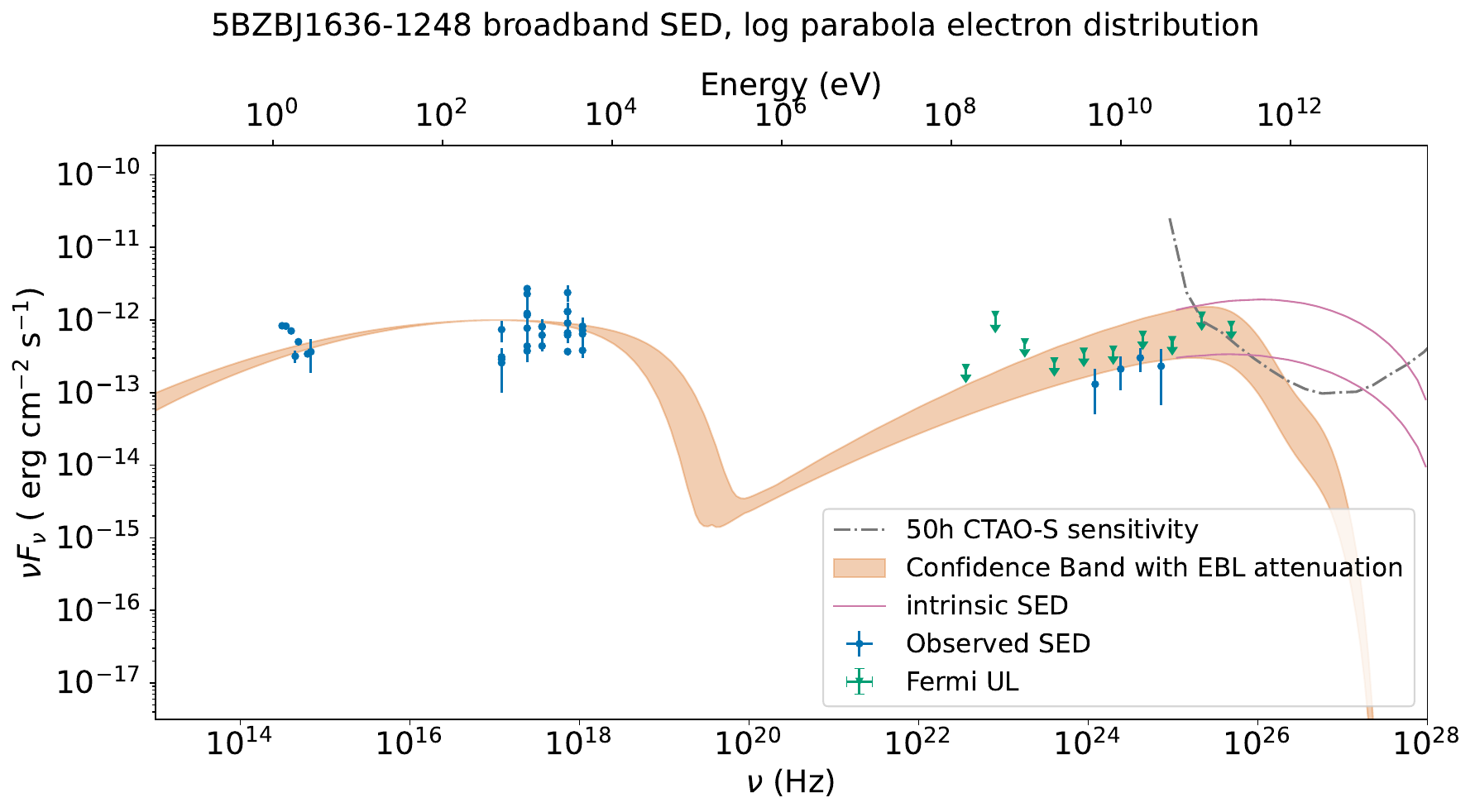} 
\end{minipage}
\begin{minipage}{0.49\textwidth} 
 \centering 
 \includegraphics[width=1\textwidth]{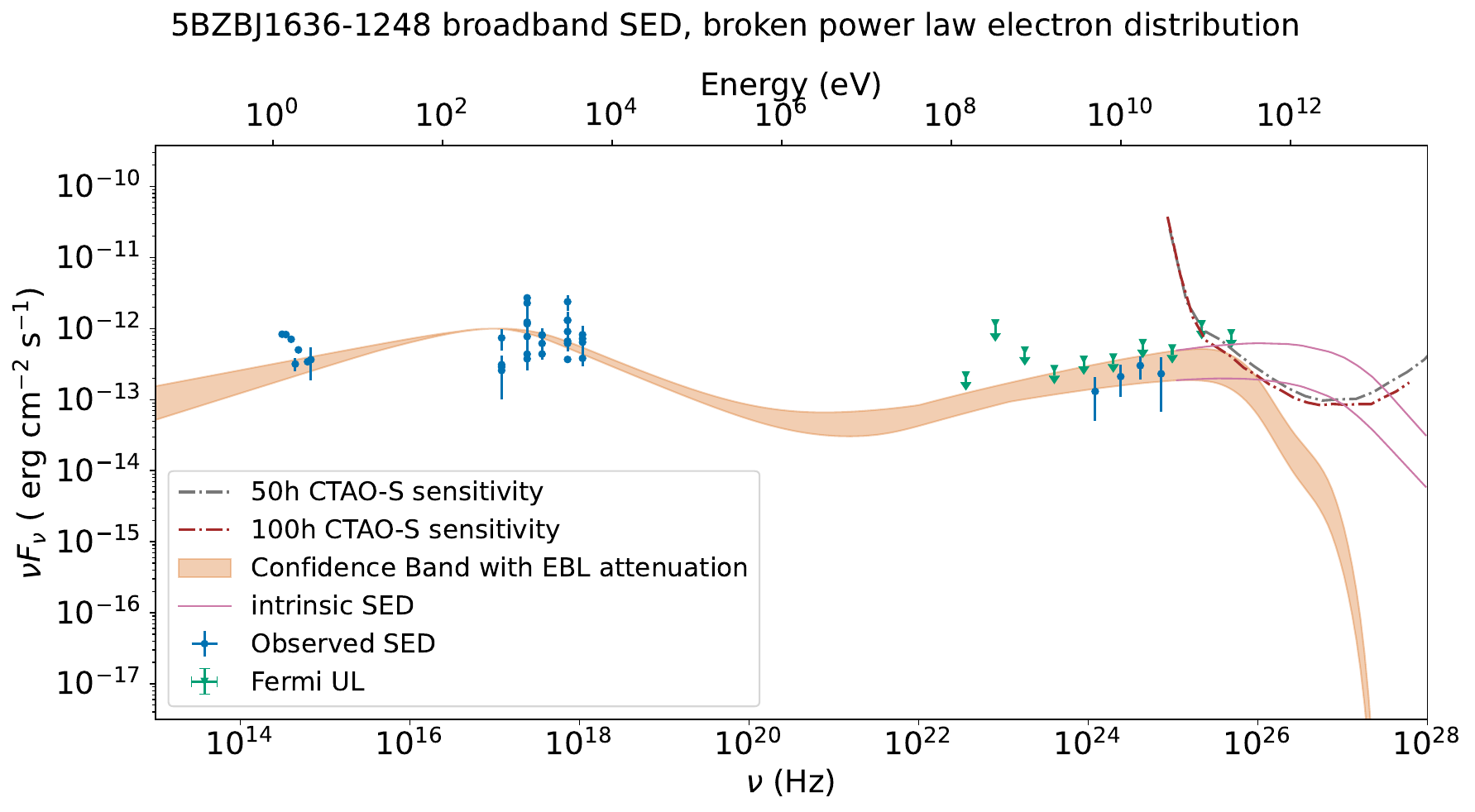} 
\end{minipage}
    \caption{ Broadband spectral energy distribution of 5BZBJ1636--1248 ($z$=0.246, $\gamma_0 \sim 10^5$, $\gamma_{break} \sim 10^6$). The data points in blue are those available from past observations, while the upper limits in green are derived from Fermi data. The regions in orange are the one-zone models assuming on the left a log parabola distribution and on the right a broken power law distribution for the electron population and EBL attenuation according to \citet{dominguez24}. Additionally, pink lines show how the intrinsic models would behave without EBL attenuation, while in grey and red we show respectively the sensitivity of the CTAO-S array in survey mode for a 50-hour and a 100-hour observation at zenith 20$^\circ$.}
    \label{fig:1636mod}
\end{figure*}

The models for the source 5BZBJ1636--1248 are shown in Figure \ref{fig:1636mod} assuming a log parabola or a broken power law electron distribution.

In both models, the synchrotron peak seems well constrained in frequency, the range in observed fluxes being very likely due to variability, since X-ray data points are obtained by multiple observations from different instruments. This object also presented datapoints in the VHE band, thanks to the BIGB catalog by \textit{Fermi} (\cite{arsioli20}). Such data points are consistent with both the $\gamma$-ray upper limits we compute in this work and our models, although lying in the lower range of the flux spectrum. Still, the source seems to have real chances of detection by CTAO-N if the log parabola model is correct and, although slightly less likely, even with the broken power law model.

\subsection{5BZBJ1251-2958}
\begin{figure*} 
\begin{minipage}{0.5\textwidth} 
 \centering 
 \includegraphics[width=1\textwidth]{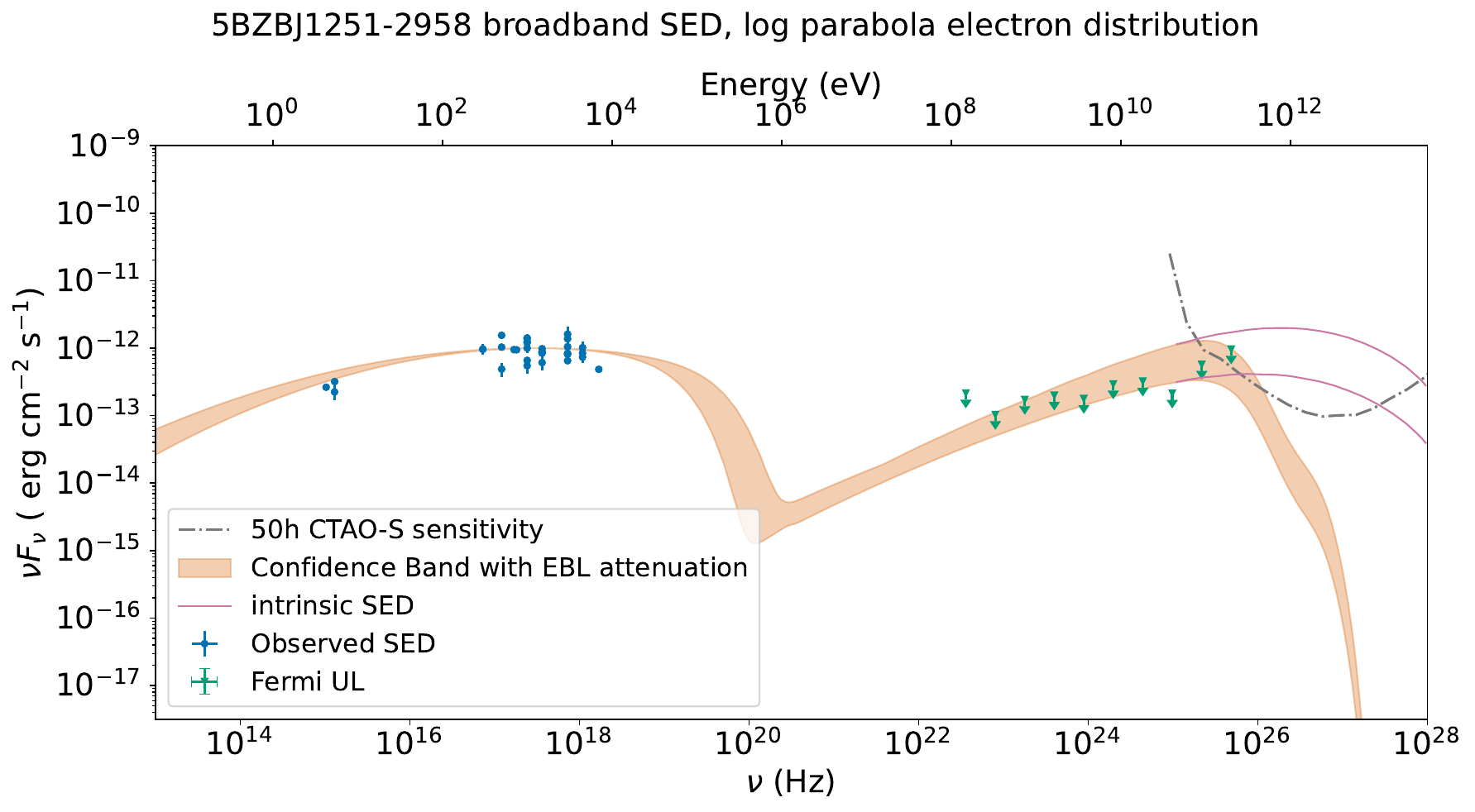} 
\end{minipage}
\begin{minipage}{0.49\textwidth} 
 \centering 
 \includegraphics[width=1\textwidth]{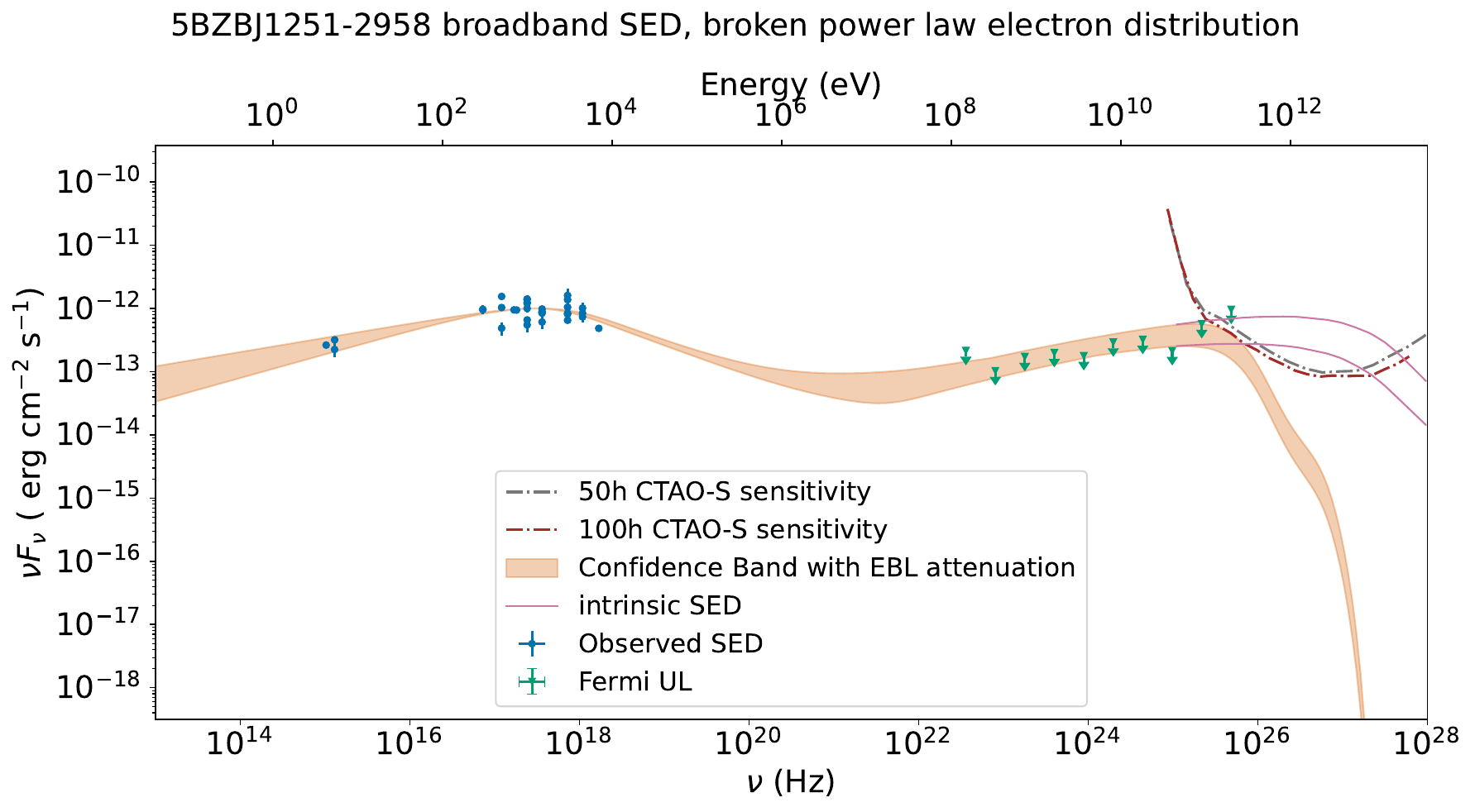} 
\end{minipage}
    \caption{ Broadband spectral energy distribution of 5BZBJ1251-2958 ($z$=0.389, $\gamma_0 \sim 10^5$, $\gamma_{break} \sim 10^6$). The data points in blue are those available from past observations, while the upper limits in green are derived from Fermi data. The regions in orange are the one-zone models assuming on the left a log parabola distribution and on the right a broken power law distribution for the electron population and EBL attenuation according to \citet{dominguez24}. Additionally, pink lines show how the intrinsic models would behave without EBL attenuation, while in grey and red we show respectively the sensitivity of the CTAO-S array in survey mode for a 50-hour and a 100-hour observation at zenith 20$^\circ$.}
    \label{fig:1251mod}
\end{figure*}

The models for the source 5BZBJ1251-2958 are shown in Figure \ref{fig:1251mod} assuming a log parabola or a power law electron distribution.

In both models, the synchrotron peak is well constrained in frequency, the range in observed fluxes being very likely due to variability, since X-ray data points are obtained by multiple observations obtained with different instruments. The $\gamma$-ray upper limits constrain the IC peak so that models with lower VHE fluxes and, therefore, with
likely higher magnetic fields are favoured. Only one upper limit seems to suggest fluxes lowers than those expected from the models. However, being this lower upper limit an isolated case, its behaviour could reasonably be considered an effect of Poissonian noise. Therefore, considering the log parabola electron distribution to be correct, there are real chances of detection with detection by CTAO-N. On the other hand, assuming the model with power law electron distribution to be correct, the source could only be detected with longer-lasting, pointed observations. 

\subsection{5BZBJ0040-2719}
\begin{figure*} 
\begin{minipage}{0.5\textwidth} 
 \centering 
 \includegraphics[width=1\textwidth]{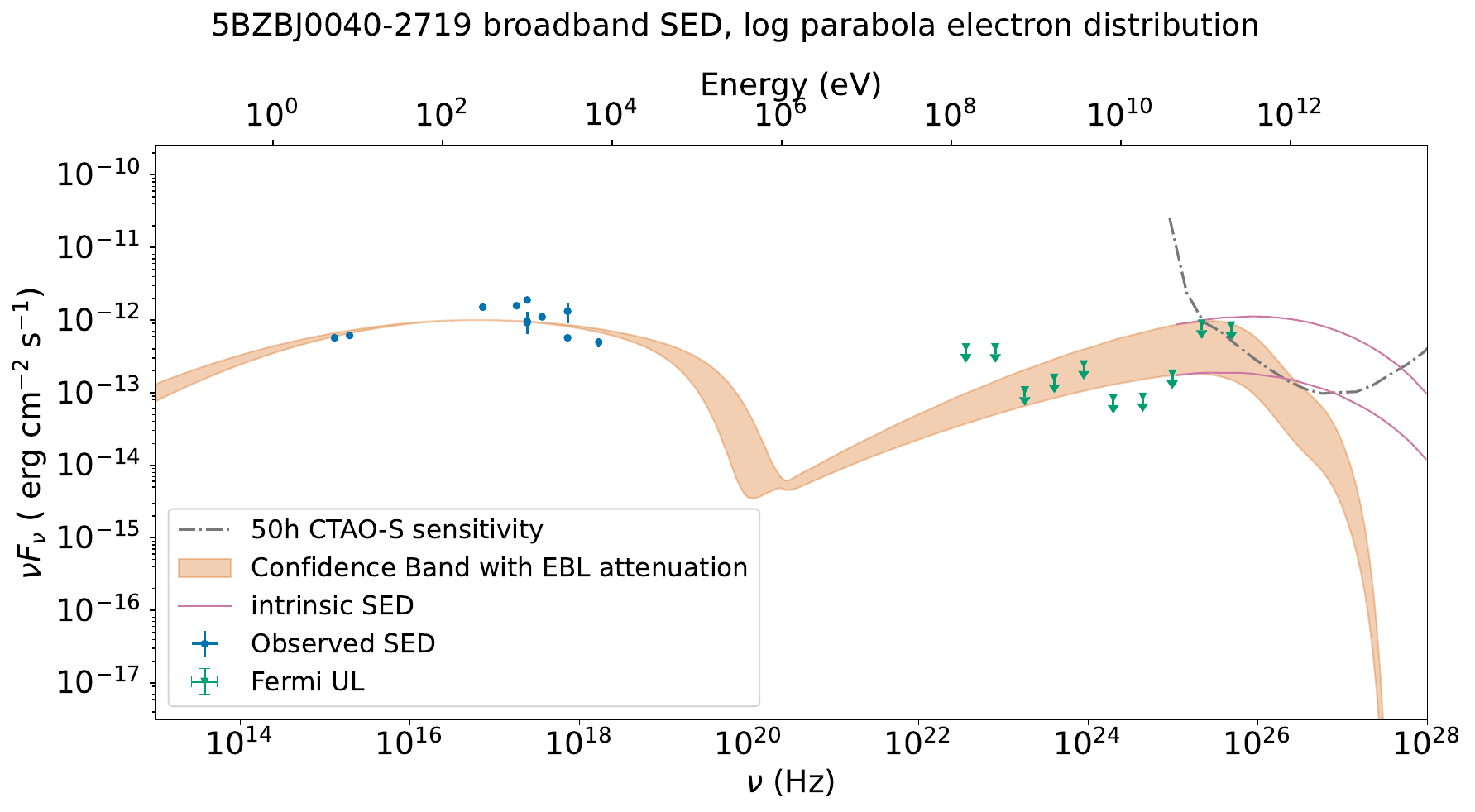} 
\end{minipage}
\begin{minipage}{0.49\textwidth} 
 \centering 
 \includegraphics[width=1\textwidth]{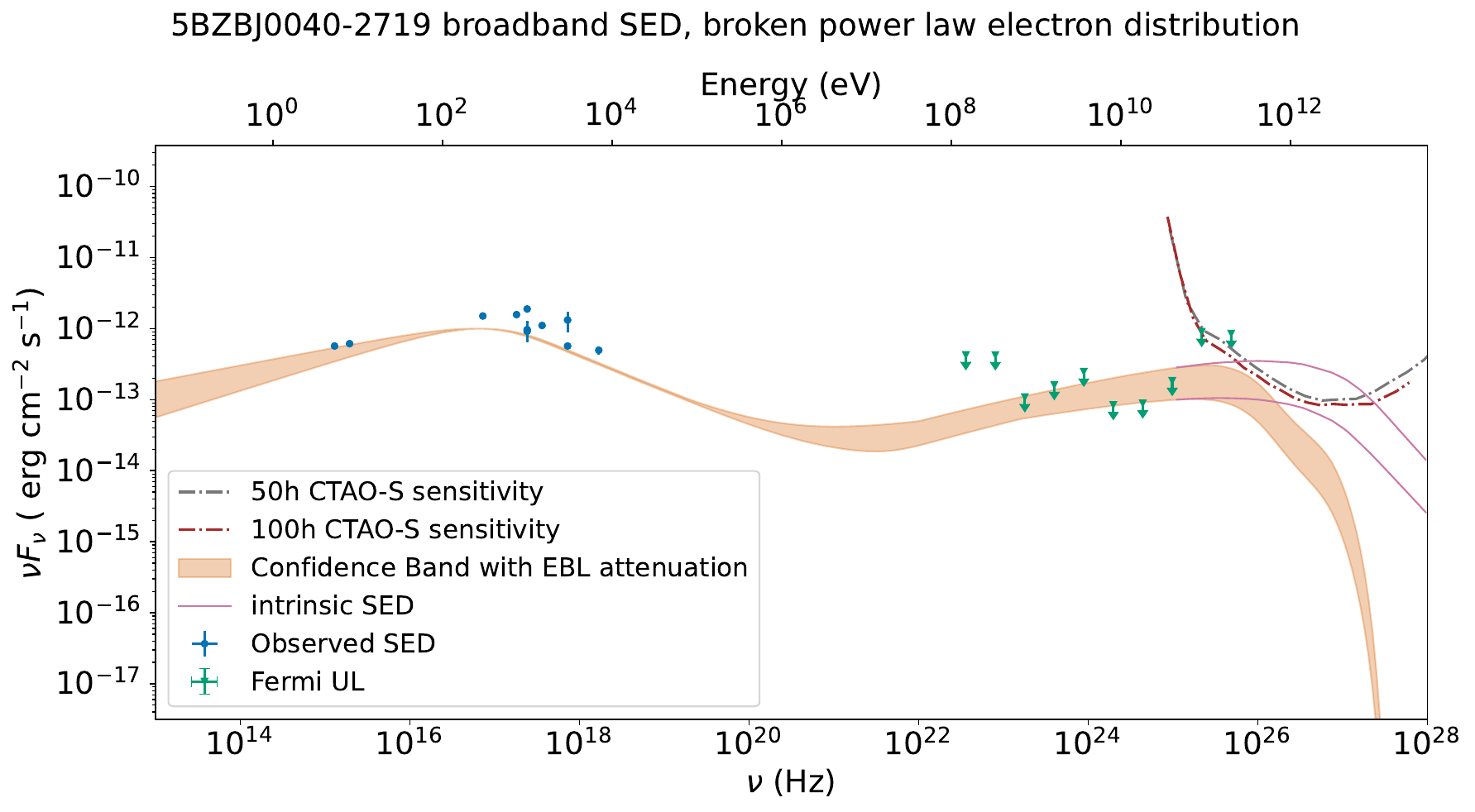} 
\end{minipage}
    \caption{ Broadband spectral energy distribution of 5BZBJ0040-2719 ($z$=0.172, $\gamma_0 \sim 10^5$, $\gamma_{break} \sim 10^5$). The data points in blue are those available from past observations, while the upper limits in green are derived from Fermi data. The regions in orange are the one-zone models assuming on the left a log parabola distribution and on the right a broken power law distribution for the electron population and EBL attenuation according to \citet{dominguez24}. Additionally, pink lines show how the intrinsic models would behave without EBL attenuation, while in grey and red we show respectively the sensitivity of the CTAO-S array in survey mode for a 50-hour and a 100-hour observation at zenith 20$^\circ$.}
    \label{fig:0040mod}
\end{figure*}

The models for the source 5BZBJ0040-2719 are shown in Figure \ref{fig:0040mod} assuming a log parabola or a broken power law electron distribution.

In both models, the synchrotron peak is well constrained in frequency, with the data following slightly better the expected behaviour from a log parabola electron distribution model. Some problems arise in the IC peak, where the expected model actually predicts fluxes above some of the Fermi upper limits in the model with the log parabola electron distribution. However, the pattern of the upper limits is unstable, and can be explained with the limited overall Fermi data quality in this region of the sky, as well as with statistical fluctuations. If this is the case, the log parabola model would suggest that the source is detectable by CTAO-S, even when taking into account the EBL attenuation. If instead we strictly adhere to the $\gamma$-ray upper limits, the data is more compatible with the broken power law electron distribution model in the VHE regime. The upper limits favour models reaching
the lowest fluxes, again suggesting a higher intensity of the magnetic field. If this model is correct, however, this source is unlikely to be detected by CTAO-S in survey mode. Some better results could be achieved with long-lasting, pointed observations.

\subsection{5BZBJ1302+5056}
\begin{figure*} 
\begin{minipage}{0.5\textwidth} 
 \centering 
 \includegraphics[width=1\textwidth]{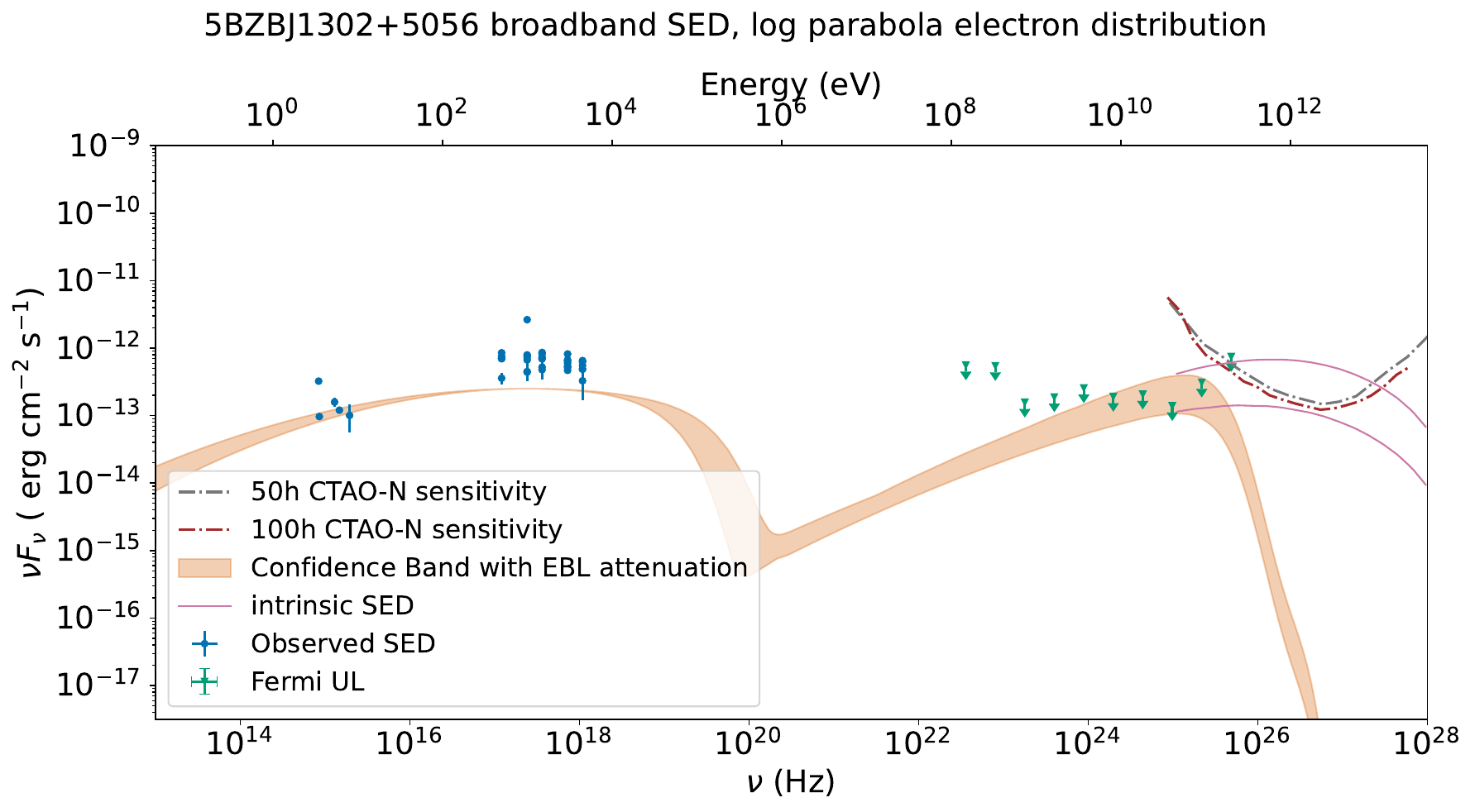} 
\end{minipage}
\begin{minipage}{0.49\textwidth} 
 \centering 
 \includegraphics[width=1\textwidth]{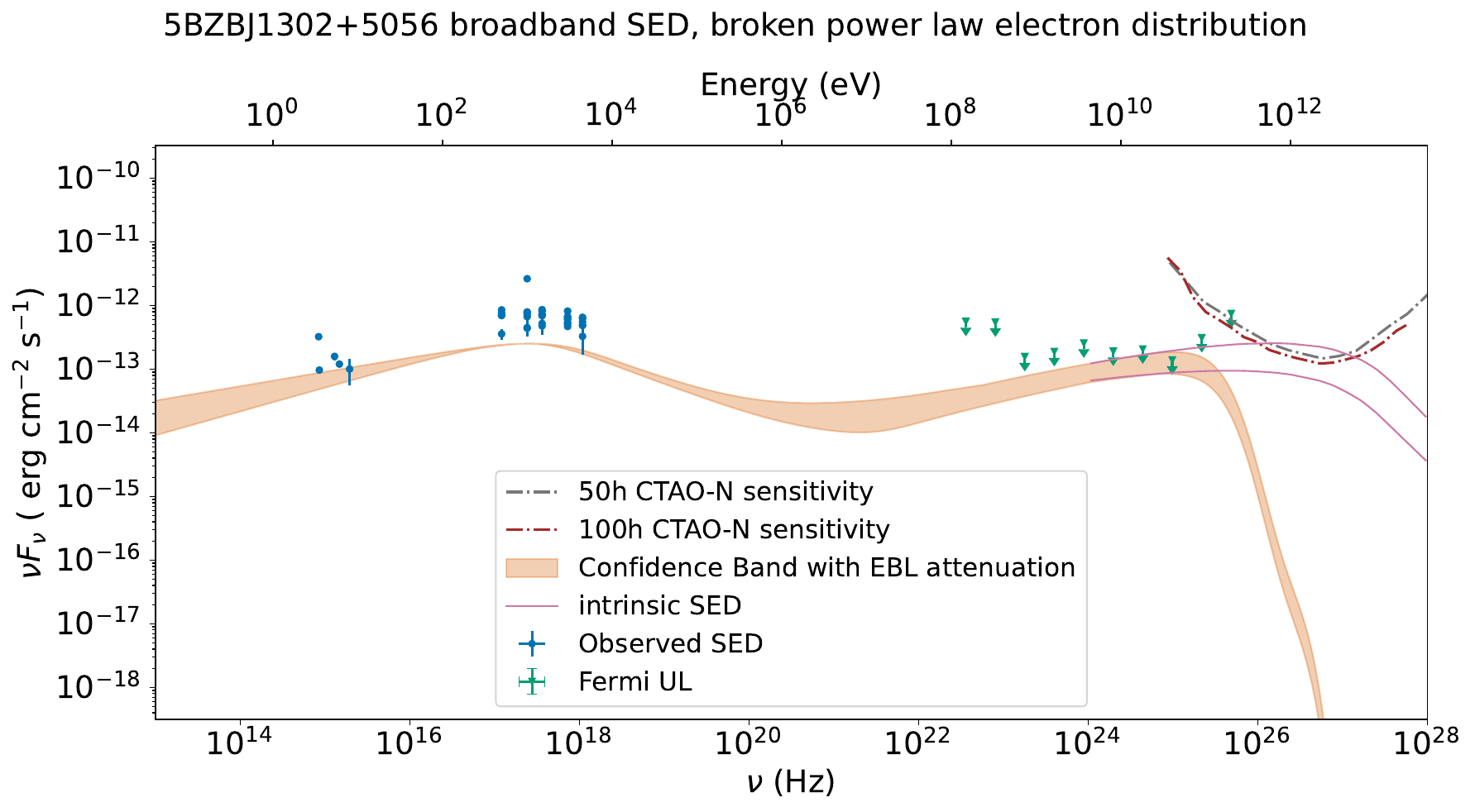} 
\end{minipage}
    \caption{ Broadband spectral energy distribution of 5BZBJ1302+5056 ($z$=0.688, $\gamma_0 \sim 10^5$, $\gamma_{break} \sim 10^6$). The data points in blue are those available from past observations, while the upper limits in green are derived from Fermi data. The regions in orange are the one-zone models assuming on the left a log parabola distribution and on the right a broken power law distribution for the electron population and EBL attenuation according to \citet{dominguez24}. Additionally, pink lines show how the intrinsic models would behave without EBL attenuation, while in grey and red we show respectively the sensitivity of the CTAO-S array in survey mode for a 50-hour and a 100-hour observation at zenith 20$^\circ$.}
    \label{fig:1302mod}
\end{figure*}

The models for the source 5BZBJ1302+5056 are shown in Figure \ref{fig:1302mod} assuming a log parabola or a broken power law electron distribution.
In both cases, the synchrotron peak flux derived by BLAST seems to be slightly underestimate with respect to the data. Nevertheless, the modelling of the IC peak seems to still be reasonable with respect to the upper limits, assuming either a log parabola or a broken power law electron distribution, with only one of the upper limits limiting the reasonable models to those with the lowest fluxes in our range. However, EBL attenuation would likely prevent chances of detection by CTAO-N in survey mode, especially if the power law distribution is assumed to be correct. 

\subsection{5BZBJ1057+2303}
\begin{figure*} 
\begin{minipage}{0.5\textwidth} 
 \centering 
 \includegraphics[ width=1\textwidth]{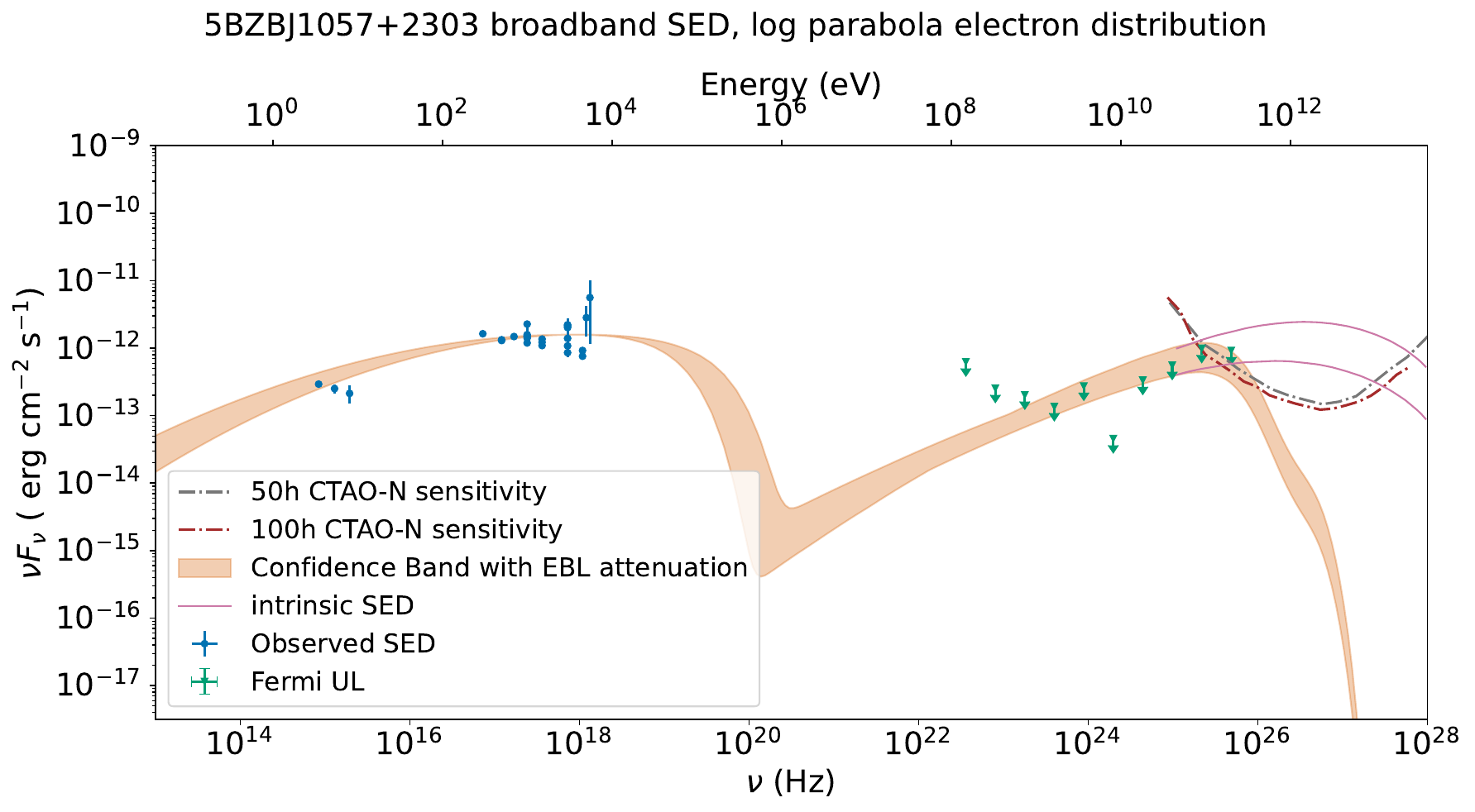} 
\end{minipage}
\begin{minipage}{0.49\textwidth} 
 \centering 
 \includegraphics[width=1\textwidth]{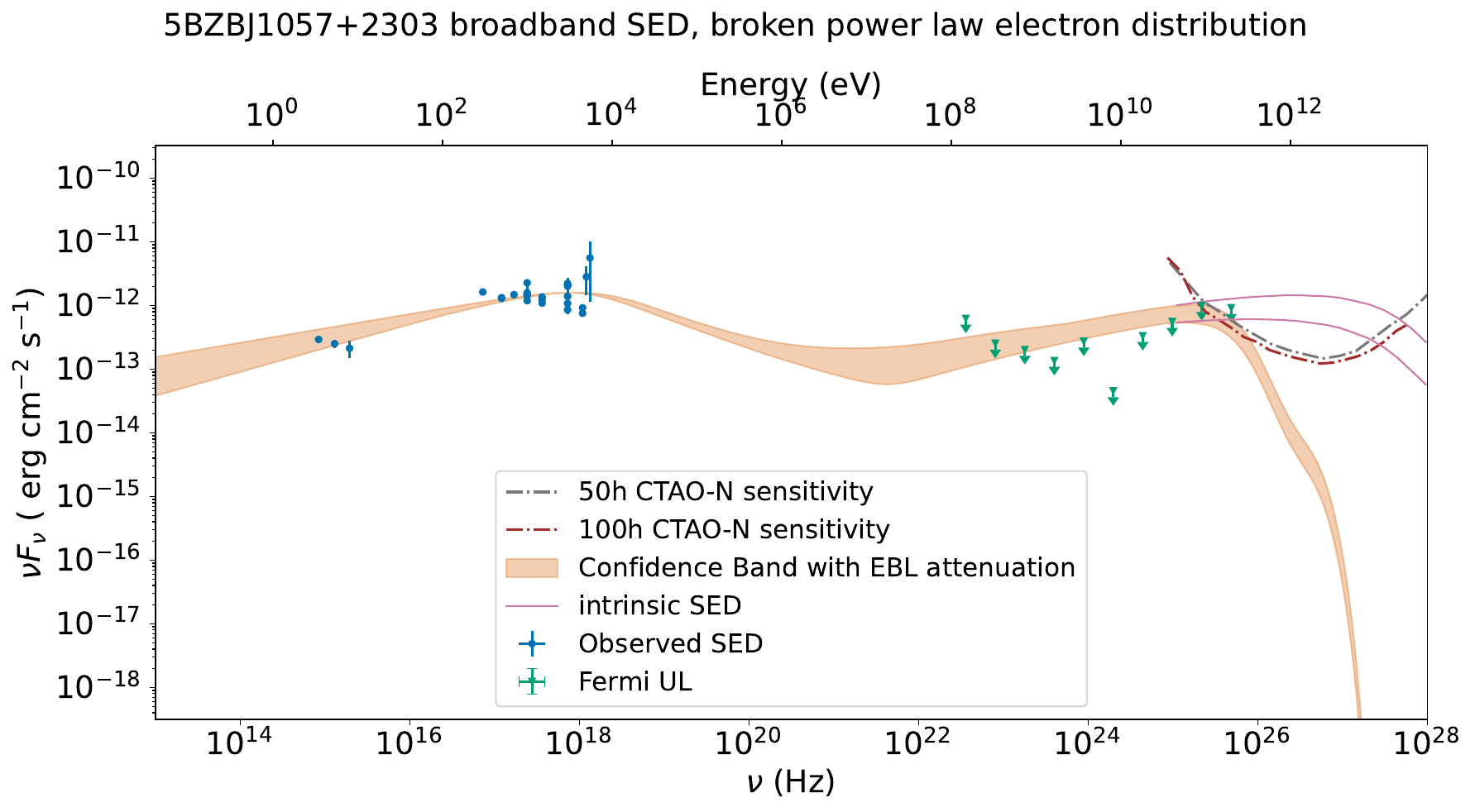} 
\end{minipage}
    \caption{ Broadband spectral energy distribution of 5BZBJ1057+2303 ($z$=0.379, $\gamma_0 \sim 10^5$, $\gamma_{break} \sim 10^6$). The data points in blue are those available from past observations, while the upper limits in green are derived from Fermi data. The regions in orange are the one-zone models assuming on the left a log parabola distribution and on the right a broken power law distribution for the electron population and EBL attenuation according to \citet{dominguez24}. Additionally, pink lines show how the intrinsic models would behave without EBL attenuation, while in grey and red we show respectively the sensitivity of the CTAO-S array in survey mode for a 50-hour and a 100-hour observation at zenith 20$^\circ$.}
    \label{fig:1057mod}
\end{figure*}

The models for the source 5BZBJ1057+2303 are shown in Figure \ref{fig:1057mod} assuming a log parabola or a broken power law electron distribution.

In both models, the synchrotron peak is well constrained in frequency, the range in observed fluxes being very likely due to variability, since X-ray data points are obtained by multiple observations from different instruments. 
In the log parabola electron distribution model, the majority of the upper limits are consistent with the models leading to lower fluxes. The one upper limit at considerably lower fluxes, being an isolated case, could reasonably be considered an effect of Poissonian noise. After accounting for the EBL, the log parabola models still is consistent with a potential detection by CTAO-N. On the other hand, the broken power law model seems to be inconsistent with most $\gamma$-ray upper limits. Nevertheless, if the model is correct, a detection with CTAO-N could be feasible. 

\subsection{5BZBJ2217--3106}
\begin{figure*} 
\begin{minipage}{0.5\textwidth} 
 \centering 
 \includegraphics[width=1\textwidth]{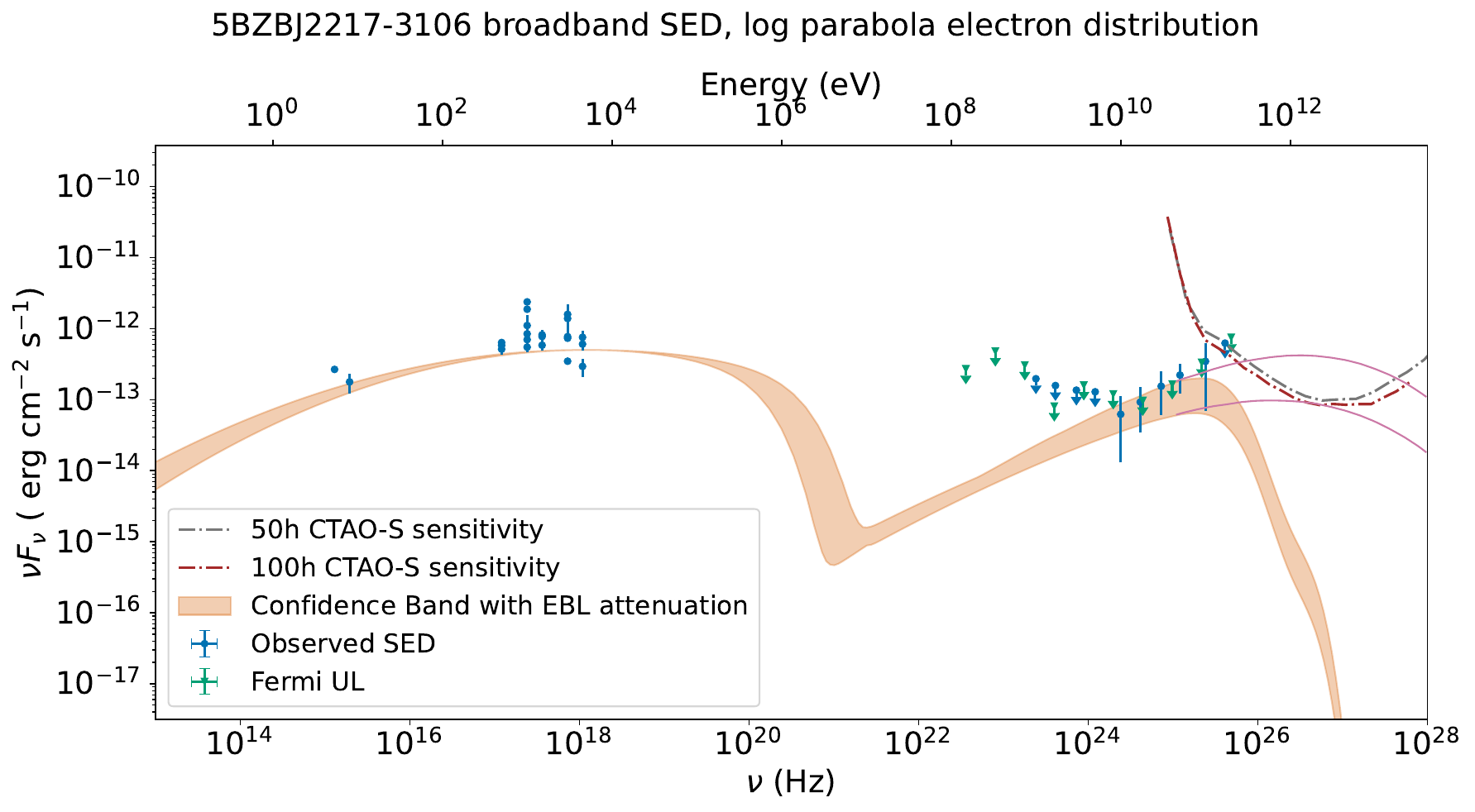} 
\end{minipage}
\begin{minipage}{0.49\textwidth} 
 \centering 
 \includegraphics[width=1\textwidth]{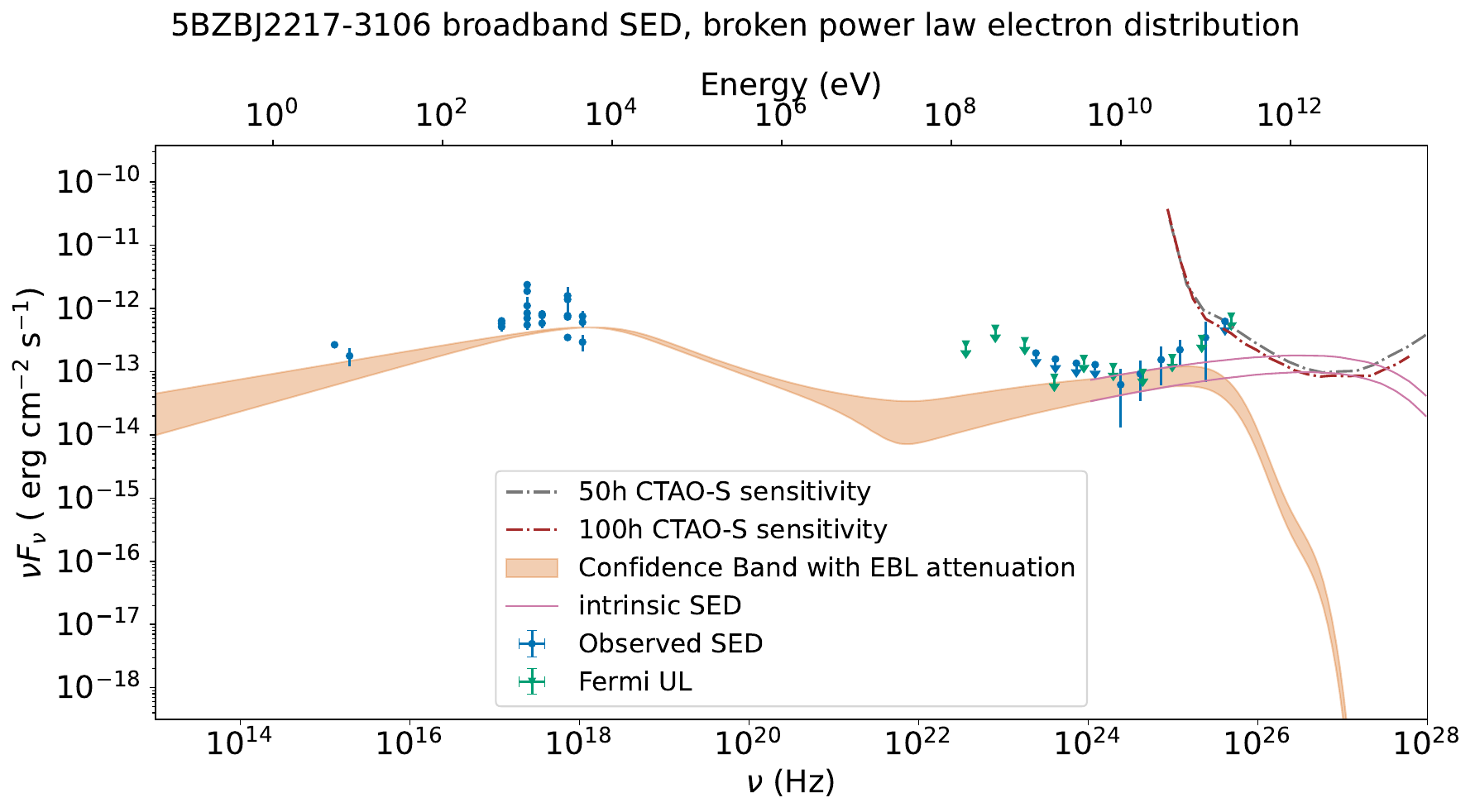} 
\end{minipage}
    \caption{ Broadband spectral energy distribution of 5BZBJ2217-3106 ($z$=0.460, $\gamma_0 \sim 10^5$, $\gamma_{break} \sim 10^6$). The data points in blue are those available from past observations, while the upper limits in green are derived from Fermi data. The regions in orange are the one-zone models assuming on the left a log parabola distribution and on the right a broken power law distribution for the electron population and EBL attenuation according to \citet{dominguez24}. Additionally, pink lines show how the intrinsic models would behave without EBL attenuation, while in grey and red we show respectively the sensitivity of the CTAO-S array in survey mode for a 50-hour and a 100-hour observation at zenith 20$^\circ$.}
    \label{fig:2217mod}
\end{figure*}

The models for the source 5BZBJ2217-3106 are shown in Figure \ref{fig:2217mod} assuming a log parabola or a broken power law electron distribution.

In both cases, the synchrotron peak is well constrained in frequency, the range in observed fluxes being very likely due to variability, since X-ray data points are obtained by multiple observations from different instruments. This is the second of the two objects in the sample where data points were also available in the VHE band, thanks to the BIGB catalog by \textit{Fermi} (\cite{arsioli20}). Such data points are consistent with both the $\gamma$-ray upper limits we compute in this work, as well as with our models. However, the EBL attenuation seems to prevent chances of detection by CTAO-S for both models.

\subsection{5BZBJ0124+0918}
\begin{figure*} 
\begin{minipage}{0.5\textwidth} 
 \centering 
 \includegraphics[width=1\textwidth]{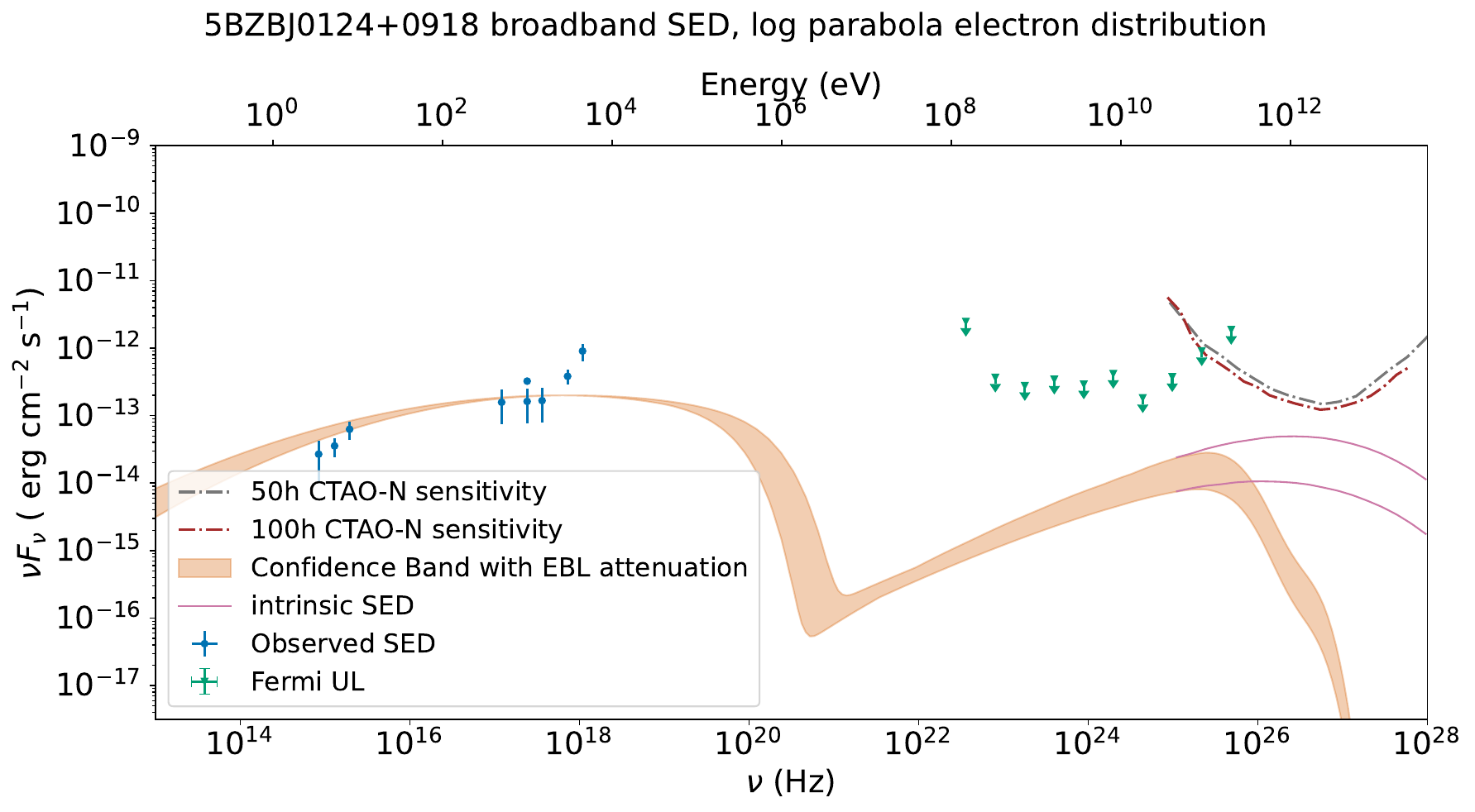} 
\end{minipage}
\begin{minipage}{0.49\textwidth} 
 \centering 
 \includegraphics[width=1\textwidth]{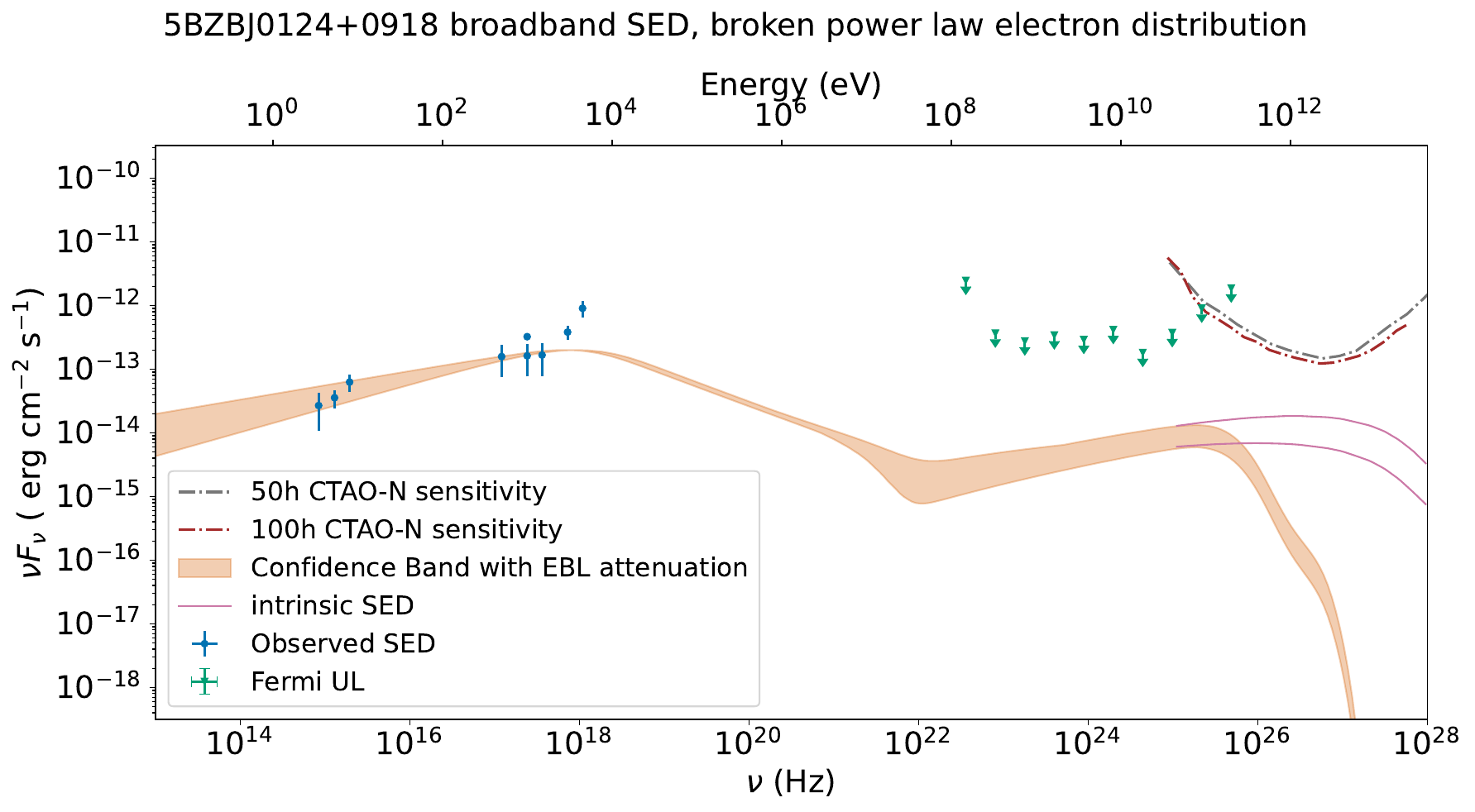} 
\end{minipage}
    \caption{ Broadband spectral energy distribution of 5BZBJ0124+0918 ($z$=0.338, $\gamma_0 \sim 10^5$, $\gamma_{break} \sim 10^6$). The data points in blue are those available from past observations, while the upper limits in green are derived from Fermi data. The regions in orange are the one-zone models assuming on the left a log parabola distribution and on the right a broken power law distribution for the electron population and EBL attenuation according to \citet{dominguez24}. Additionally, pink lines show how the intrinsic models would behave without EBL attenuation, while in grey and red we show respectively the sensitivity of the CTAO-S array in survey mode for a 50-hour and a 100-hour observation at zenith 20$^\circ$.}
    \label{fig:0124mod}
\end{figure*}

The models for the source 5BZBJ0124+0918 are shown in Figure \ref{fig:0124mod} assuming a log parabola or a broken power law electron distribution.

In both cases, assuming a correct extrapolation of the synchrotron peak by BLAST, the model suggests flux values of the IC peak that are compatible with the derived upper limits, but considerably lower than other targets and in any case too low to be detected by CTAO-N. 

\subsection{5BZBJ1258+0134}
\begin{figure*} 
\begin{minipage}{0.5\textwidth} 
 \centering 
 \includegraphics[width=1\textwidth]{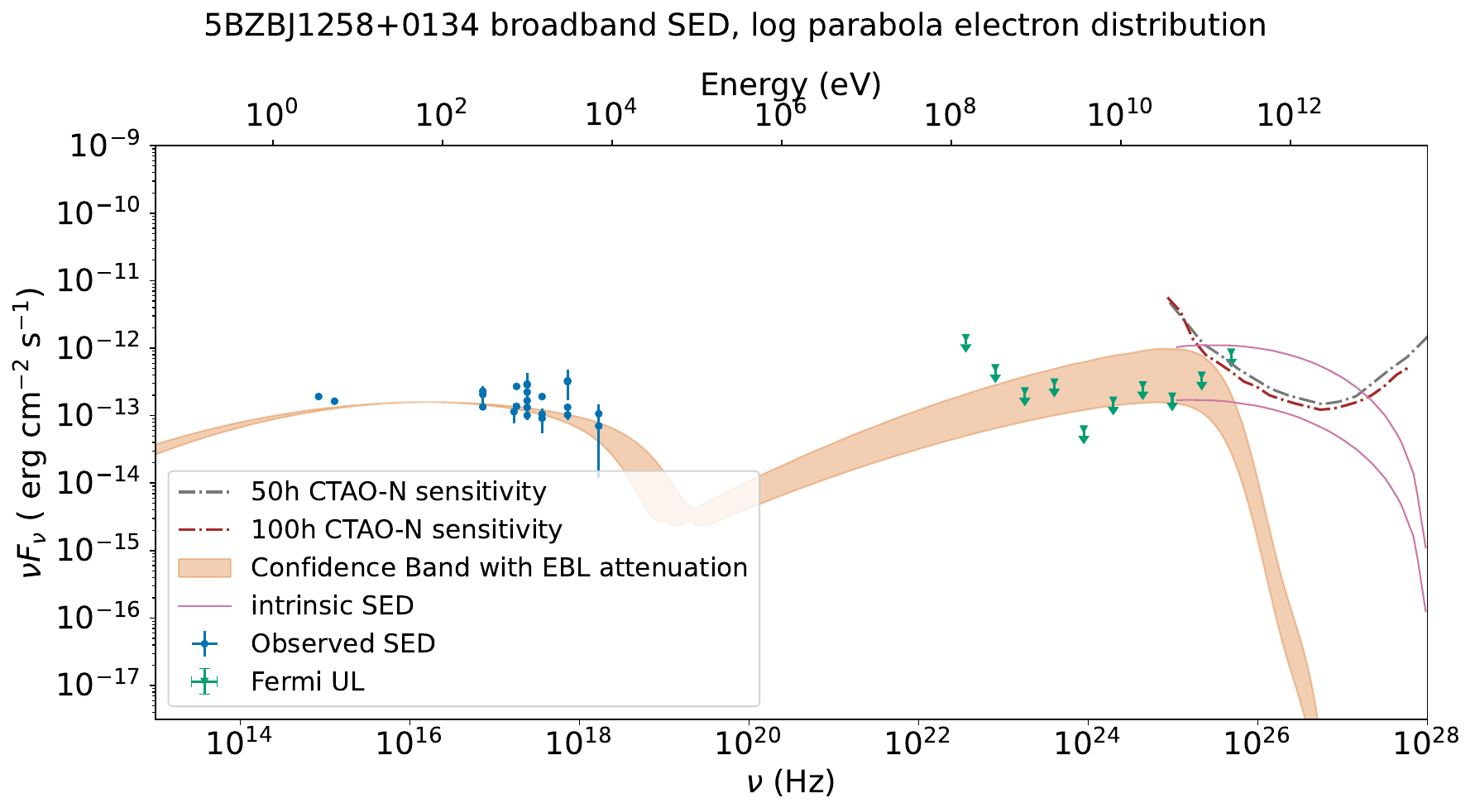} 
\end{minipage}
\begin{minipage}{0.49\textwidth} 
 \centering 
 \includegraphics[width=1\textwidth]{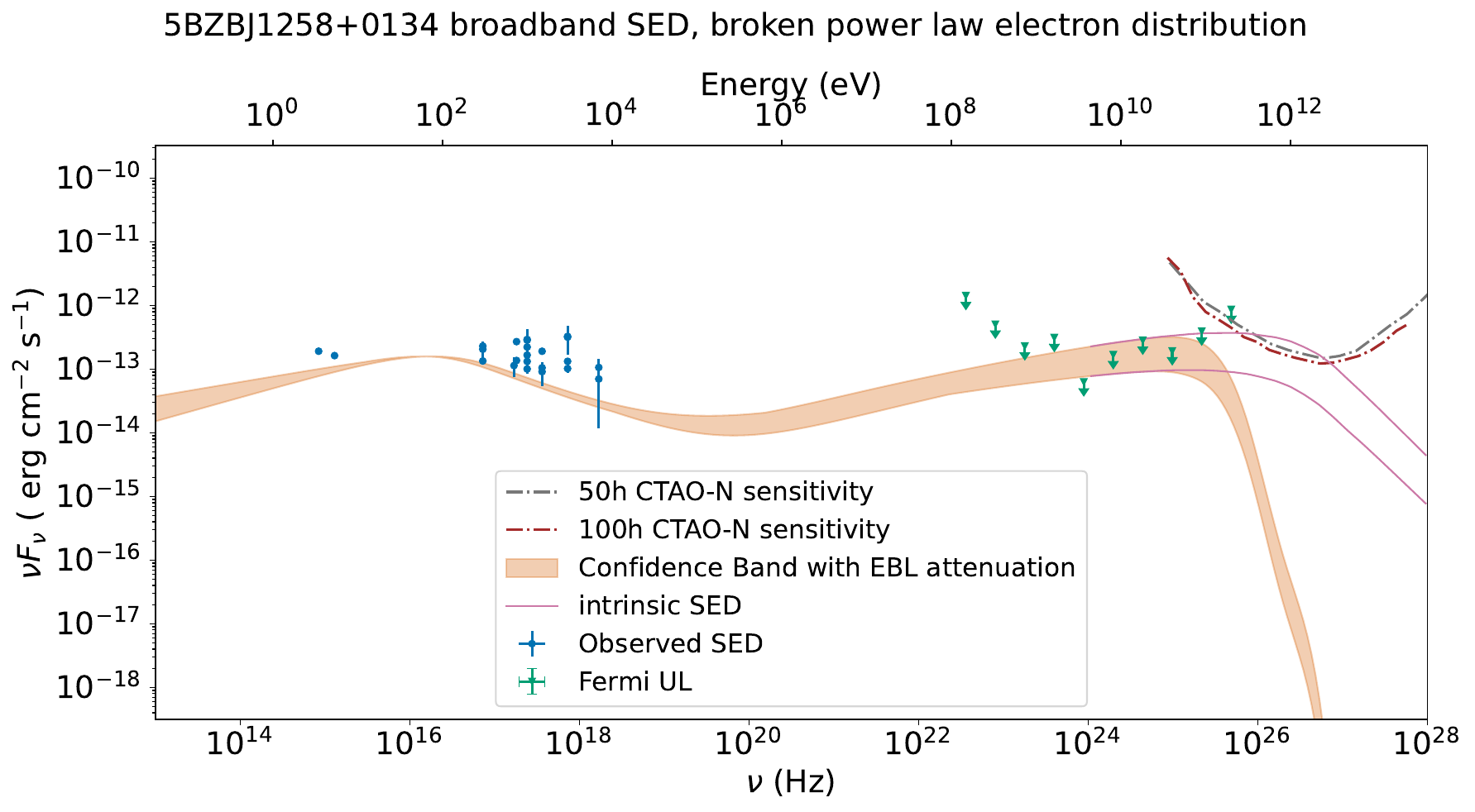} 
\end{minipage}
    \caption{ Broadband spectral energy distribution of 5BZBJ1258+0134  ($z$=0.688, $\gamma_0 \sim 10^4$, $\gamma_{break} \sim 10^5$). The data points in blue are those available from past observations, while the upper limits in green are derived from Fermi data. The regions in orange are the one-zone models assuming on the left a log parabola distribution and on the right a broken power law distribution for the electron population and EBL attenuation according to \citet{dominguez24}. Additionally, pink lines show how the intrinsic models would behave without EBL attenuation, while in grey and red we show respectively the sensitivity of the CTAO-S array in survey mode for a 50-hour and a 100-hour observation at zenith 20$^\circ$.}
    \label{fig:1258mod}
\end{figure*}

The models for the source 5BZBJ1258+0134 are shown in Figure \ref{fig:1258mod} assuming a log parabola or a broken power law electron distribution.

In both models, the synchrotron peak seems to be actually at lower frequencies than those probed by X-ray data, so that it cannot actually be directly observed, and it is hard to assess if the peak flux has been correctly constrained. Still, the post-peak decrease seems to be consistent with both models. The behaviour of the IC peak is harder to constrain, since the $\gamma$-ray upper limits tend to have an unstable behaviour. They are broadly consistent with the lowest flux ends of the modelled regions, with only one upper limit slightly detached from the others. Furthermore, in both models, the EBL attenuation seems to preclude chances of observation with CTAO-N, though, if the log parabola electron distribution model is correct there are some chances that the source could be detected with long-lasted, pointed observations.

Finally, it must be reminded that no information concerning the redshift of this source was available, and therefore it was simply assumed to have the same redshift as the farthest object in our sample (5BZBJ1302+5056, $z= 0.688$). This, combined with the fact that this source is the one with lowest log($\rm \nu_{peak}$)=16.2 among those analysed in this Chapter, makes the estimate of the IC peak less reliable than the ones we obtained for the other sources. Consequently, this source should be given lower priority for a potential follow-up campaign with CTAO, unless a spectroscopic redshift is obtained in the meantime.

\end{document}